\documentclass[longauth]{aa} % for the long lists of affiliations 
\usepackage{graphicx}
\usepackage{txfonts}
\usepackage{lscape}
\usepackage{xcolor}
\usepackage{rotating}
\usepackage{graphicx} 

\usepackage{natbib}
\bibpunct{(}{)}{;}{a}{}{,} % to follow the A&A style

\begin{document}

\title{The Gaia-ESO survey: Calibrating a relationship between Age and the [C/N] abundance ratio with open clusters\thanks{Based on observations collected with the FLAMES instrument at VLT/UT2 telescope (Paranal Observatory, ESO, Chile), for the Gaia- ESO Large Public Spectroscopic Survey (188.B-3002, 193.B-0936).}}
%\subtitle{<your subtitle>}
\author{G. Casali\inst{1,2},
L. Magrini\inst{2},
E. Tognelli\inst{3,4},
R. Jackson\inst{5}, 
R. D. Jeffries\inst{5},
N. Lagarde\inst{6}, 
G. Tautvai\v{s}ien\.{e}\inst{7},
T. Masseron\inst{8,9},
S. Degl'Innocenti\inst{3,4}, 
P. G. Prada Moroni\inst{3,4},  
G. Kordopatis\inst{10}, 
E. Pancino\inst{2,1	1}, 
S. Randich\inst{2}, 
S. Feltzing\inst{12}, C. Sahlholdt\inst{12},  
L. Spina\inst{13}, 
E. Friel\inst{14}, 
V. Roccatagliata\inst{2,4},
N. Sanna\inst{2}, 
A. Bragaglia\inst{15}, 
A. Drazdauskas\inst{7}, 
\v{S}. Mikolaitis\inst{7}, 
R. Minkevi\v{c}i\={u}t\.{e}\inst{7}, 
E. Stonkut\.{e}\inst{7}, 
Y. Chorniy\inst{7}, 
V. Bagdonas\inst{7},   
F. Jimenez-Esteban\inst{16}, 
S. Martell\inst{17,18}, 
M. Van der Swaelmen\inst{2},
G. Gilmore\inst{19},
A. Vallenari\inst{20},  
T. Bensby\inst{12}, 
S. E. Koposov\inst{21}, 
A. Korn\inst{22}, 
C. Worley\inst{19}, 
R. Smiljanic\inst{23},  
M. Bergemann\inst{24}, 
G. Carraro\inst{25}, 
F. Damiani\inst{26}, L. Prisinzano\inst{26}, R. Bonito\inst{26},
E. Franciosini\inst{2},  
A. Gonneau\inst{19}, 
A. Hourihane\inst{19},  
P. Jofre\inst{27},
J. Lewis\inst{19},   
L. Morbidelli\inst{2},  
G. Sacco\inst{2},  
S. G. Sousa\inst{28},
S. Zaggia\inst{20},
A.~C. Lanzafame\inst{29},
U. Heiter\inst{30},
A. Frasca\inst{31},
A. Bayo\inst{32} 
}
\institute{Dipartimento di Fisica e Astronomia, Università degli Studi di Firenze, via G. Sansone 1, 50019 Sesto Fiorentino (Firenze), Italy  \email{casali@arcetri.astro.it} 
\and 
INAF - Osservatorio Astrofisico di Arcetri, Largo E. Fermi, 5, I-50125 Firenze, Italy 
\and
INFN, Sezione di Pisa, Largo Bruno Pontecorvo 3, I-56127 Pisa, Italy
\and
Dipartimento di Fisica, Universit\'a di Pisa, Largo Bruno Pontecorvo 3, I-56127, Pisa, Italy
\and 
Astrophysics Group, Keele University, Keele, Staffordshire ST5 5BG, UK 
\and
Institut UTINAM, CNRS UMR6213, Univ. Bourgogne Franche-Comté, OSU THETA Franche-Comté-Bourgogne, Observatoire de Besan\c{c}on, BP 1615, 25010 Besan\c{c}on Cedex, France
\and
Institute of Theoretical Physics and Astronomy, Vilnius University, Saul\.{e}tekio av. 3, 10257 Vilnius, Lithuania
\and
Instituto de Astrofísica de Canarias, E-38205 La Laguna, Tenerife, Spain
\and
Departamento de Astrofísica, Universidad de La Laguna, E-38206 La Laguna, Tenerife, Spain
\and
Universit\'e C\^{o}te d'Azur, Observatoire de la C\^{o}te d'Azur, CNRS, Laboratoire Lagrange, France
\and
Space Science Data Center - Agenzia Spaziale Italiana, via del Politecnico, s.n.c., I-00133, Roma, Italy
\and
Lund Observatory, Department of Astronomy and Theoretical Physics, Box 43, SE-221 00, Lund, Sweden
\and
Monash Centre for Astrophysics, School of Physics and Astronomy, Monash University, VIC 3800, Australia
\and
Department of Astronomy, Indiana University, Bloomington, IN, USA
\and
INAF - Osservatorio Astronomico di Bologna, via Gobetti 93/3, 40129, Bologna, Italy
\and
Departamento de Astrofísica, Centro de Astrobiología (CSIC-INTA), ESAC Campus, Camino Bajo del Castillo s/n, 28692 Villanueva de la Cañada, Madrid, Spain
\and
School of Physics, University of New South Wales, Sydney, NSW 2052, Australia
\and
Center of Excellence for Astrophysics in Three Dimensions (ASTRO-3D), Australia
\and
Institute of Astronomy, Madingley Road, University of Cambridge, CB3 0HA, UK
\and
INAF-Osservatorio Astronomico di Padova, vicolo Osservatorio 5, 35122 Padova, Italy
\and
McWilliams Center for Cosmology, Department of Physics, Carnegie Mellon University, 5000 Forbes Avenue, Pittsburgh, PA 15213, USA
\and
Observational Astrophysics, Division of Astronomy and Space Physics, Department of Physics and Astronomy, Uppsala University, Box 516, SE-751 20 Uppsala, Sweden
\and
Nicolaus Copernicus Astronomical Center, Polish Academy of Sciences, ul. Bartycka 18, 00-716, Warsaw, Poland
\and
Max-Planck Institute for Astronomy, D-69117, Heidelberg, Germany
\and
Dipartimento di Fisica e Astronomia, Università di Padova, Vicolo dell’Osservatorio 3, 35122 Padova, Italy
\and
INAF - Osservatorio Astronomico di Palermo G. S. Vaiana, Piazza del Parlamento 1, 90134 Palermo, Italy
\and
N\'ucleo de Astronom\'{i}a, Facultad de Ingenier\'{i}a, Universidad Diego Portales, Av. Ej\'ercito 441, Santiago, Chile
\and
Instituto de Astrofisica e ciencias do espac\c o - CAUP, Universidade do Porto, Rua das Estrelas, 4150-762 Porto, Portugal
\and
Dipartimento di Fisica e Astronomia, Sezione Astrofisica, Università di Catania, via S. Sofia 78, 95123, Catania, Italy
\and
Observational Astrophysics, Department of Physics and Astronomy, Uppsala University, Box 516, 75120 Uppsala, Sweden
\and
INAF-Osservatorio Astrofisico di Catania, via S. Sofia 78, 95123, Catania, Italy
\and
Instituto de F\'isica y Astronomi\'ia, Universidad de Valparai\'iso, Chile
}

 \abstract
   { In the era of large high-resolution spectroscopic surveys, like Gaia-ESO and APOGEE, high-quality spectra can contribute to our understanding of the Galactic chemical evolution, providing abundances of elements belonging to the different nucleosynthesis channels, and also providing constraints to one of the most elusive astrophysical quantities, i.e. stellar age. }
   { Some abundance ratios, such as [C/N], have been proven to be excellent indicators of stellar ages. We aim at providing an empirical relationship between stellar ages and [C/N] using, as calibrators,  open star clusters observed by both the Gaia-ESO and APOGEE surveys.   
   }
   {We use stellar parameters and abundances from the Gaia-ESO Survey and APOGEE Survey of the Galactic field and open cluster stars. Ages of star clusters are retrieved from the literature sources  and validated using a common set of isochrones. We use the same isochrones to determine, for each age and metallicity, the surface gravity at which the first dredge-up and red giant branch bump occur.  
   We study the effect of extra-mixing processes in our sample of giant stars, and we derive the mean [C/N] in evolved stars,  including only stars without evidence of extra-mixing. Combining the Gaia-ESO and APOGEE samples of open clusters, we derive a linear relationship between [C/N] and (logarithmic) cluster ages.  }
   { We apply our relationship to selected giant field stars in both Gaia-ESO and APOGEE surveys. 
   We find an age separation between thin and thick disc stars and age trends within their populations, with an increasing age towards lower metallicity populations.}
   {With such empirical relationship, we are able to provide an age estimate for giant stars in which C and N abundances are measured. For giant stars, the isochrone fitting method is indeed less sensitive than for dwarf stars at the turn off.  
   The present method can be thus considered as an additional tool to give an independent estimate of the age of giant stars, with uncertainties in their ages comparable to those obtained using isochrone fitting for dwarf stars.      
    }

\authorrunning{Casali, G. et al.}
\titlerunning{Age-[C/N] relation with Open Clusters}

   \maketitle

\section{Introduction}

  High-resolution spectroscopic surveys, like, e.g., Gaia-ESO \citep{Gil, randich13},  APOGEE \citep{Hol}, GALAH \citep{gala, galah2018} are providing us an extraordinary data-set of radial velocities, stellar parameters and elemental abundances for large samples of stars belonging to different Galactic components: from the thin and thick discs to the halo and bulge, including also many globular and open star clusters. 
Line-of-sight distances and proper motions obtained with the {\em Gaia} satellite \citep[e.g.][]{gaia1, gaia2, gaia3}, coupled with the information from spectroscopic surveys, are improving our knowledge of the spatial distribution of stellar populations,  constraining their properties through our Galaxy \citep[e.g.][]{monari, randich18, Li2018, galah2018}, as well as resolving multiple populations in young clusters \citep[e.g. in Gamma Velorum and Chameleon~I in][respectively]{francio18, roccatagliata18}. 

An important step forward to study how our Galaxy formed and evolved to its present-day structure is the determination of ages of individual stars to disentangle time-scales of 
the formation within the different Galactic components. However, the ages of stars are elusive and they cannot be directly measured. Measuring stellar ages is indeed one of the most difficult tasks of astrophysics \citep[see, e.g.][and references therein]{soderblom14, randich18}. 

The most commonly used technique to compute stellar ages is the comparison of quantities related to observations with the results of stellar evolution models.
This comparison can be done in several {\em planes}, i.e., using directly observed quantities, as magnitudes and colours, and with derived quantities, as surface gravities, log~$g$, and effective temperatures, T$_{\rm eff}$. 
This method provides better results in regions of the planes where isochrones have a good separation, mainly close to the turn-off of the main sequence. On the other hand, isochrones of different ages almost overlap on the red giant branch and on the lower main sequence. Therefore, small uncertainties on T$_{\rm eff}$ and log~$g$, or on magnitude and colour, correspond 
to large uncertainties on the age.
This method is more effective for stars belonging to clusters, for which we can observe several coeval member stars in different evolutionary stages, putting thus stronger constraints to 
the comparison with the isochrones. For instance, the combination of the Gaia-ESO survey  and {\em Gaia}-DR1  data allowed a comparison of the observed sequences of open clusters with stellar evolutionary models, providing an accurate determination of their ages and paving the way to their exploitation as age calibrators to be adopted by other methods \citep{randich18}.  

In addition to the above described classical procedure, there are several alternative methods to estimate stellar ages. For instance, the review of \citet{soderblom14} presented several ways to estimate the ages of young stars, among them the Lithium Depletion Boundary, kinematics ages, pulsation and astroseismology, rotation and activity. 
 Moreover, in the last few years, several chemical indicators have been considered as possible age tracers. Among them there are elemental abundance ratios dependent on the Galactic chemical evolution, such as for instance [Y/Mg], [Ba/Fe] and [Y/Al] \citep[e.g.,][]{tuccimaia16, nissen17, feltzing17, slumstrup17, spina18} and those related to stellar evolution, such as [C/N] \citep{salaris15, masseron15, martig16, ness16, ho17, feuillet18}.

In the present paper, we focus on the use of [C/N] measured in red giant stars as an age indicator. Carbon and nitrogen are processed through the CNO-cycle in previous evolutionary phases and taken towards the surface by the penetration of the convective stellar interior. The phase in which the convection reaches its maximum penetration is called the first dredge-up (FDU, hereafter). As a result of that convective mixing, the atmosphere shows a variation in the chemical composition, changing in particular the abundance ratio [C/N]. Since the penetration of the convection in the inner regions, and therefore the abundances of C and N brought up to in the stellar surface, depends on the stellar mass and the mass is related to the age, then the [C/N] ratio can be used to estimate stellar ages \citep[e.g.,][]{salaris15, lagarde17}.

The aim of the present paper is to calibrate an empirical relation between the [C/N] ratio and stellar age using open clusters..
Starting from the ages determined with isochrone fitting for the open clusters observed by Gaia-ESO and by APOGEE, we calibrate a relationship between cluster age and the [C/N] ratio in their evolved stars, accurately selecting them among post-FDU stars and studying the occurrence of non-canonical mixing.  
In Sec.~\ref{sec_data}, we present our datasets, describing our pre-selection of the Gaia-ESO and APOGEE catalogues and the sample of open clusters observed in both surveys. In Sec.~\ref{sec_criteria} we discuss the choice of the evolved stars to be used to compute [C/N], and in Sec.~\ref{sec_rel} we present the relationship between cluster ages and [C/N] abundances. 
In Sec.~\ref{sec_theory} we compare our relationship with theoretical predictions. 
The application of the relationship to the field stars of Gaia-ESO and APOGEE high-resolution samples is instead analysed in Sec.~\ref{sec_application}. 
Finally, in Sec.~\ref{sec_conclusions} we draw our summary and conclusions.

\section{The data samples}
\label{sec_data}

\subsection{The Gaia-ESO sample}
The Gaia-ESO survey \citep[][hereafter GES]{Gil,randich13} is a high-resolution spectroscopic survey observing about 10$^5$ stars whose spectra were collected with FLAMES (Fiber Large Array Multi-Element Spectrograph) multi-fiber facility \citep{pasquini02} at VLT (Very Large Telescope). It has two different observing modes, using GIRAFFE, the medium-resolution spectrograph (R$\sim$20000), and UVES, the high-resolution spectrograph (R$\sim$47000).\\
The data were reduced as described in \citet{sacco14} and Gilmore et al. (in preparation) for UVES and GIRAFFE, respectively.
The stellar atmospheric parameters of the stars considered in the present work, UVES FGK stars,  were determined as described in \citet{smi14}.   
The calibration and homogenisation of stellar parameters and abundances obtained by the different working groups were performed as described by \citet{pancino17} and Hourihane et al. (in preparation), respectively.
All data used in the present work are included in the fourth and fifth internal GES data releases ({\sc idr4} and {\sc idr5}). Since the C and N abundances of the stars in common to both releases were not re-derived in {\sc idr5}, we adopt the C and N values present in {\sc idr4} for those stars\footnote{Hereafter we will not specify the GES release because we used the {\sc idr5}, but where for stars in common we adopted the C and N abundances of {\sc idr4}, not present in {\sc idr5}}. 
In particular, the method used for deriving the abundances discussed in the present paper, carbon and nitrogen\footnote{C and N are derived by one of the Nodes of analysis, the Vilnius Node}, is described in \citet{tau15}. 
 Here, we briefly recall the main step of the analysis. C and N abundances were derived from molecular lines of C$_{2}$ \citep{broke13} and CN respectively \citep{sneden14}. 
In the analysis of the optical stellar spectra, the $^{12}$C$^{14}$N molecular bands in the spectral range 6470$-$6490 \AA, the C$_{2}$ Swan (1,0) band head at 5135 \AA, the C$_{2}$ Swan (0,1) band head at 5635.5 \AA\, are used. 
The selected C$_{2}$ bands do not suffer from non-local thermodynamical equilibrium (NLTE) deviations, and thus are better suited for abundance studies than [C I] lines.
All molecular bands and atomic lines are analysed through spectral synthesis with the code BSYN \citep{tau15} and all synthetic spectra have been calibrated to the solar spectrum of \citet{kurucz05}.\\

The selected clusters are shown in Table~\ref{tab:clustersGES}, where we summarise their basic properties from the literature: coordinates, Galactocentric distances, heights above the plane, mean RV of cluster members, median metallicity, ages and the references for ages and distances.

We make a first quality check on the GES catalogue, excluding from our sample stars with highly uncertain stellar parameters (i.e., including only those stars with T$_{\rm eff}>0$, $e$T$_{\rm eff} < 500$, log~$g > -0.5$, $e$log~$g < 0.5$) and with signal-to-noise ratio (SNR)$<$20. 
We select giant stars in open clusters and in the Galactic field. 
We perform a membership analysis of stars in clusters with a Bayesian approach, taking into account both GES and {\em Gaia} information.
Membership probabilities are estimated from the radial velocities RV (from GES) and proper motion velocities (from {\em Gaia}) of stars observed with GIRAFFE. 
A maximum likelihood method is used to determine the probability of cluster
membership by fitting two quasi-Gaussian distributions to the velocity data. 
The basic supposition is that the set of stars that we observed as potential
cluster members  are draw from two populations. 
One distribution defines cluster members, whereas a second much broader
distribution represents the background population. 
These two populations show
different velocity distributions, the first one has velocities close
to the average velocity of the cluster, the second one shows a much wider
range of velocities.
The mean value and standard deviation of the background population are fixed for the final
calculation of cluster membership.
If substructure is evident in the velocity distributions then the cluster is
simultaneously fitted with two independent populations and targets
are assigned a membership probability for each population. Most clusters, especially the older ones, present a single population. More details on the membership estimation are in Jackson et al. (in prep.). For our analysis, we select stars with a minimum membership
probability of 0.8 giving an average probability P$_{memb.}^{mean}$ for cluster members, shown in Table~\ref{tab:cnall}.

\begin{table*}[h]
\caption{Parameters of the open clusters in the GES sample}
\begin{center}
\tiny{
\begin{tabular}{lcccccclc}
\hline
Open Clusters & R.A.$^{a}$ & Dec.$^{a}$ & R$_{GC}$ & z & RV$^{b}$ &[Fe/H]$^{a}$ & log(Age[yr])&  Ref. Age \& Distance\\ 
              & (J2000) &&    (kpc) & (pc) &(km s$^{-1}$) &(dex) &  &  \\
\hline
Berkeley~31      & 06:57:36 & +08:16:00 & 15.16$\pm$0.40 & +340$\pm$30         & $+56.99\pm0.14$    &$-$0.27$\pm$0.06      & $9.40_{-0.05}^{+0.05}$  &     \citet{cign11} \\ 
Berkeley~36      & 07:16:06 & $-$13:06:00 & 11.30$\pm$0.20  & $-$40$\pm$10     & $+62.70\pm0.13$    &$-$0.16$\pm$0.10    & $9.84_{-0.03}^{+0.03}$  &    \citet{donati12}\\ [2pt]
Berkeley~44      & 19:17:12 & +19:33:00 & 6.91$\pm$0.12 & +130$\pm$20          & $-8.71\pm0.19$     &+0.27$\pm$0.06    &  $9.20_{-0.09}^{+0.07}$  & \citet{jacobson16} \\ [2pt]
Berkeley~81      & 19:01:36 & $-$00:31:00 & 5.49$\pm$0.10 & $-$126$\pm$7       & $+48.14\pm0.26$    &+0.22$\pm$0.07     &  $8.93_{-0.05}^{+0.05}$ &    \citet{magrini15} \\
M67                   & 08:51:18 & +11:48:00 & 9.05$\pm$0.20 & +405$\pm$40     & $+34.7\pm0.9^{a}$      &$-$0.01$\pm$0.04    & $9.63_{-0.05}^{+0.05}$   &   \citet{salaris04}\\
Melotte~71       & 07:37:30 & $-$12:04:00 & 10.50$\pm$0.10 & +210$\pm$20       & $+50.8\pm1.3^{a}$      &$-$0.09$\pm$0.03   & $8.92_{-0.11}^{+0.08}$   &   \citet{salaris04}\\ [2pt]
NGC~2243          & 06:29:34 & $-$31:17:00 & 10.40$\pm$0.20 & +1200$\pm$100    & $+59.65\pm0.05$    &$-$0.38$\pm$0.04 & $9.60_{-0.15}^{+0.11}$   & \citet{bratosi06}\\ [2pt]
NGC~6005          & 15:55:48 & $-$57:26:12 & 5.97$\pm$0.34 & $-$140$\pm$30     & $-24.75\pm0.35$    &+0.19$\pm$0.02    & $9.08_{-0.12}^{+0.10}$   &   \citet{piatti98}\\ [2pt]
NGC~6067          & 16:13:11 & $-$54:13:06 & 6.81$\pm$0.12 & $-$55$\pm$17      & $-38.65\pm0.28$    & +0.20$\pm$0.08    & $8.00_{-0.30}^{+0.18}$   &   \citet{alsant}\\ [2pt]
NGC~6259         & 17:00:45 & $-$44:39:18 & 7.03$\pm$0.01 & $-$27$\pm$13       & $-32.98\pm0.49$    & +0.21$\pm$0.04    &  $8.32_{-0.07}^{+0.06}$   &   \citet{mer01}\\ [2pt]
NGC~6705          & 18:51:05 & $-$06:16:12 & 6.33$\pm$0.16 & $-$95$\pm$10      & $+35.67\pm0.19$    & +0.16$\pm$0.04    & $8.48_{-0.05}^{+0.04}$   &   \citet{cantat14}\\ [2pt]
NGC~6802          & 19:30:35 & +20:15:42 & 6.96$\pm$0.07 & +36$\pm$3           & $+13.44\pm0.60$    & +0.10$\pm$0.02     & $9.00_{-0.10}^{+0.08}$   &    \citet{jacob} \\ [2pt]
Rup~134           & 17:52:43 & $-$29:33:00 & 4.60$\pm$0.10 & $-$100$\pm$10     & $-40.78\pm0.15$    & +0.26$\pm$0.06  & $9.00_{-0.01}^{+0.01}$   &  \citet{carraro06}\\
Pismis~18         & 13:36:55 & $-$62:05:36 & 6.85$\pm$0.17 & +12$\pm$2         & $-27.94\pm0.29$    & +0.22$\pm$0.04    & $9.08_{-0.06}^{+0.05}$   &    \citet{piatti98}\\ [2pt]
Trumpler~23      & 16:00:50 & $-$53:31:23 & 6.25$\pm$0.15 & $-$18$\pm$2        & $-60.74\pm0.38$    & +0.21$\pm$0.04     & $8.90_{-0.06}^{+0.05}$   &  \citet{jacob} \\ [2pt]
NGC~4815          & 12:57:59 & $-$64:57:36 & 6.94$\pm$0.04 & $-$95$\pm$6       & $-29.53\pm0.21$    & +0.11$\pm$0.01     & $9.18_{-0.04}^{+0.04}$  & \citet{friel14}\\
Trumpler~20      & 12:39:32 & $-$60:37:36 & 6.86$\pm$0.01 & +134$\pm$4         & $-39.82\pm0.14$    & +0.15$\pm$0.07    & $8.48_{-0.08}^{+0.07}$   & \citet{donati14}\\
 \hline
\end{tabular}
}
\tablefoot{Ref: $^{a}$\citet{magrini18}. $^{b}$ Radial velocities (RVs)  are determined with the GIRAFFE sample of high probability cluster members.}
\end{center}
\label{tab:clustersGES}
\end{table*}

\subsection{The APOGEE sample}
The Apache Point Observatory Galactic Evolution Experiment  \citep[APOGEE,][]{zasowski13,majewski17} is a large-scale high-resolution and high-signal-to-noise spectroscopic survey in the near infrared (H-band) that observed 
over 10$^5$ giant stars in the bulge, bar, discs, and halo of our Galaxy.
The first set of observations of the APOGEE Survey was executed from September 2011 to July 2014  with a spectral resolution R$\sim$ 22500 on the Sloan Foundation 2.5m Telescope of Apache Point Observatory (APO). 
The second instalment will continue the data acquisition at APO until summer 2020. Other observations will be taken with the APOGEE-South spectrograph on the Irénée du Pont 2.5m Telescope of Las Campanas Observatory. In the present work, 
we use the {\sc dr14} release\footnote{\url{https://www.sdss.org/dr14/irspec/spectro_data/}}, which is the second data release of the fourth phase of the Sloan Digital Sky Survey \citep[SDSS-IV][] {york00,blanton17}.
The APOGEE {\sc dr14} spectra were reduced with the pipeline ASPCAP \citep[APOGEE Stellar Parameter and Chemical Abundances Pipeline,][]{garcia2016}. The {\sc dr14} release contains data for approximately 263,000 stars.\\ 
As for the GES sample, we select stars based on the quality of their measurements, according to the following criteria: T$_{\rm eff}>0$, $e$T$_{\rm eff} < 500$, log~$g > -0.5$, $e$log~$g < 0.5$ and SNR  $> 20$.

To identify the open clusters in the APOGEE catalogue, we start our search by defining a master list of known open clusters. We used the \citet{khar13} catalogue, where coordinates and angular radii of clusters are listed. 
 For most of the cluster in our list, a cluster membership has been recently published \citep{donor18}. For those clusters, we thus select the member stars of \citet{donor18}, whereas for the four clusters that were not available, we compute the membership adopting a similar approach as in \citet{donor18}.  
We perform the membership selection by retaining all stars within a circle with a radius three times 
the reference radius and centred on the cluster coordinates. 
We select as member stars those within $\pm 2 \sigma$ from the mean metallicity and radial velocity RV. 
The open clusters selected for the present work are shown in Table ~\ref{tab:clustersAPO}, where we give the following parameters: coordinates, galactocentric distances, heights from the Galactic plane, distances from the Sun, mean RV,  metallicities, and logarithmic ages.

The C and N abundances are computed within ASPCAP, which compares the observed spectra to a grid of synthetic spectra to determine stellar parameters. A $\chi^{2}$ minimisation finds the best fit spectrum, and the corresponding stellar parameters are assigned to the observed star. The abundances of carbon and nitrogen are measured from molecular lines of CO and CN \citep{garcia2016}.  \\

 \begin{table*}[h]
 \caption{Open clusters in the APOGEE sample.} 
 \tiny{
\begin{center}
\begin{tabular}{lcccccclc}
\hline
Open Clusters & R.A.$^{f}$ &Dec.$^{f}$  &   R$_{GC}$ &  z     &        RV               & [Fe/H]$^{b}$      &    log(Age[yr])    &     Ref. Age \\       
              &  (J2000)   &  (J2000)   &    (kpc) & (pc)  &       (km s$^{-1}$)        & (dex)             &                    &                       \\
\hline
 Berkeley 17  &  05:20:37  & +30:35:24  &    9.79  &$-$114  &    $-$73.7$\pm$0.8$^{a}$   &   $-0.11\pm0.03 $ &    $9.98_{-0.05}^{+0.05}$  &     \citet{friel05}\\ 
 Berkeley 53  &  20:55:57  & +51:03:36  &    8.66  &  216   &      7.5$^{b}$               &   $+0.00\pm0.02 $ &    $9.09_{-0.02}^{+0.02}$   &     \citet{maciej09}\\
 Berkeley 66  &  03:04:04  & +58:44:24  &   14.07  &   22   &     $-$50.6$\pm$0.2$^{c}$   &   $-0.13\pm0.02 $ &    $9.54_{-0.12}^{+0.12}$   &     \citet{phelps96}\\ 
 Berkeley 71  &  05:40:57  & +32:15:58  &   11.25  &   51   &    $-$25.5$\pm$6.0$^{d}$   &   $-0.20\pm0.03 $ &    $9.00_{-0.15}^{+0.15}$   &     \citet{mn07}\\
 FSR0494      &  00:25:41  & +63:45:00  &   11.43  &   91   &    $-$63.3$\pm$1.5$^{b}$   &   $+0.01\pm0.02 $ &    $8.70_{-0.10}^{+0.08}$   &     This work \\ [2pt]
 IC166        &  01:52:23  & +61:51:54  &   11.68  &$-$13   &      $-$40.5$\pm$1.5$^{b}$   &   $-0.06\pm0.02 $ &    $9.00_{-0.10}^{+0.08}$ & \citet{schiappacasse18}\\
 King 5       &  03:14:46  & +52:41:49  &    9.85  &$-$164  &      $-$52.0$\pm$12.0$^{e}$  &   $-0.11\pm0.02 $ &    $9.10_{-0.10}^{+0.10}$   &     \citet{maciej07}\\
 King 7       &  03:59:07  & +51:46:55  &   10.35  &$-$47   &      $-$11.9$\pm$2.0$^{b}$   &   $-0.05\pm0.02 $ &    $8.82_{-0.07}^{+0.06}$ & \citet{dias02}\\
 NGC~188       &  00:47:24  & +85:15:18  &    9.13  &  761   &    $-$42.4$\pm$0.1$^{l}$   &   $ 0.14\pm0.01 $ &    $9.87_{-0.04}^{+0.04}$   &     \citet{fornal07}\\ 
 NGC~1193      &  03:05:53  & +44:22:48  &   13.30  &$-$1264 &      $-$82.0$\pm$0.39$^{f}$  &   $-0.22\pm0.02^{g}$  & $9.70_{-0.10}^{+0.10}$   &     \citet{kye08}\\ 
 NGC~1245      &  03:14:48  & +47:15:11  &   10.60  &$-$464  &     $-$29.7$\pm$1.1$^{h}$   &   $-0.06\pm0.02 $ &    $8.95_{-0.05}^{+0.05}$  &     \citet{subramaniam03}\\ 
 NGC~1798      &  05:11:38  & +47:41:42  &   13.05  &  443   &      $-$2.0$\pm$10.0$^{i}$   &   $-0.18\pm0.02 $ &    $9.15_{-0.09}^{+0.09}$   &     \citet{park99}\\
 NGC~2158      &  06:07:26  & +24:05:31  &   12.75  &  148   &       26.9$\pm$1.9$^{h}$     &   $-0.15\pm0.02 $ &    $9.28_{-0.11}^{+0.11}$   &     \citet{salaris04}\\ 
 NGC~2420      &  07:38:23  & +21:34:01  &   10.61  &  967   &      73.6$\pm$0.2$^{m}$     &   $-0.12\pm0.02 $ &    $9.47_{-0.17}^{+0.17}$   &     \citet{pancino10}\\ 
 NGC~2682 (M67)&  08:51:23  & +11:48:54  &    8.63  &  470   &       33.6$\pm$0.1$^{n}$     &   $-0.05\pm0.03 $ &    $9.63_{-0.05}^{+0.05}$   &     \citet{salaris04}\\ 
 NGC~6705      &  18:50:59  & -06:16:48  &    6.50  &  $-$84   &     35.1$\pm$0.3$^{m}$     &   $+0.16\pm0.02 $ &    $8.48_{-0.07}^{+0.07}$   &     \citet{cantat14}\\ 
 NGC~6791      &  19:20:53  & +37:46:48  &    7.80  &  932   &      $-$47.4$\pm$0.1$^{o}$   &   $+0.42\pm0.05 $ &    $9.90_{-0.02}^{+0.02}$   &     \citet{wu14}\\ 
 NGC~6811      &  19:37:22  & +46:23:42  &    7.86  &  255   &     6.7$\pm$0.1$^{p}$     &   $-0.01\pm0.02 $ &    $9.00_{-0.07}^{+0.07}$  &     \citet{janes13}\\ 
 NGC~6819      &  19:41:17  & +40:11:42  &    7.69  &  348   &      2.3$\pm$0.1$^{q}$     &   $+0.11\pm0.03 $ &    $9.28_{-0.02}^{+0.02}$   &     \citet{wu14}\\ 
 NGC~6866      &  20:03:57  & +44:09:36  &    7.88  &  158   &   13.7$\pm$0.1$^{m}$     &   $+0.07\pm0.07^{r}$  &$8.85_{-0.10}^{+0.10}$   &     \citet{janes14}  \\
 NGC~7789      &  23:57:25  & +56:43:48  &    8.91  &$-$168  &     $-$54.7$\pm$1.3$^{h}$   &   $+0.05\pm0.03 $ &    $9.21_{-0.12}^{+0.12}$   &     \citet{pancino10}\\ 
 Teutsch 51   &  05:53:50  & +26:49:48  &   11.78  &   31   &    2.7$^{b}$               &   $-0.28\pm0.02 $ &    $8.90_{-0.10}^{+0.10}$   &     \citet{dias02}\\ 
 Trumpler 5   &  06:36:29  & +09:28:12  &   10.59  &   49   &      $-$49.7$\pm$1.9$^{s}$   &   $-0.40\pm0.01^{s}$ &$9.45_{-0.04}^{+0.04}$   &     \citet{kim09}\\ 
\hline
\end{tabular}
\end{center}
}

\tablebib{$^{a}$\citet{friel05}, $^{b}$\citet{donor18}, $^{c}$\citet{villa05}, $^{d}$\citet{zhang15}, $^{e}$\citet{friel02}, $^{f}$\citet{khar13}, $^{g}$\citet{neto16}, $^{h}$\citet{jacob11}, $^{i}$\citet{carr12}, $^{l}$\citet{gao14}, $^{m}$\citet{merm08}, $^{n}$\citet{gell15}, $^{o}$\citet{toff14}, $^{p}$\citet{molenda14}, $^{q}$\citet{dias02}, $^{r}$\citet{frasca16}, $^{s}$\citet{donati15}. \label{tab:clustersAPO}}
\end{table*}

\subsection{The age scale of open clusters} 
Star clusters offer the unique opportunity with respect to field stars since they have well-determined ages through isochrone fitting thanks to the many members observed across the cluster sequence. Thus they are powerful tools to calibrate relations between stellar ages and chemical properties of stars. 
 
A large sample of clusters is observed in the GES and APOGEE surveys, covering sizable ranges in ages, distances and metallicities. 
The latest available GES data release, {\sc idr5}, contains 38 Galactic open clusters, and in 15 of them 
red giant stars have been observed, while there are 23 clusters in APOGEE with red giant stars (see Tables~\ref{tab:clustersGES} and \ref{tab:clustersAPO}).

In the present work, we adopt cluster ages from the recent literature. 
Since a reliable and homogeneous age determination of star clusters is important for our project, 
we check for consistency the literature ages against a common set of isochrones. 

We compare the log~$g$-T$_{\rm eff}$ diagrams of member stars of GES and APOGEE clusters with the PISA isochrones \citep[for details see][]{dellomo12,tognelli18}, corresponding to ages and metallicities in Tables~\ref{tab:clustersGES} and \ref{tab:clustersAPO}. 

For the open clusters in the GES sample we also add stars observed with GIRAFFE, mainly in the main sequence for a  better characterisation of the cluster sequence. The agreement between the literature ages and metallicities and the corresponding isochrones is remarkably good, since most of them were recently re-determined by the Gaia-ESA consortium in a homogeneous way \citep[see, e.g.,][]{donati14, friel14, jacobson16, magrini17, overbeek17, tang17, randich18}. 
For the APOGEE sample, most of the clusters are in agreement with the PISA isochrones computed with the literature ages and metallicities. 
We only redetermine the age of FSR0494, for which the PISA isochrone for the age given by
\citet{khar13} is inconsistent with the present data. The new age is listed in 
Table~\ref{tab:clustersAPO}.
In the left panels of Fig.~\ref{fig:lithium1}  in Sec.~\ref{sec_criteria} and in Figs.~\ref{fig:lithium:all1}, \ref{fig:lithium:all2}, \ref{fig:lithium:all3} in the Appendix, 
we show the log~$g$-T$_{\rm eff}$ diagrams of member stars of each GES cluster with the corresponding PISA isochrones. Instead, the colour-magnitude diagrams of NGC~4815, NGC~6705 and Trumpler~20 are shown in \citet{tau15}.
In Fig~\ref{fig:apo} and in Figs.~\ref{fig:apo:all1}, \ref{fig:apo:all2}, \ref{fig:apo:all3}, \ref{fig:apo:all4}, \ref{fig:apo:all5}, (left panels) the log~$g$-T$_{\rm eff}$ diagrams of APOGEE member stars and the PISA isochrones are shown.

\section{The selection criteria}
\label{sec_criteria}

To build-up a relationship between the cluster [C/N] abundance ratios and their ages, 
it is necessary to select member stars in the red-giant branch (RGB) that have already modified their carbon and nitrogen surface abundances, passing through the FDU event. 

\subsection{The selection of red giant stars beyond the FDU}

We derive a quantitative criterion to select among the member stars of our sample open clusters those which have passed the FDU. To identify the location in the log~$g$-T$_{\rm eff}$ diagram at which the FDU occurs for clusters of different ages and metallicities, in particularly the surface gravity log~$g$, we use the same set of isochrones employed to check cluster ages. We consider isochrones in the age and metallicity ranges of our clusters, i.e., with [Fe/H] from $-0.4$ to $+0.4$ dex and with ages from 0.1 to 10~Gyr. For these isochrones, we obtain the surface abundances of N and C.

The surface gravity that corresponds to the FDU has been identified by selecting a proper value where the [C/N] starts to monotonically decrease. Indeed, during the FDU the surface [C/N] abundance progressively decreases as surface convection reaches deeper and deeper regions inside the star. We tried two different approaches: 1) the point where [C/N] starts to monotonically decrease or 2) the point where the derivative $d[C/N]/d\log g$ is maximum. We tested these two criteria on clusters with ages larger than 2~Gyr finding that the difference in the location of the FDU in the log~$g$ plane is quite negligible for our purposes (about 0.1~dex).

On the other hand, the second method (the point where $d[C/N]/d \log g$ is maximum) is not suitable for clusters with ages below 500~Myr/1~Gyr. The reason is that for such ages, the extension of the convective envelope and consequently the surface [C/N] abundance changes slowly with log~$g$, causing a significant difference between the $\log g$ at FDU estimated using the two methods.

\begin{figure}[h]
\center
\includegraphics[scale=0.5]{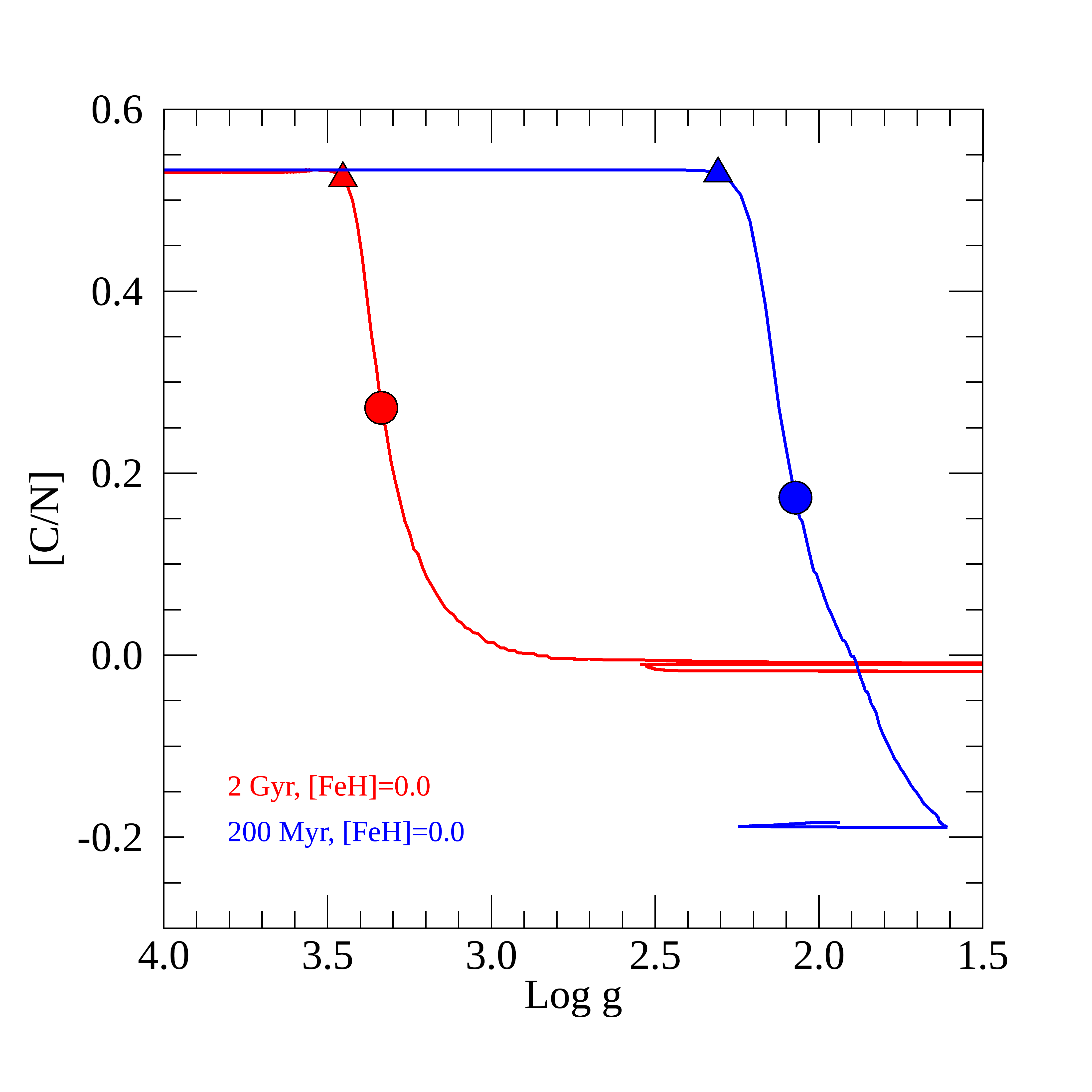}
\caption{[C/N] vs log~$g$ of 200~Myr (red line) and 2~Gyr (blue line) [Fe/H]=$0.0$~dex PISA isochrones. The triangle represents the point where [C/N] starts to monotonically decrease, while the circle represents the point where the derivative $d[C/N]/d\log g$ is maximum. \label{fig:isoFDU}}
\end{figure}

Figure~\ref{fig:isoFDU} shows, as an example, the variation of [C/N] vs $\log g$ for isochrones of 200~Myr and 2~Gyr (for solar metallicity). For the 2~Gyr case the interval of $\log g$ where [C/N] starts/stops decreasing is $\Delta \log g \approx 0.3$-0.4~dex, while for young clusters the interval is 
larger, of about 0.6-0.7~dex. The maximum of the derivative $d[C/N]/d \log g$ corresponds to about the median point of the interval. Given the large interval of $\log g$ during the giant branch in young clusters, the adoption of the maximum derivative underestimates $\log g$ at the FDU by about 0.3~dex with respect to the alternative criterion. 
For this reason, we decided to use the first approach and, therefore, identify the $\log g$ FDU as the point where [C/N] starts to monotonically decrease, in all the analyzed clusters. 

\begin{figure}[h]
\centering
\includegraphics[scale=0.50]{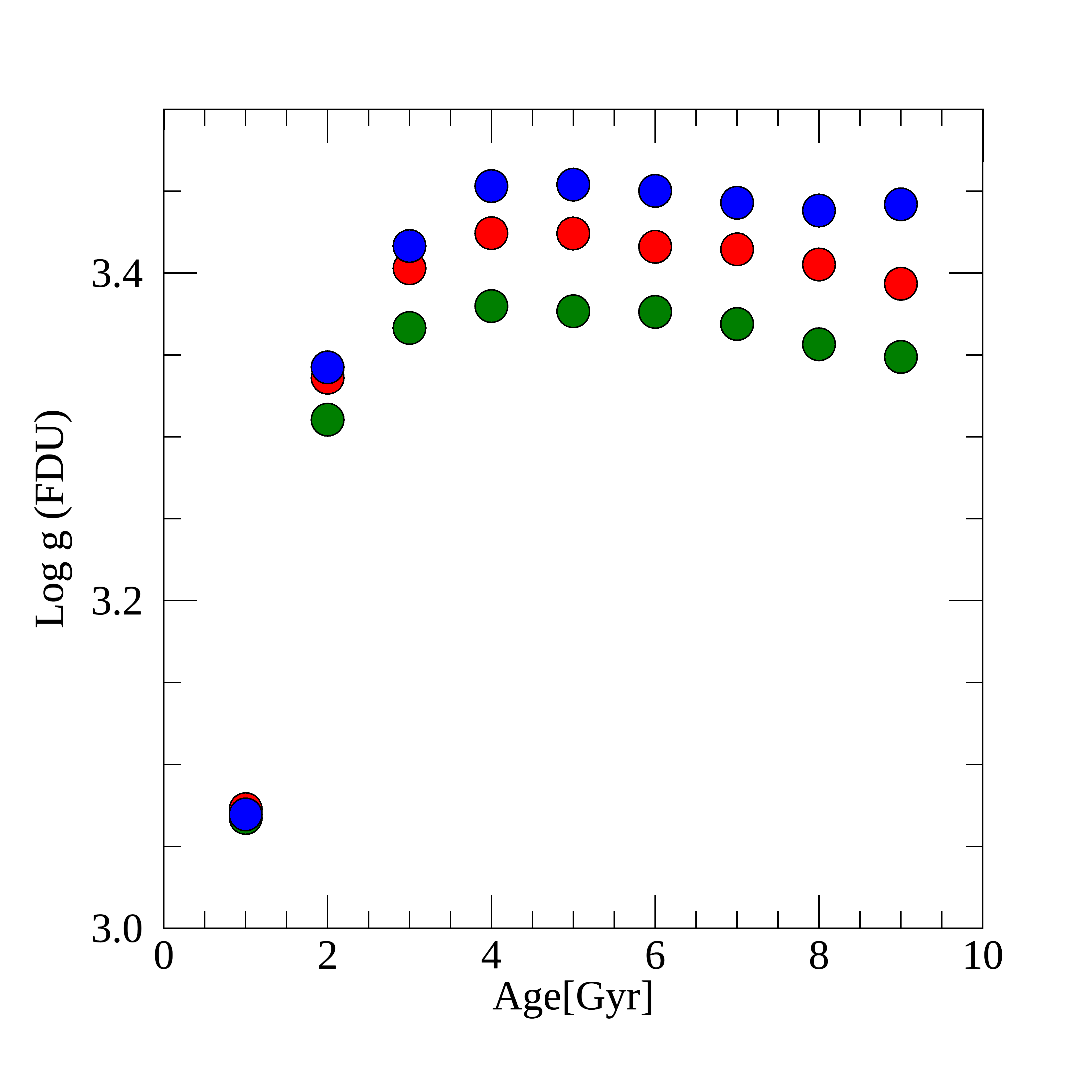}  
\caption{Theoretical log~$g$ at the FDU as a function of age and metallicity. Three different colours indicate three different metallicities: [Fe/H]=0 (red), [Fe/H]=$-$0.15 (green) and [Fe/H]=+0.15 (blue). \label{fig:loggdred}}
\end{figure}

For each cluster in both GES and APOGEE samples, the values of log~$g$ corresponding to the FDU are shown in Table ~\ref{tab:loggcut}. 
To estimate the uncertainties in log~$g$ at which the FDU happens, we consider how the uncertainties on the age and metallicity reflect on the position of the FDU in the age vs. log~$g$(FDU) plane, as shown in Figure~\ref{fig:loggdred}.  
At younger ages, the uncertainty on the location of the FDU is mainly due to the uncertainties on the age. 
For the youngest clusters, with age $<$ 1~Gyr, the uncertainty in the determination of the log~$g$ of the FDU can be very large, since the log~$g$ range in which the FDU occurs is wide ($\sim$1~dex in log~$g$). 
Typical uncertainties of clusters with ages of about 1 Gyr ranges from 0.1 to 0.3 Gyr (see Tables~1 and 2), thus implying an uncertainty in the determination of the log~$g$ of the FDU of $\sim$0.05~dex. For the oldest clusters, the main variation in the FDU position is related to the uncertainty in metallicity: from 2 to 9 Gyr, a variation of [Fe/H] of $\pm$0.15~dex\footnote{a typical 2 or 3-$\sigma$ uncertainty on cluster [Fe/H]} implies a change in the FDU log~$g$ from 0.05 to 0.1~dex, respectively. 
To select stars which have passed the FDU, we take into account both the uncertainties on the estimation of the theoretical position of the FDU and on the derived spectroscopic gravities. These uncertainties are added to FDU log~$g$ in order to have a lower limit in the selection of stars beyond the first dredge-up. 

 \begin{table}[h]
 \caption{Surface gravity, log~$g$, at FDU for the open clusters in the GES and APOGEE samples.} 
\begin{center}
\begin{tabular}{lc}
\hline
Open Clusters &         log~$g$ at FDU   \\
              &         (dex)       \\  
\hline  
 GES clusters & \\
 \hline
  Berkeley 31      &      3.5\\
  Berkeley 36      &      3.5\\
  Berkeley 44      &      3.4\\
  Berkeley 81      &      3.1\\
  Melotte~71       &      3.1\\
  NGC~2243         &      3.5\\
  NGC~6005         &      3.3\\
  NGC~6067   &      2.1  \\ 
  NGC~6259   &      2.4 \\ 
  NGC~6802         &      3.2    \\
  Rup134           &      3.2      \\
  Pismis~18        &      3.3  \\
  Trumpler~23      &      3.1 \\
  \hline \hline
APOGEE clusters &\\
\hline
 Berkeley~17     &   3.5       \\
 Berkeley~53     &   3.3       \\
 Berkeley~66     &   3.5       \\
 Berkeley~71     &   3.2       \\
 FSR0494         &   2.8 \\
 IC166           &   3.2\\
      King~5     &   3.3       \\
      King~7     &   3.0\\
    NGC~1193     &   3.5       \\
    NGC~1245     &   3.2       \\
    NGC~1798     &   3.3       \\
     NGC~188     &   3.6       \\
    NGC~2158     &   3.4       \\
    NGC~2420     &   3.5       \\
    NGC~6791     &   3.6       \\
    NGC~6811     &   3.2       \\
    NGC~6819     &   3.4       \\
    NGC~6866     &   3.1       \\
  Teutsch~51     &   3.1       \\
  Trumpler~5     &   3.4       \\
     NGC~7789    &   3.4       \\
     \hline \hline
GES and APOGEE clusters &\\  
\hline
M67      &      3.4 \\
NGC~6705     &   2.6       \\ 
\hline \hline
\end{tabular}
\end{center}
\label{tab:loggcut}
\end{table}

\subsection{C and N variations beyond the FDU: the effect of extra-mixing}

Selecting stars which have passed the FDU defines indeed a broad class of giant stars: 
lower-RGB stars, i.e. stars just after the phase of sub-giant, at the beginning of the RGB; 
stars in the red clump (RC), and stars in the upper-RGB. 
After the FDU, star evolves along the RGB where more mixing can occur, further decreasing the C/N ratio \citep[see, for instance][]{gratton00, martell08} due to non-canonical mixing possibly driven by thermohaline mechanisms \citep{lagarde12}. These effects are mainly expected along the upper-RGB, after the RGB bump phase, and they are more important at low metallicity and for low-mass stars \citep[see the lower panel of Fig.1 of][]{masseron15,shetrone19}.   
On the other hand, at solar metallicity and for massive stars, the effect is quite small and almost negligible compared to the precision of measured C/N, unless other non-canonical extra-mixing processes modified them in specific stars. 
\citet{masseron17} confirmed the universality of extra-mixing along the upper part of the RGB for low mass stars. They found a N depletion between the RGB tip and the He-burning phase/clump evolutionary stages, likely due to the mixing with the outer envelope during the He flash. 
This effect is particularly appreciable and strong at low metallicity, while at solar metallicity
RGB and RC stars present similar N abundances \citep{shetrone19}.

Observational evidences of mixing processes happening during the RGB-bump phase  have been the detection of a further depletion of Li abundance along the RGB in globular clusters \citep[e.g.,][]{lind09, mucciarelli11}.  
During the bump, indeed, the thermohaline convection is expected to be more efficient and to rapidly transport surface Li in the internal hotter regions where it is destroyed. The same mechanism can modify C and N abundances. 

Open clusters usually do not reach very low metallicities and thus are not expected to
show strong extra mixing. Nevertheless, thanks to the GES data, we can
check cluster by cluster whether this extra mixing occur by looking at  Li abundance, A(Li)=$\log$(Li/H), and [C/N]. Then we can safely exclude the stars with strong
extra-mixing with lower [C/N] and which would affect our age
determination. In our sample, the lowest metallicities are those of Br~31 ([Fe/H]=$-$0.27$\pm$0.06) and NGC~2243 ([Fe/H]=$-$0.38$\pm$0.04) in the GES sample and of Trumpler~5 ([Fe/H]=$-$0.40$\pm$0.01) and Teutsch~51 ([Fe/H]=$-$0.28$\pm$0.03) in the APOGEE sample. 
As shown in Fig.1 of \citet{masseron15}, [C/N] from lower RGB to RC should be unchanged at solar metallicity, while it is expected to be modified in the upper-RGB, but at lower metallicity. 
Although from a theoretical point of view, we do not expect strong extra-mixing effects in the metallicity range of open clusters, we
can use Li abundance available in the GES sample to check for correlations between Li, [C/N] abundances and the evolutionary phase of member stars of the same clusters. We can thus identify possible extra-mixing processes in a metallicity range so far poorly investigated to this respect.

In stars belonging to the same cluster, we expect to have a large dispersion of lithium abundance even before the RGB bump. 
Such dispersion can be explained with several mechanisms: rotational mixing \citep{Pinsonneault92, eggenberger10}, mass loss \citep{SF92}, diffusion \citep{michaud86}, gravity waves \citep{MS00, CT05}, and overshooting \citep{xD09}.
For instance, stars having a range of rotation velocities might have 
depleted Li at different rate. Rotation velocity is, indeed, 
responsible for the transport of chemical species and a higher rotation allows a deeper dredge-up \citep[e.g.,][]{Balachandran90,lebre05}, which modify the Li content on the main sequence (such as the Li dip).
However, the efficiency of such rotation-induced mixing can explain the spread of A(Li), but its effect is too small to reproduce the drop of
Li in some stars beyond the bump \citep[see, e.g.][]{mucciarelli11}. 
A non-canonical mixing is, thus,  necessary to explain Li depletion beyond the bump. 
Li depletion caused by thermohaline mixing is expected to be combined with a further decline of C and an enhancement of N, with a global decrease of [C/N] in lithium-depleted stars.

In Figure~\ref{fig:lithium1} (in the Appendix we present all clusters in Figs.~\ref{fig:lithium:all1}, \ref{fig:lithium:all2}, \ref{fig:lithium:all3}) we show two examples of clusters in the GES survey. In these Figures, we present the log~$g$-T$_{\rm eff}$ diagrams (left panels), A(Li) vs [C/N] (central panels), and  [C/N] abundance vs T$_{\rm eff}$ (right panels) of giant stars that have passed the FDU. 
We consider a star to be Li-depleted if A(Li)$<$0.4~dex or if its A(Li) is an upper limit. For most clusters, Li-depletion happens in stars beyond the RGB bump (but not for all clusters, see, for instance, M67). 
In many cases, Li-depletion is associated with a variation of [C/N] with respect 
to the bulk of  stars with  A(Li)$>$0.4~dex. 
In the rightmost panel, we show also [C/N] as a function of T$_{\rm eff}$. 
For our analysis, we conservatively remove stars with A(Li)$<$0.4~dex to maintain a sample of giant stars with homogeneous conditions and similar [C/N] abundances 
(with the exception of the stars of M67, for which we have only Li-depleted stars whose [C/N] is in good agreement with previous literature results \citet{bertellimotta17}).
However, we remind that the final A(Li) during the giant phase is strongly related to the initial Li abundance during the main sequence: a main sequence star with a strong rotational velocity might deplete its surface Li abundance more than a star rotating at a lower rate. For stars with low A(Li), but with normal [C/N] with respect to the bulk of stars of the cluster (see, e.g., Pismis~18, NGC6067), the origin of the Li under-abundance might be related to their condition during the main sequence. 

We also consider the location of stars in the log~$g$-T$_{\rm eff}$ diagram, excluding stars beyond the RGB-bump if their [C/N] differs more than 1-$\sigma$ from the average [C/N] of the stars located before the bump.

\begin{figure*}[h]
\center
\includegraphics[scale=1.4]{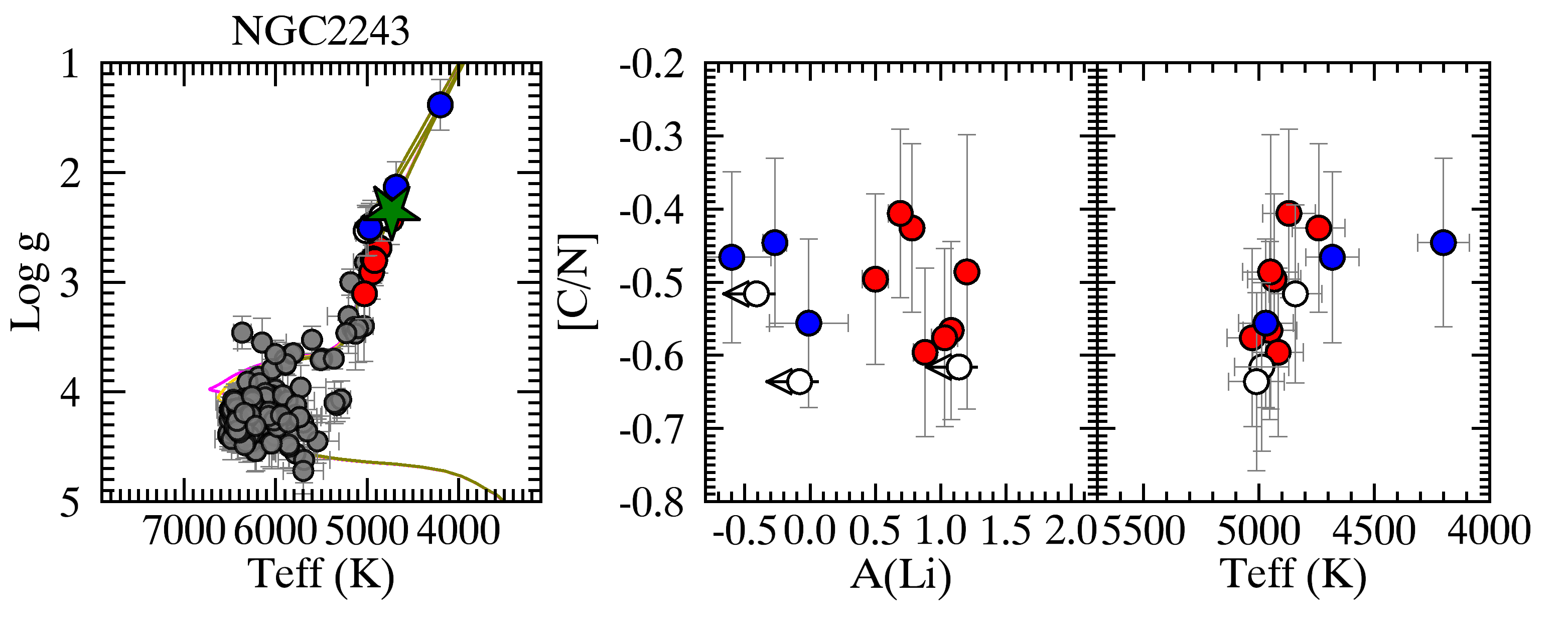} 
\includegraphics[scale=1.4]{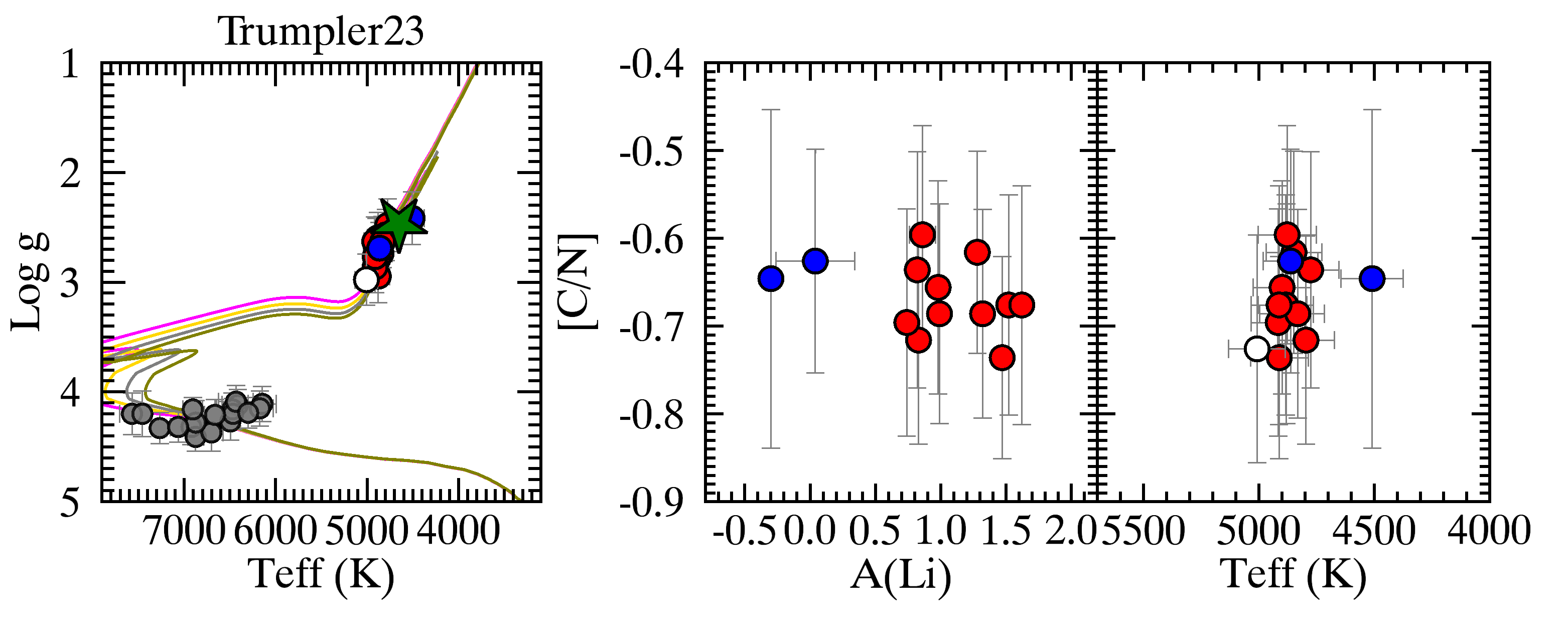} 
\caption{GES clusters NGC~2243 and Trumpler~23: log~$g$-T$_{\rm eff}$ diagrams with PISA isochrones 3.7, 4.0, 4.3, 4.5 Gyr and 0.7, 0.8, 0.9, 1.0 Gyr, respectively (left panels), and member stars beyond the FDU, A(Li) vs [C/N] (central panel), and [C/N] abundance vs T$_{\rm eff}$ (right panel). Stars with A(Li)$>$0.4 are shown with red circles, stars with A(Li)$<$0.4 with blue circles, stars with upper limits measurements of A(Li) are indicated with empty circles and stars no-RGB or without C and N are shown with grey circles. The green star represents the position of the RGB bump. \label{fig:lithium1}}
\end{figure*}

\begin{figure*}[h]
\center
\includegraphics[scale=1.4]{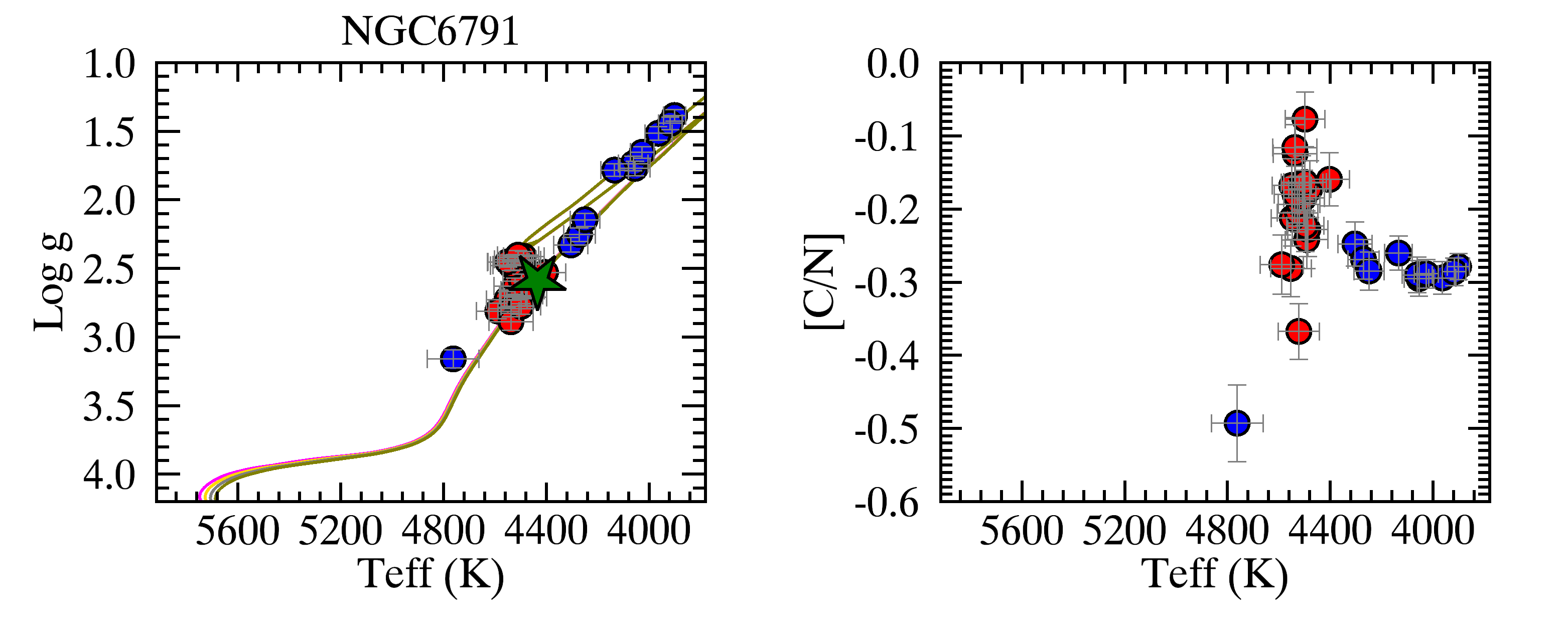}
\includegraphics[scale=1.4]{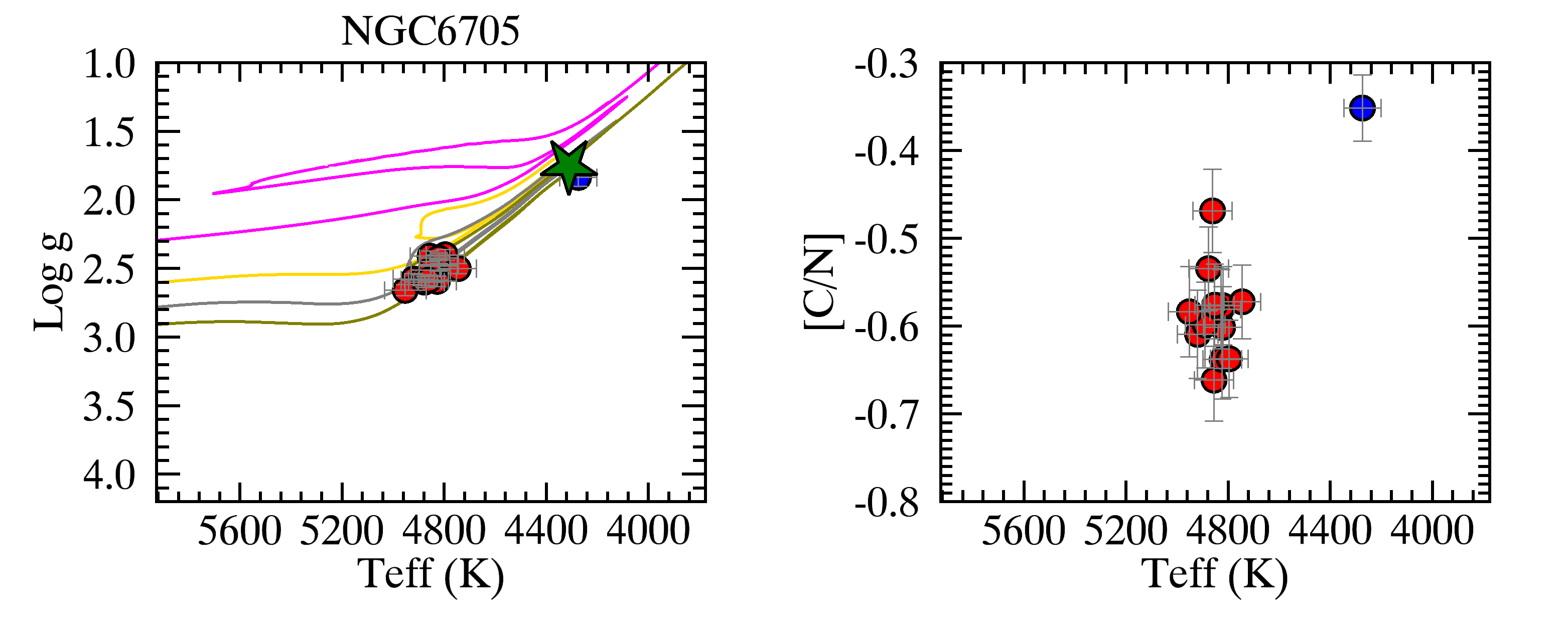}
\caption{APOGEE clusters NGC~6791 and NGC~6705: log~$g$-T$_{\rm eff}$ diagrams with PISA isochrones 7.3, 7.6, 7.9, 8.2 Gyr for NGC~67091 and 0.1, 0.2, 0.3, 0.4 Gyr for NGC~6705 (left panels) and  [C/N] abundance vs T$_{\rm eff}$ (right panels). Symbols and colours as in Figure~\ref{fig:lithium1}.  \label{fig:apo}}
\end{figure*}

In Figure~\ref{fig:apo} (in Figs.~\ref{fig:apo:all1}, \ref{fig:apo:all2}, \ref{fig:apo:all3}, \ref{fig:apo:all4}, \ref{fig:apo:all5} in the Appendix we present all clusters) we show the log~$g$-T$_{\rm eff}$ diagram (left panel) and  [C/N] abundance vs T$_{\rm eff}$ (right panels) in giant stars of the APOGEE sample of open clusters for two example clusters. 
For these clusters, we do not have information on Li abundance. We thus use the information on the location on the log~$g$-T$_{\rm eff}$ diagram and the variation of [C/N] to exclude stars for our final sample: basically we exclude stars beyond the RGB-bump but only if their [C/N] differs more than 1-$\sigma$ from the average [C/N] of the stars located before the bump. \\
 
From stellar evolution theory, we would expect that more evolved stars than the FDU to have lower [C/N] values if extramixing processes are in place \citep[see, e.g.][]{lagarde12}. For some clusters we observe an opposite effect with at least one star more evolved along the isochrone having a higher [C/N] abundance (see, e.g., NGC6705, NGC6866) than the stars located at the clump. This unexpected result might be due to incorrect stellar parameters or abundances. These stars are not considered in the final average values.  
In addition, in many clusters (see, e.g. NGC~2420, Tr5, NGC~6819) we removed several stars (blue circles) which are probably non members, both for their location, quite far from the best fit isochrone, and for their discrepant [C/N]. They are not considered to compute the final average value.  

Globally, the effect of the extramixing seems to be very limited in the metallicity range of open clusters: there is not a clear and systematic correlation between the location of a RGB star in the log~$g$-T$_{\rm eff}$ diagram and the [C/N] value. 
There are some clusters for which this effect is appreciable, as for instance NGC~6791, in which many stars above the bump are observed and they have all systematically lower [C/N] abundances, while in other clusters, as, e.g., NGC~6819, the stars above the bump show similar [C/N] as the clump stars. 
In addition, where Li abundance is available, there is not
a strong correlation of Li depletion with lower [C/N] values, as expected as a consequence of an extramixing episode. 
Only for the  oldest clusters, we observe a [C/N] decrease for stars with A(Li)$<$0.4~dex, resulting from a more efficient dredge-up or thermohaline extra-mixing (see, e.g., 
Berkeley~31 and Berkeley~26). 
On the other hand, for younger clusters A(Li)$<$0.4~dex is often coupled with higher values of [C/N] with respect to stars with A(Li)$>$0.4~dex (see, for instance,  NGC~6005, Rup~134, Pismis~18, Trumpler~23). 
This can be related to the higher mass of their evolved stars in which 
thermohaline instability does not occur \citep{lagarde18}. 
Thus, the mechanisms of Li-depletion can be uncorrelated to those that modify the C and N abundances. The latter can be related to rotation-induced mixing that has an impact on the internal chemical structure of main sequence stars, observable during the RGB phase
\citep[see. e.g.][]{CL10}. 
The anti-correlation in young clusters between A(Li) and [C/N] in Li-depleted stars cannot easily be explained by the current models and deserve further investigation and it is, perhaps, associated to the Li abundance owned in the main sequence and inherited during the following evolutionary phases.

\section{The Age-[C/N] relationship using open clusters}
\label{sec_rel}

\begin{table}[h]
\caption{Mean post-FDU [C/N] of the open clusters in the GES and APOGEE samples.} 
\begin{center}
\tiny{
\begin{tabular}{llcc}
\hline
Open Clusters & [C/N] & \# member stars & P$_{memb.}^{mean}$\\
             & (dex)  &  &\\
\hline
GES  & & &\\
\hline      
Berkeley~31  &  $-0.32\pm0.12$  &   1              & 0.87      \\
Berkeley~36  &  $-0.30\pm0.26$        &   1       & 0.89     \\ 
Berkeley~44  &  $-0.38\pm0.12$        &   2        & 1.00     \\ 
Berkeley~81  &  $-0.64\pm0.04$ (0.07) &  11      & 0.99          \\ 
 Melotte71  &   $-0.65\pm0.09$        &  2    & \\            
    NGC~2243  &  $-0.51\pm0.05$ (0.08) &   7     & 1.00\\        
    NGC~6005  &  $-0.64\pm0.04$ (0.05) &  9    &   0.97 \\    
    NGC~4815$^a$ & $-0.70\pm0.11$ (0.11)&    5  & \\
    NGC~6067  &  $-0.87\pm0.04$ (0.08) &     9 & 0.99\\ 
    NGC~6259  &  $-0.84\pm0.04$ (0.04) &     12 & 0.99 \\
    NGC~6802  &  $-0.67\pm0.04$ (0.08) &      7  & 1.00\\ 
     Rup134  &  $-0.63\pm0.05$ (0.06) &    16 & 0.98\\ 
   Pismis18  &  $-0.74\pm0.05$ (0.03) &       5 & 0.99\\  
 Trumpler23  &  $-0.68\pm0.04$ (0.04) &        9 & 1.00\\  
 Trumpler20$^a$&$-0.60\pm0.12$ (0.12) &       42  &  \\  
\hline\hline                           
APOGEE & &  &\\                         
\hline                                     
 Berkeley 17  & $-0.24\pm0.02$ (0.06)   &   7   &\\
 Berkeley 53  & $-0.52\pm0.03$ (0.05)   &   4   &\\
 Berkeley 66  & $-0.36\pm0.03$ (0.08)   &   4   &\\
 Berkeley 71  & $-0.66\pm0.04$ (0.18)   &   7   &\\
 FSR0494      & $-0.60\pm0.04$ (0.09)   &   5   &\\
 IC166        & $-0.69\pm0.02$ (0.20)   &   15   &\\
 King 5       & $-0.57\pm0.03$ (0.01)   &    5   &\\
 King 7       & $-0.53\pm0.03$ (0.04)   &    3   &\\
 NGC~1193      & $-0.44\pm0.04$ (0.09)   &      3 &  \\
 NGC~1245      & $-0.64\pm0.02$ (0.11)   &     23  & \\
 NGC~1798      & $-0.44\pm0.02$ (0.06)   &      8   &\\
 NGC~188       & $-0.32\pm0.01$ (0.04)   &    10   &\\
 NGC~2158      & $-0.41\pm0.02$ (0.13)   &     18   &\\
 NGC~2420      & $-0.40\pm0.02$ (0.07)   &     11  & \\
 NGC~6791      & $-0.20\pm0.01$ (0.06)   &     19   &\\
 NGC~6811      & $-0.68\pm0.03$ (0.01)   &      4   &\\
 NGC~6819      & $-0.49\pm0.01$ (0.03)   &     14  & \\
 NGC~6866      & $-0.69\pm0.03$ (0.06)   &      6  & \\
 Teutsch 51   & $-0.58\pm0.06$ (0.04)   &    4  & \\
 Trumpler 5   & $-0.37\pm0.02$ (0.04)   &    9   &\\
 NGC~7789      & $-0.52\pm0.01$ (0.03)   &     17   &\\
 \hline\hline
 GES and APOGEE & & \\
 \hline
 M67-GES                      &   $-$0.37$\pm$0.07 (0.06)  &  3 & \\ 
 M67-APOGEE {\sc dr14}        &   $-$0.57$\pm$0.01 (0.07) &  23 &\\
 M67-APOGEE {\sc dr14}$^b$    &   $-$0.50$\pm$0.09         &  21 &\\
 M67-adopted ({\sc dr14}$^b$-GES) &   $-$0.43$\pm$0.11        & $-$ &\\
 \hline
 NGC~6705-GES$^a$              &     $-$0.69$\pm$0.09      &  27 &\\ 
 NGC~6705-APOGEE {\sc dr14}    &    $-$0.60$\pm$0.01 (0.03)      & 12 & \\
 NGC~6705-adopted              &    $-$0.64$\pm$0.09                & $-$ &\\
 \hline\hline
\end{tabular}
}
\end{center}
\tablefoot{They derived in the present work, with the exception of ${a}$ which are from \citet{tau15} and ${b}$ from \citet{souto19}.  
P$_{memb.}^{mean}$ is the membership probability.  }
\label{tab:cnall}
\end{table}

In Table~\ref{tab:cnall}, we present the results of our analysis: the mean [C/N] abundance ratio of the member giant stars (with selection criteria described in the previous sections) and the number of used stars for each cluster in both the GES and APOGEE samples.  
The uncertainties in Table~\ref{tab:cnall} are the formal errors on the weighted mean and, within brackets, the standard deviation $\sigma$, and they take into account of uncertainties on C and N abundances. 
Using the solar values from \citet{grevesse07}, we compute the average [C/N] ratio
in each cluster in the GES sample, while for APOGEE\footnote{ Solar abundances in APOGEE survey are adopted from \citet{asplund2005}, which is consistent, for most elements, with \citet{grevesse07}, including C and N.} clusters the solar-scaled abundance ratios are already given in their database.
For three clusters of the GES sample, Trumpler~20, NGC~6705 and NGC~4815,  we use their literature [C/N] estimated only for post-FDU stars from \citet{tau15} based on GES results (C and N abundances are not re-derived in next data releases, therefore they are not present in GES {\sc idr4} and {\sc idr5}, used in the present work). 

There are two clusters in common between the two surveys, M67 and NGC~6705, that can be used to cross-calibrate the [C/N] abundances in the two surveys, together with few other stars in common, mainly benchmark stars.  

\begin{figure}[h]
\center
\includegraphics[scale=0.5]{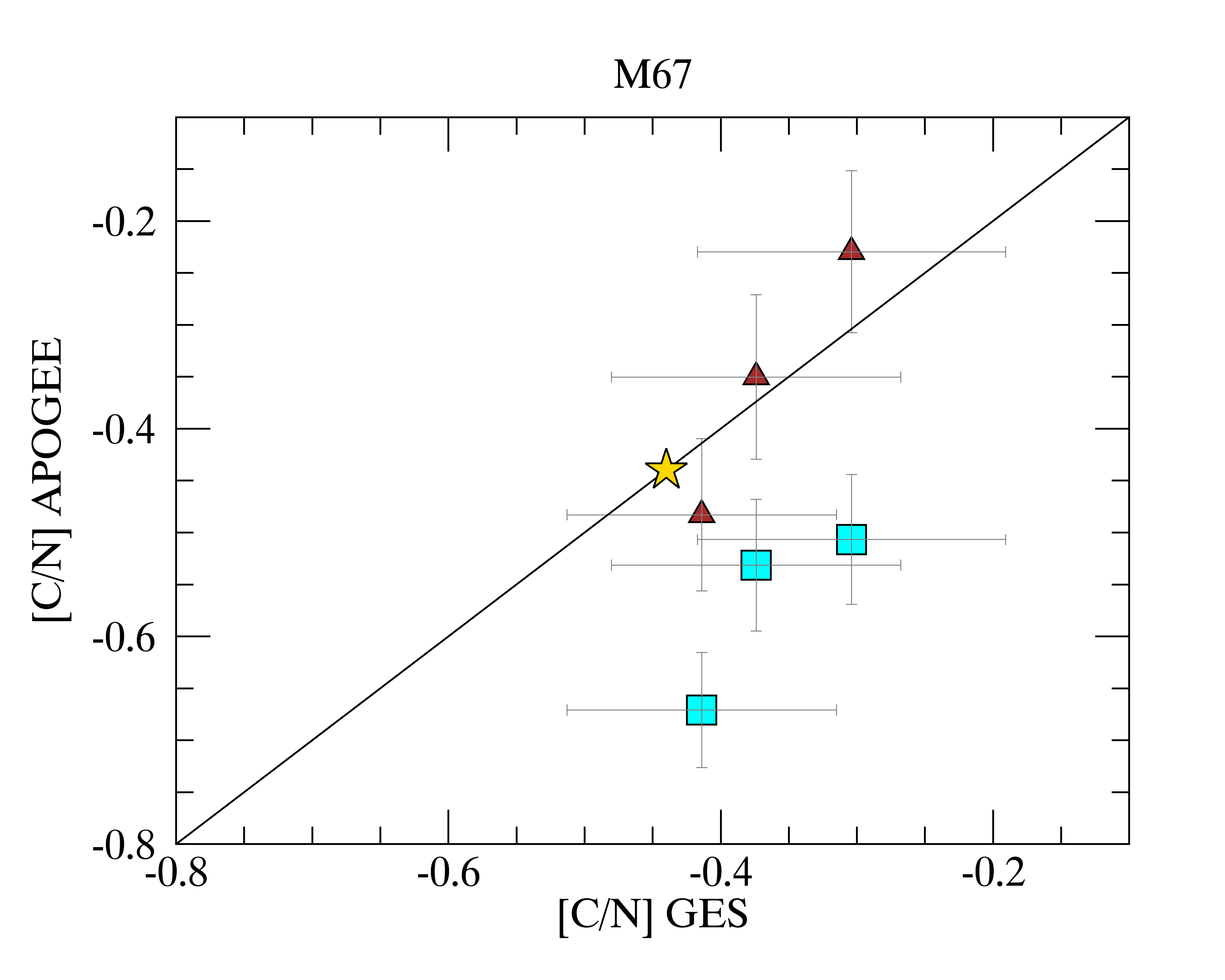} 
\caption{Stars in common between GES and APOGEE ({\sc dr12}-- brown triangles and {\sc dr14} --cyan squares) in M67. The continuous line is the 1-to-1 relation. The yellow stars is the theoretical location of the FDU in the models of \citet{salaris15} for the age and metallicity of M67. \label{fig:m67_stars_ic} }
\end{figure}

In M67, there are three stars in common between GES and APOGEE with [C/N] abundances. 
The average post-FDU abundance in M67, using APOGEE {\sc dr14}, is [C/N]=$-$0.57$\pm$0.01, while  
the GES results give [C/N]=$-$0.37$\pm$0.07. 
The C and N abundances of stars in M67 along the evolutionary sequence have been studied in details by \citet{bertellimotta17} using APOGEE {\sc dr12}. Their post-FDU stars have an average [C/N]= $-$0.46$\pm$0.03, which is in good agreement with the expectation of theoretical models for stars of mass and metallicity of the evolved stars of M67 \citep{salaris15}, as shown in their Figure~9. 
The results for M67 in APOGEE {\sc dr14} are slightly higher,  due to an offset in the [C/N] abundances from APOGEE {\sc dr12} to the latest APOGEE {\sc dr14} release. 
In Fig.~\ref{fig:m67_stars_ic} we compare [C/N] in three stars in common between APOGEE ({\sc dr12}, {\sc dr14}) and GES. The [C/N] values of APOGEE {\sc dr14} are systematically lower than those in {\sc dr12}. The values of GES and APOGEE {\sc dr12} are in better agreement (close to the 1-to-1 relationship).
We notice that a recent paper \citep{souto19} re-analyzed 83 APOGEE {\sc dr14} spectra of stars in M67. The analysis performed in this paper is more accurate than the standard one (K. Cunha, S. Hasselquist, private communication). There are five stars in common between GES and the sample of \citet{souto19}, two of them with C and N abundances. For these stars the agreement in [C/N] is within 0.01~dex. 
We thus use the re-analysis of \citet{souto19} to compute the mean [C/N] of giant stars after the FDU in M67, which is [C/N]=$-$0.50$\pm$0.09. 
In the following analysis, we adopt for M67 an average between the GES\footnote{the three stars of M67 have A(Li)$<$0.4~dex, but their [C/N] is consistent with the other stars in APOGEE and with the theoretical model of \citet{salaris15}. For these reasons we include them in the analysis. } and the results of \citet{souto19}. 

\begin{figure}[h]
\center
\includegraphics[scale=0.5]{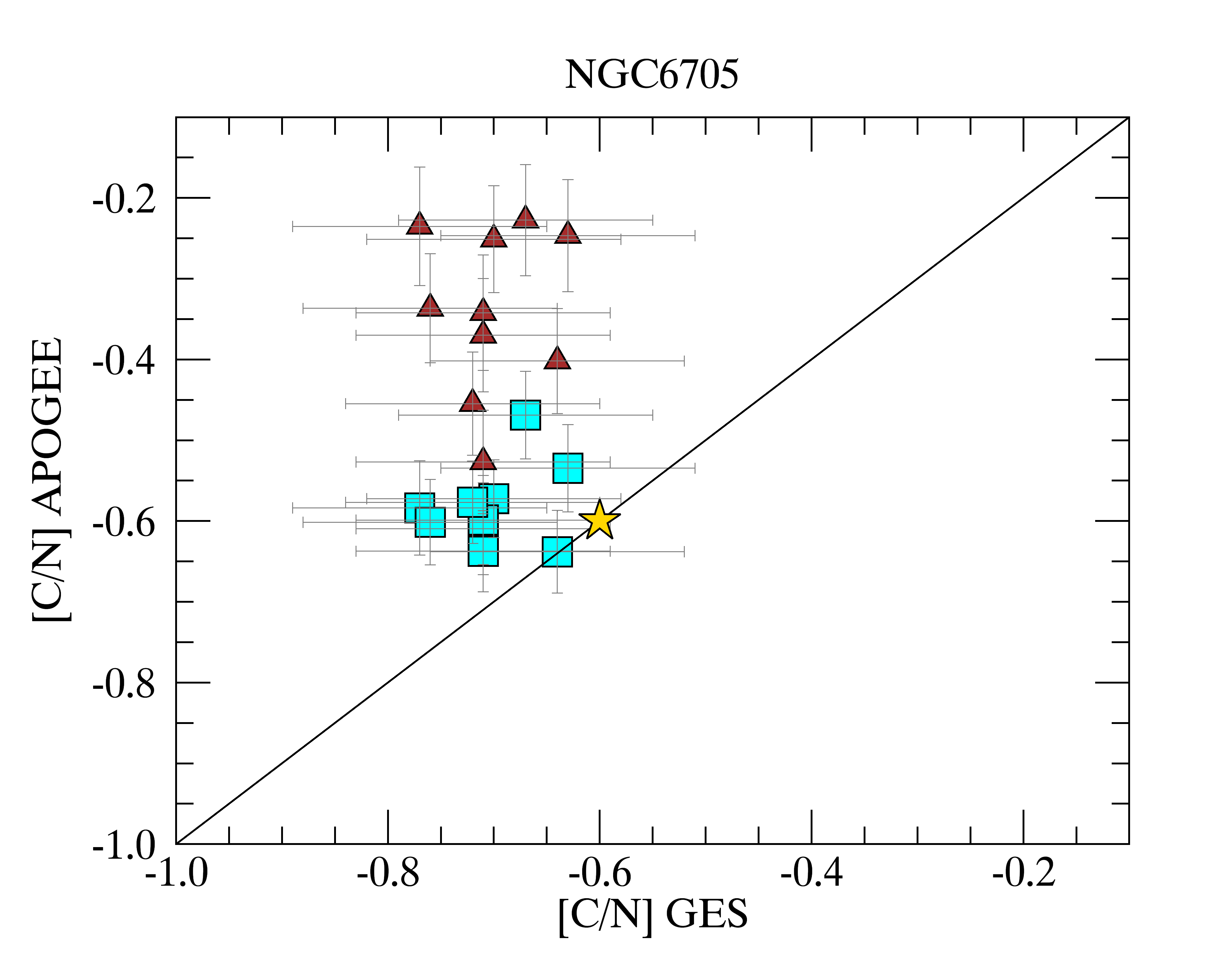}
\caption{Stars in common between GES and APOGEE ({\sc dr12}-- brown triangles and {\sc dr14} --cyan squares) in NGC~6705. The continuous line is the 1-to-1 relation. The yellow stars is the theoretical location of the FDU in the models of \citet{salaris15} for the age and metallicity of NGC~6705. \label{fig:NGC6705_stars_ic} }
\end{figure}

The other cluster in common between the two surveys is NGC~6705. 
In Fig.~\ref{fig:NGC6705_stars_ic} we show the [C/N] abundances in stars in common between APOGEE ({\sc dr12} and {\sc dr14}) and GES. As for M67, the [C/N] values of APOGEE {\sc d14} are systematically lower than those in {\sc dr12}. On the other hand, for this cluster the values of {\sc dr14} are in much better agreement with the GES results.

In the following analysis, we adopt for NGC~6705 an average between the GES from \citet{tau15} and the APOGEE {\sc dr14} results. 

\begin{figure}[h]
\center
\includegraphics[scale=0.5]{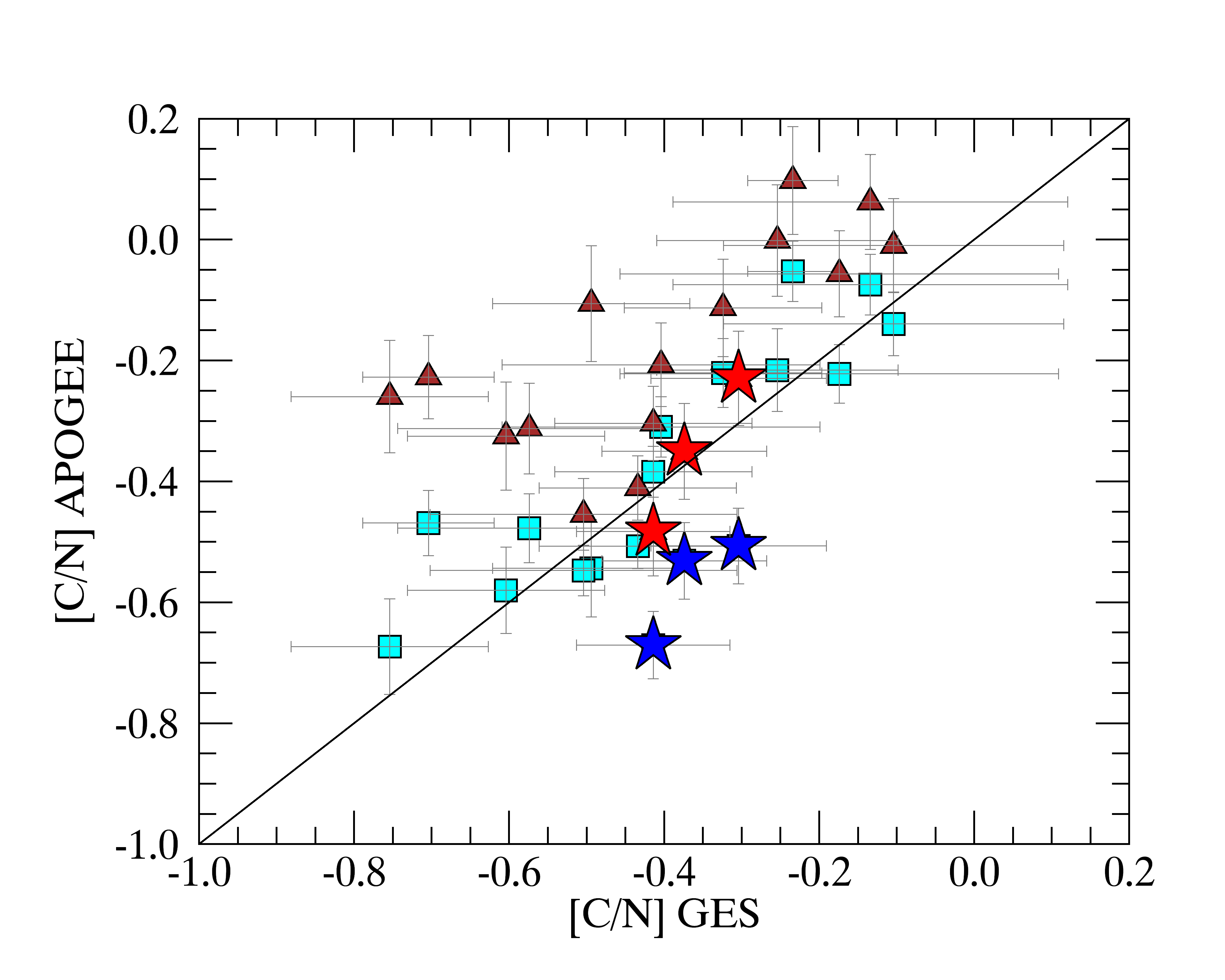}
\caption{Stars in common between GES and APOGEE ({\sc dr12}-- brown triangles and {\sc dr14} --cyan squares). The continuous line is the 1-to-1 relation. The red stars are the abundances in M67 for APOGEE {\sc dr12} and the blue stars for APOGEE {\sc dr14}. \label{fig:all_ges_apogee} }
\end{figure}

To identify possible systematic offsets or trends, we make a global cross-match between APOGEE {\sc dr12} and {\sc dr14} and GES. 
In Fig.~\ref{fig:all_ges_apogee} we show the [C/N] abundances of stars in common between GES and the two APOGEE releases. 
We notice a general better agreement of APOGEE {\sc dr14} with GES, while the [C/N] abundances in APOGEE {\sc dr12} are usually offset of $\sim$0.2 dex. 

In the plot of Fig.~\ref{fig:all_ges_apogee}, we highlight the abundances of the three stars of M67. These stars are indeed outliers in the APOGEE {\sc dr14}, with a negative offset 
with respect to GES. For the stars we use the improved analysis of \citet{souto19} as suggested by the APOGEE team (K. Cunha, S. Hasselquist, private communication).  

For the other stars, there are no systematic  offsets and the agreement between the two sets of results is, within the errors, good. 
This allows us to securely combine the GES and APOGEE {\sc dr14} samples of clusters (with the exception of the results of M67, for which we adopt the \citet{souto19}'s results).


\begin{figure*}[h]
\center
\includegraphics[scale=1.0]{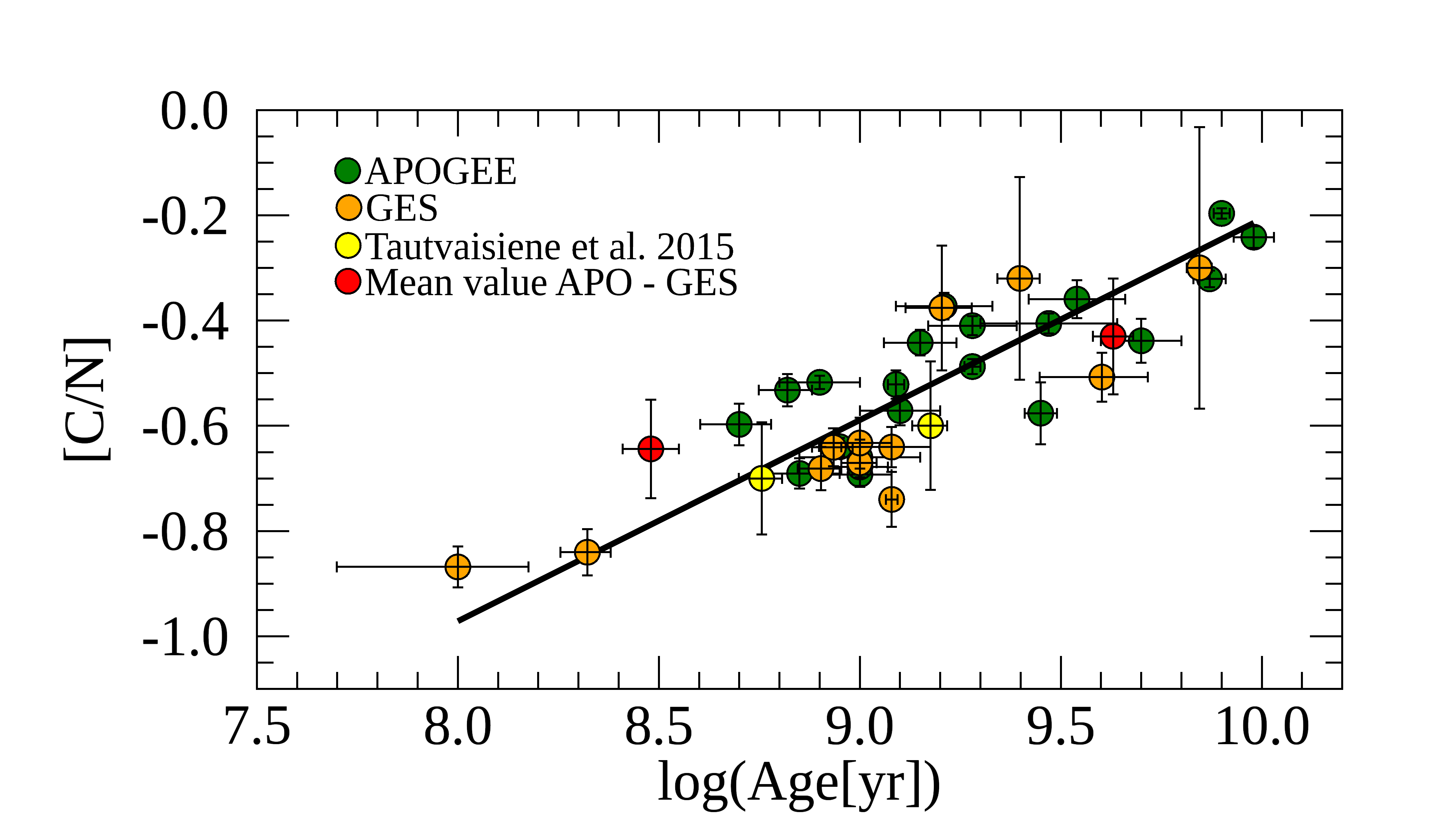}
\caption{Average [C/N] ratios of cluster member stars with $\log$~g cut as a function of $\log$(Age[yr]) for the GES (orange circles) and the APOGEE (green circles) surveys.  The yellow circles are the GES literature results by \citet{tau15}. The red circles represent the mean value [C/N] for NGC~6705 and M67. The straight line is the linear weighted least-squares fit of data. \label{fig:relationOC}}
\end{figure*}

In Figure~\ref{fig:relationOC}, we show the relationship between [C/N] in GES and APOGEE clusters and their logarithmic ages.

With the ages expressed in logarithmic form the relationship is linear, with a Pearson coefficient R=0.85, and it has the following expression
\begin{equation}
\log[\rm Age(yr)]=10.54(\pm0.06)+2.61(\pm0.10){\rm [C/N]}
\label{eq1}    
\end{equation}
This relationship has important implication, because, if we know the evolutionary stage of a giant star and we can measure its [C/N] abundance ratio, we can infer its age.  
However, there are some caveats that need to be explicitly discussed here: 
{\em i)} the relationship in Eq.\ref{eq1} is empirical and it is built using open clusters whose metallicity range is $-$0.4$<$[Fe/H]$<$+0.4. Thus, its application outside this [Fe/H] range is not correct. 
{\em ii)} At low metallicity, the effect of non-canonical extra-mixing in post-FDU stars increases and thus the measured [C/N] loses its dependence on stellar mass (and thus age). 
{\em iii)} Although the ages of giant stars are poorly constrained by isochrone fitting, and thus 
their age estimate through Eq.~\ref{eq1} is a step forwards, the uncertainties on [C/N]-ages are still high, as we discuss in the next paragraphs.

\subsection{Error estimate} 
It is known that the measurements of stellar ages are usually affected by large uncertainties.
However, accurate knowledge of stellar ages is a fundamental constraint for the scenarios of the formation and evolution of the different components of our Galaxy. 
Astroseismology data from CoRoT and Kepler have improved our knowledge of 
stellar ages, but they are limited to still small samples of stars. 
In this framework, chemical clocks, as [C/N] but also [Y/Mg] and [Y/Al] \citep{tuccimaia16, spina16, feltzing17, spina18,  delgado18}, 
might be important auxiliary tools because they can be obtained for large numbers of stars.
However, it is important to have a clear idea of the uncertainties related to the use of abundance ratios to derive the age of a star. 

From the GES results\footnote{Results estimate by Vilnius Node}, the typical uncertainties on the C and N abundances are 0.05~dex and 0.065~dex, respectively, with a typical uncertainties on the abundance ratio [C/N] of $\sim$0.10~dex. 
They affect the uncertainties on $ \rm \log(Age[yr])$ giving a mean value of about $\sim$0.26 dex on the logarithm of the age. These uncertainties translate into $\sim$55-60\% for the linear ages. 
For the APOGEE results, the typical uncertainties on the C and N abundances are $\sim$0.04~dex and $\sim$0.04~dex, respectively, with a typical uncertainty on the abundance ratio [C/N] of $\sim$0.06~dex. As for GES, this implies a mean uncertainty on  logarithm of the age of $\sim$0.17~dex and of $\sim$40-45\%  for the linear ages. 
Typical uncertainties are shown in Table~\ref{tab:error}. 

 \begin{table}[h]
 \caption{Typical uncertainties on logarithmic and linear age in different range of logarithmic age.}
\begin{center}
\begin{tabular}{ccc}
\hline
log(Age[yr]) &   e(log(Age[yr]))  &     e(Age)\%  \\
 \hline
 GES & & \\
 \hline
8.0-8.5        &     0.24     & 55\% \\
8.5-9.0        &     0.26     & 60\% \\
9.0-9.5        &     0.26     & 60\% \\
9.5-10.0      &     0.26     & 60\% \\
\hline 
APOGEE & & \\
\hline
8.0-8.5         &     0.19     & 45\% \\
8.5-9.0         &     0.19     & 45\% \\
9.0-9.5         &     0.17     & 40\% \\
9.5-10.0        &     0.17     & 40\% \\
\hline
\end{tabular}
\end{center}
\label{tab:error}
\end{table}

However, another important part in the error budget is the uncertainty in the parameters of the fit of the linear relationship of Eq.~\ref{eq1}. Combining both GES and APOGEE sample, we lower the uncertainties on the slope and on the intercept of the 
linear relationship. At the moment, the uncertainties on the parameters of the fit are of the order of $\sim$0.06 and $\sim$0.10 on the intercept and slope, respectively. We expect to have a more accurate fit of the relationship at GES completion, when about 70 open clusters will be available.
At the moment, the age estimate with [C/N] abundance can be considered as an additional tool to give  an independent estimate of the age of field stars, in particular for giant stars whose age determination through isochrone fitting is not appropriate due to the little separation of isochrones of different ages.

\subsection{Comparison with previous results}

A relationship between logarithmic age and [C/N] was already provided for the APOGEE catalogue: 
\citet{martig16} presented an empirical relationship between [C/N] and stellar masses (and thus stellar ages) based on the 
asteroseismic mass estimates from the APOKASC survey \citep[a spectroscopic follow-up by APOGEE of stars with
asteroseismology data from the Kepler Asteroseismic Science Consortium,][]{borucki10}.
To compare with our results, we cross-matched the APOKASC catalogue of \citet{martig16} with APOGEE {\sc dr14}, and we find the mean offset between the {\sc dr12} and {\sc dr14} abundances. 
We apply the offset to the {\sc dr12} APOKASC abundances to have them on the same scale as our clusters. 

\begin{figure*}[h]
\center
\includegraphics[scale=1.0]{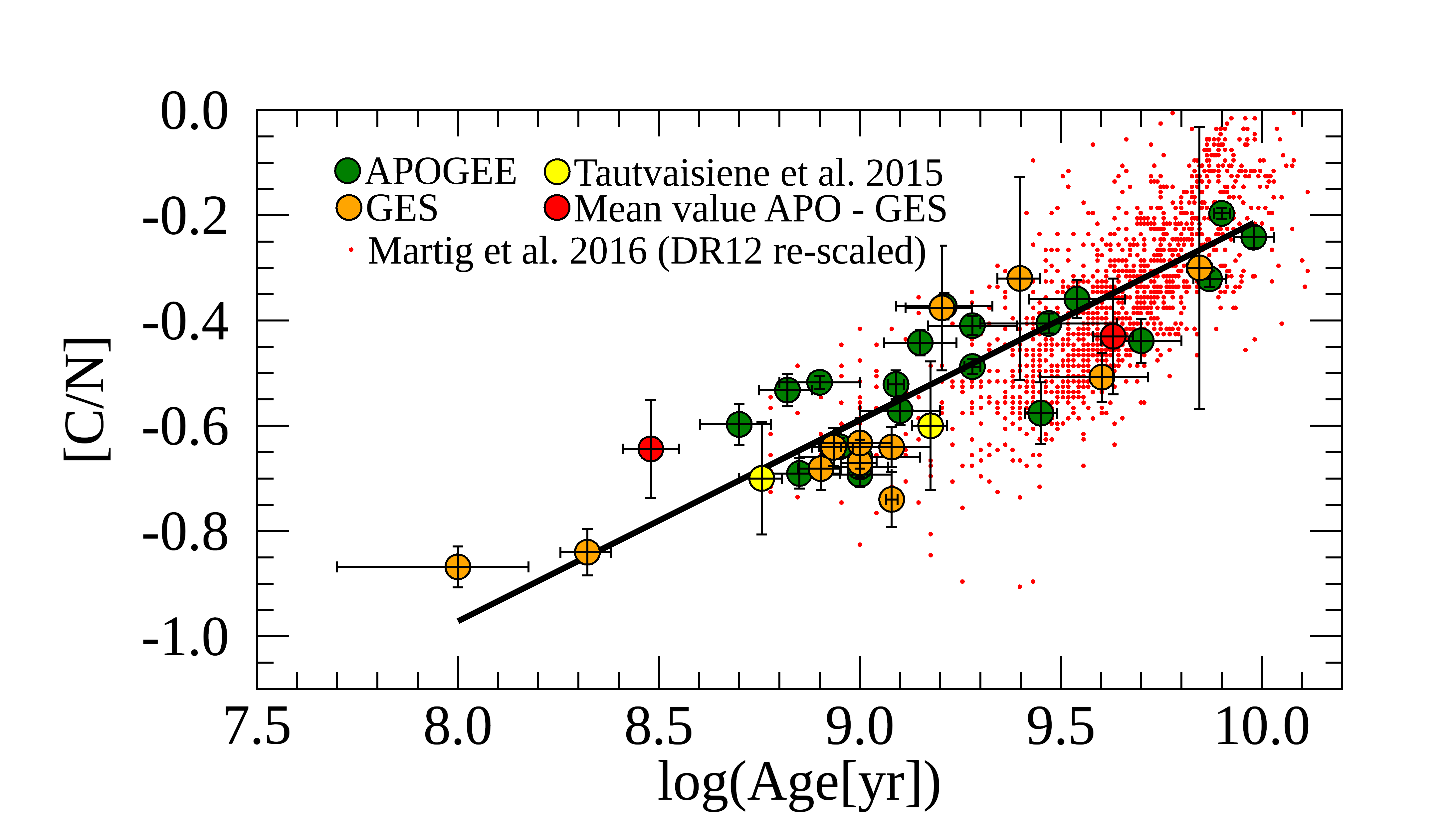}
\caption{Comparison between the field stars in APOKASC with [C/N] vs age computed by \citet{martig16} ({\sc dr12} small red dots and {\sc dr14} small blue dots) and those obtained in the present work using open clusters in the APOGEE and GES surveys (symbols and colour-codes as in Fig.~\ref{fig:relationOC}). 
\label{fig:MartigAPO}}
\end{figure*}

 In Fig.~\ref{fig:MartigAPO} we plot the ages from \citet{martig16} and [C/N] measurements from {\sc dr12} (scaled to {\sc dr14}), together with the results of our sample of the open clusters and the relation of Eq.\ref{eq1}.
We note that the two samples have different slopes at older ages, where the effect of extra-mixing might be higher. In the oldest regime ($\log$[Age]$>$9.5), the ages obtained by \citet{martig16} are $\sim$ 0.2 dex younger than ours at fixed [C/N]. \citet{mackereth2017} showed that \citet{martig16} underestimated the ages in the older regime by up to a factor of 2 when compared with those based on asteroseismological masses. Therefore, our relationship provide ages in better agreement with those of \citet{mackereth2017}.

For the youngest ages, at a given [C/N] our ages correspond to [C/N] values close to the APOGEE {\sc dr12} re-scaled ones, although with a large dispersion. 
This is true in the region where the two samples overlap, even if the APOKASC sample does not populate the youngest-ages regions, where 
many star clusters are located, reaching a lower limit of about $\log$[Age]$\sim$8.8.

The combination of ages from asteroseismology with those from isochrone fitting in young open clusters deserve to be investigated and to be fully exploited to obtain a calibration of the [C/N]-Age relation in a wider age range.   

\section{Comparison with theoretical models}
\label{sec_theory}

Later phases of stellar evolution affect the abundances of C and N, and
therefore the abundances of these elements do not trace the initial chemical composition of
stars, but they are modified by stellar nucleosynthesis and by internal mixing. 
In this framework, it is of interest to compare with predictions of various theoretical models of stellar evolution,
having in mind the limits of such comparison, such as: {\em i)} not perfect correspondence between the stellar models used for the isochrones adopted to derive the ages of open clusters and those adopted to predict surface [C/N] abundances; {\em ii)} possible offsets in [C/N] between models and observations ; {\em iii)} comparison of different kinds of stars in model and observations.

We compare our results with models in two different ways: a classical comparison with a set of standard stellar evolution models, computed for different metallicities \citep{salaris15} and 
a more innovative approach that combines a stellar population synthesis model of the Galaxy with a complete grid of stellar evolution models  \citep{lagarde17}. 

\begin{figure}[h]
\center
\includegraphics[scale=0.5]{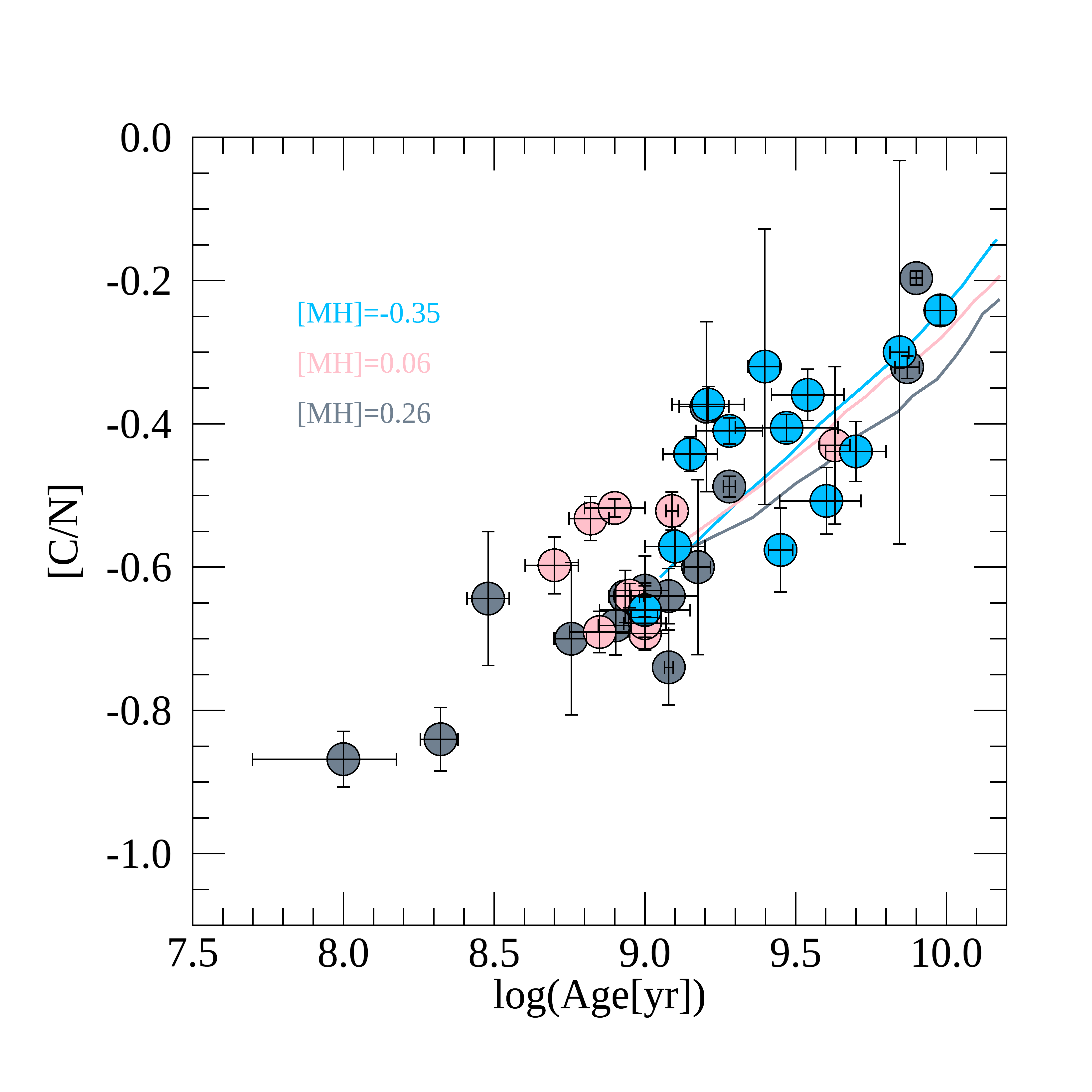}
\caption{Comparison of [C/N] and age between the theoretical models by \citep{salaris15} and the open clusters in GES and APOGEE surveys. The models are computed for different metallicities. Their colour-codes are shown in the legend. The clusters are divided in three metallicity bins: [Fe/H]$ \leq-0.10$ (light blue circles), $-0.1\leq$ [Fe/H] $ \leq +0.1$ (pink circles), and [Fe/H] $ \geq+0.1$ (grey circles). \label{fig:Salaris}}
\end{figure}

In Fig.~\ref{fig:Salaris}, we compare the observed [C/N] ratio at the FDU vs. age  with theoretical predictions of \citet{salaris15}. 
We recall that in our observations we select giant stars beyond the FDU, avoiding stars after the RGB bump with measurable effects of extra-mixing in their [C/N] abundances and/or Li-depleted.   
The models of \citet{salaris15} are computed with the BaSTI stellar library \citep{pietr04,pietr06} and they include the effect of the overshooting. 
The theoretical curves give the [C/N] abundance after the FDU at a given stellar age. They are provided for different metallicities in the age range 1-10 Gyr. We select three curves covering the metallicity range of our sample clusters. 

The data of open clusters follow the general trend of the models, 
confirming the predicted increase in [C/N] with age. 
In the age range where both observations and models are available we have colour-coded them in the same way to facilitate the association of the observations with the corresponding curves. 
There is a general agreement with model curve and observations within the uncertainties, but no trend with metallicity can be seen for the observations.

\begin{figure}[h]
\centering
\includegraphics[scale=0.5]{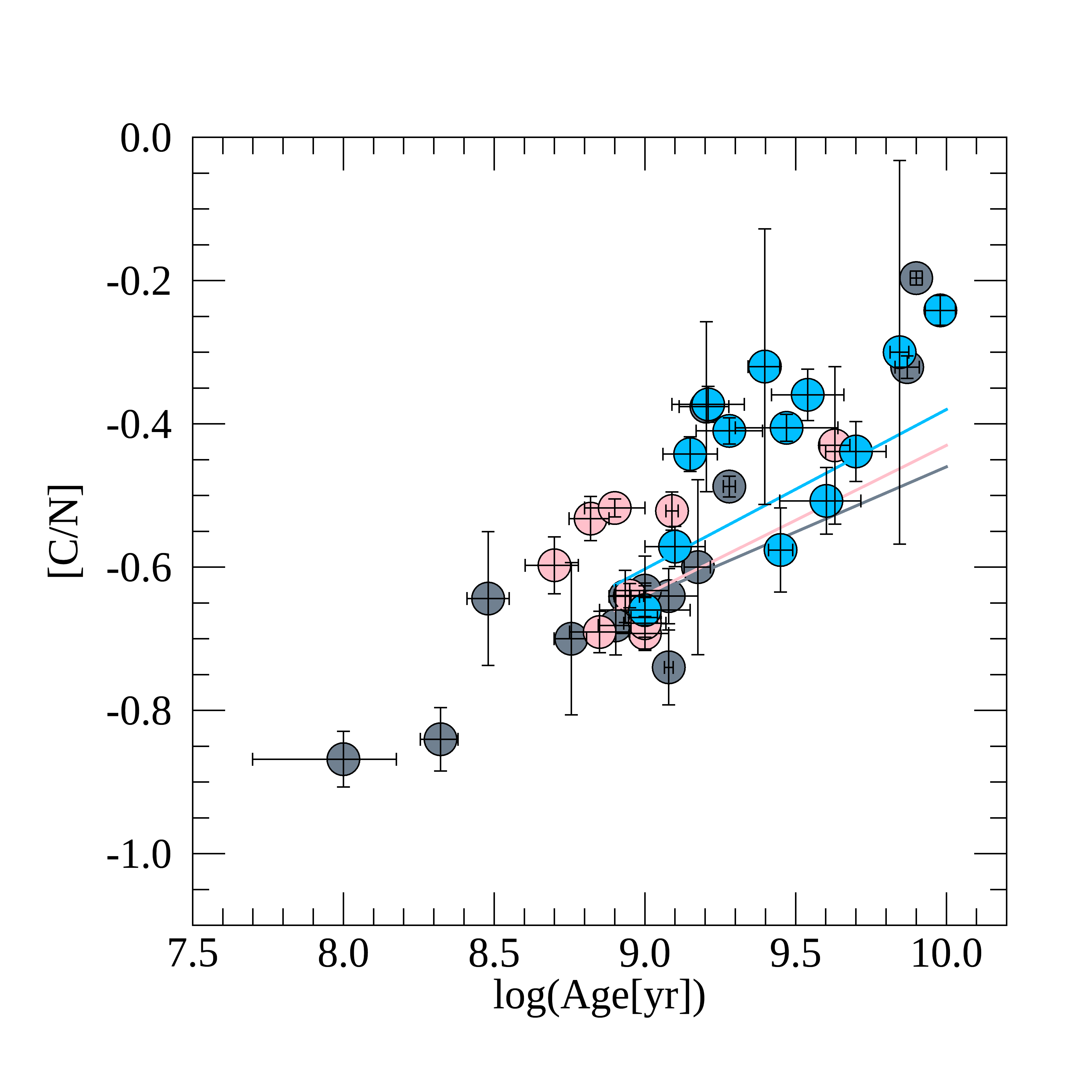}
\caption{Comparison between age and [C/N] of the thin-disc RC giant modelled by \citet{lagarde17} (continuous lines - light blue, pink and grey) and the clusters in GES and APOGEE. The clusters and models are divided in three metallicity bins: [Fe/H]$ \leq-0.10$ (light blue circles), $-0.1\leq$ [Fe/H] $ \leq +0.1$ (pink circles), and [Fe/H] $ \geq+0.1$ (grey circles). \label{fig:Lagarde} }
\end{figure}

In Fig.~\ref{fig:Lagarde} we compare our data with the results of the model of \citet{lagarde17} for three ranges of metallicities. The models of \citet{lagarde12} have been applied to an improved stellar population model of the Galaxy by \citet{lagarde17}, providing  global, chemical and seismic properties of the thin-disc stellar population. They estimate the effect of the thermohaline mixing in thin-disc giants, which produce measurable effects on the [C/N] ratios of stars of different metallicities. 
They derive mean relations between [C/N] and age, usable to estimate ages for thin-disc red-clump giants as a function of their [Fe/H]. 
Since the open clusters are a thin-disc population and we mainly select low-RGB and RC stars in them, we can compare our results with the models of \citet{lagarde17}.  

The data are in qualitative agreement with the model curves of \citet{lagarde17}, which, however, predict lower [C/N] values for the oldest stars. 
On the other hand, the open cluster data are in better agreement with the model curves of \citet{salaris15}, whose models show an increasing [C/N] abundance with age which better follows the observed trend in the oldest clusters. 
Both models are computed for ages$>$1~Gyr, and for different metallicities (in the plot we show the models computed for the metallicity range of our sample of clusters).
In the data, at a given age, we do not distinguish between trends of clusters with different metallicities. However, in the metallicity range spanned by the data, model predictions would differ only of $\sim$0.1~dex considering the two metallicity extremes. This difference is  comparable with typical [C/N] uncertainty.

\section{Application to field stars}
\label{sec_application}

As mentioned by \citet{masseron15}, passing from [C/N] measurements to assign masses and ages is tempting, but at the same time it is risky because of several unknown or underestimated effects, such as, for instance, the exact evolutionary stage of each observed stars, the metallicity dependence of C and N abundances after the FDU, and the effect of the extra-mixing at different ages and metallicities \citep[cf.][]{lagarde17}. 

Keeping in mind these limits, we combine the sample of giant stars in the GES and APOGEE databases with C and N abundances aiming at identifying age trends along the 
disc populations. 
Among giant stars, we select stars which have passed the FDU. To identify the maximum log~$g$ at which the FDU can appear, we consider our grid of isochrones \citep[Pisa isochrones,][]{dellomo12} at different ages and metallicities. For each of them we derive the value of log~$g$ corresponding to the point where [C/N] starts to monotonically decrease, i.e., log~$g$ (FDU). 
For most of the field stars, which are usually older than the cluster stars, 
the mean value of surface gravity at the FDU is log~$g$ $\sim$3.4, with a little dependence on metallicity (see Fig.~1). We also excluded stars above the RGB-bump (considering the most conservative case of stars of 2-3~Gyr, the age of the bulk of the stars in our sample, for which the RGB-bump is located at log~$g$ $\sim$2.4).  
Therefore, in the following analysis, we select giant field stars with 2.4$<$ log~$g$ $<$3.4. 
In addition, we use the line-of-sight distances derived by \citet{Bailer-Jones18} with {\em Gaia} {\sc dr2}. 
They use a distance prior that varies smoothly as a function of Galactic longitude and latitude according to a Galaxy model.  
We apply the relationship of Eq.~\ref{eq1} to giant field stars taking into account 
its limits of applicability:  $-$0.4$<$[Fe/H]$<$+0.4 and 0.1~Gyr$<$age$<$10~Gyr, range of metallicity and age of clusters studied in this work.\\
There are about 52 000 giant stars -- of which $\sim$ 264 belonging to GES -- in the our combined sample for which we can estimate the age. Thanks to the distance obtained with {\em Gaia}, we can locate them in the Galaxy. With our determination of stellar ages, we can investigate the age trends in the thin- and thick-disc populations.\\

\begin{figure*}[h]
\hspace{-0.5cm}
\includegraphics[scale=0.8]{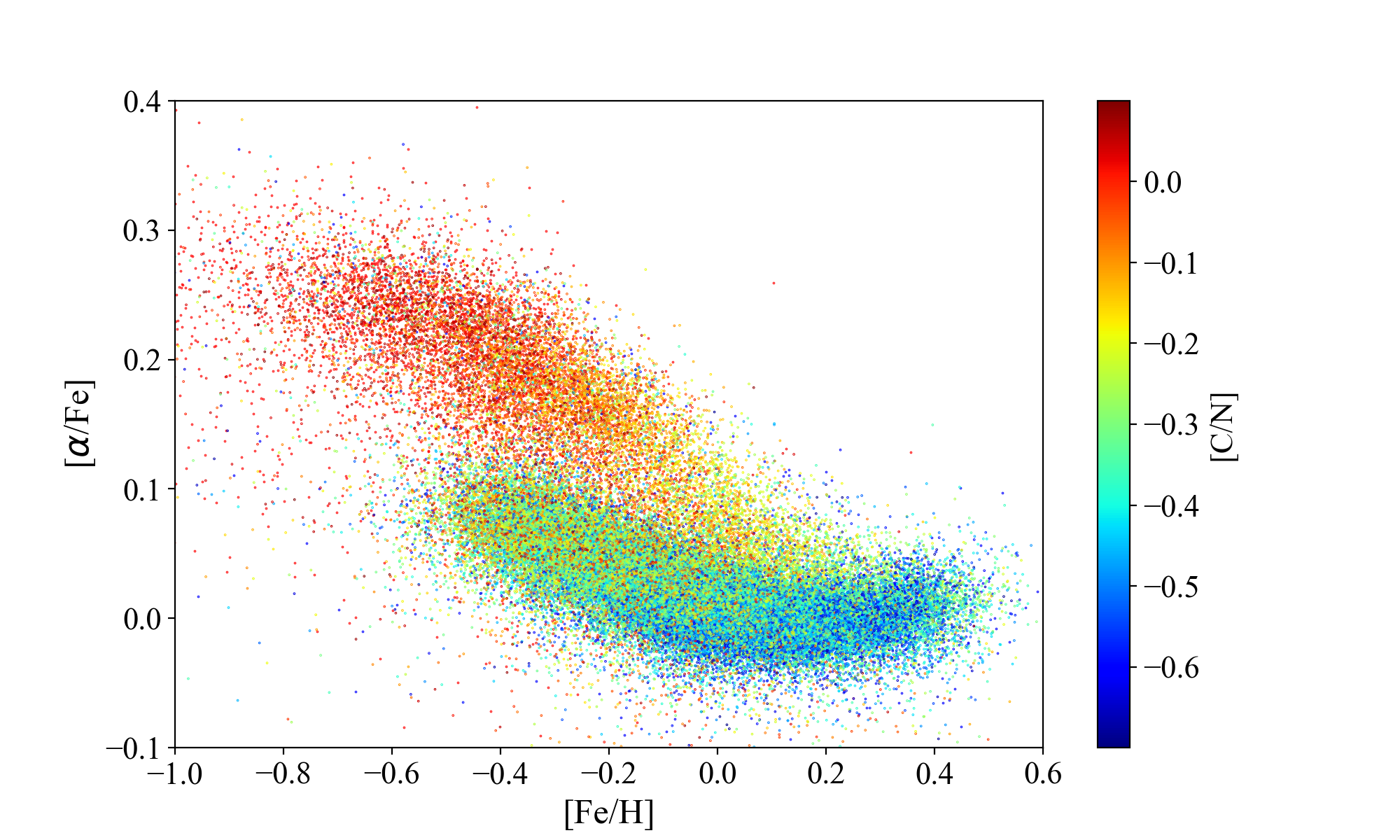}
\caption{[$\alpha$/Fe] as a function of [Fe/H] for field stars in the APOGEE {\sc dr14} and a GES samples. The stars are colour-coded by [C/N]. \label{fig:DiscsCNAPO}}
\end{figure*}

\begin{figure*}[h]
\hspace{-0.5cm}
\includegraphics[scale=0.8]{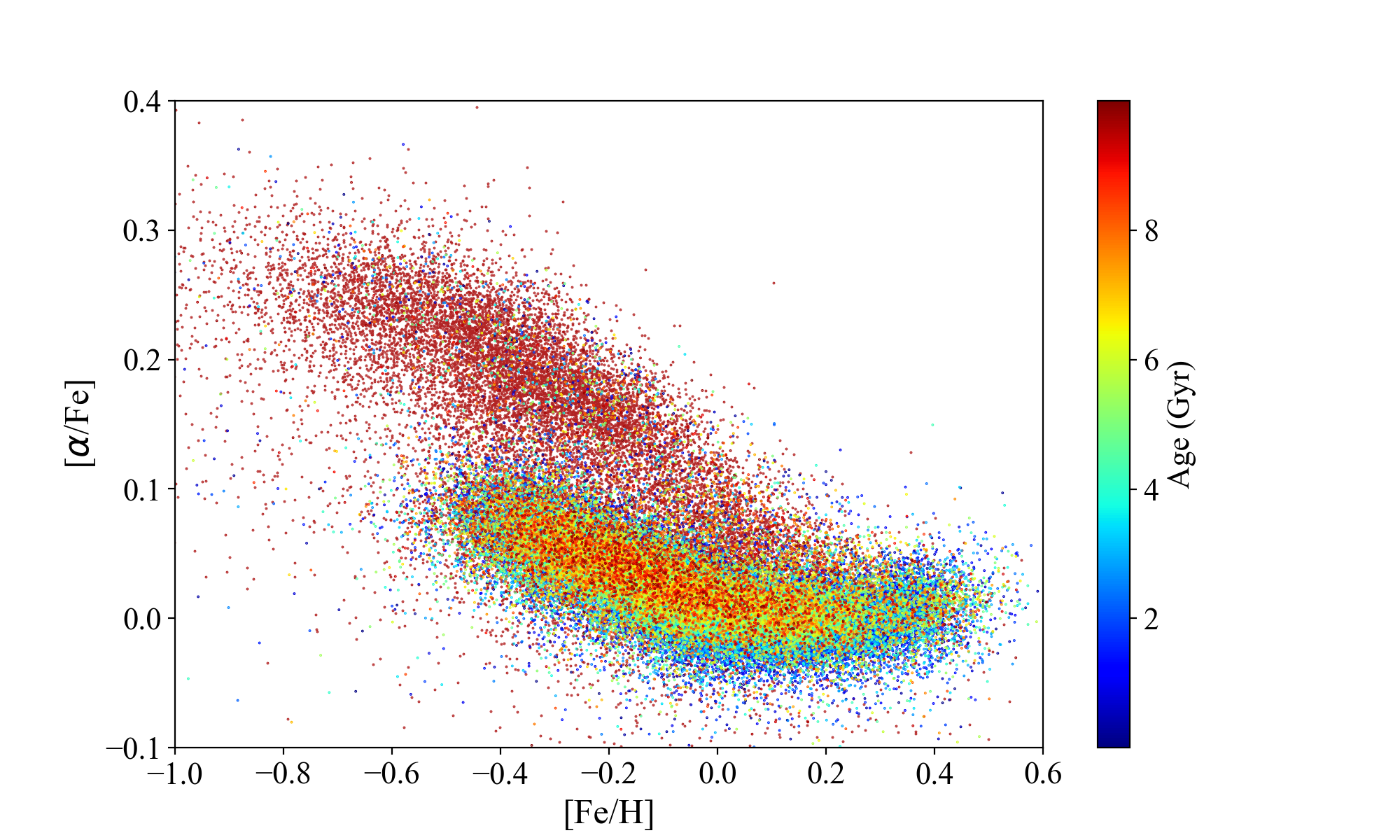}
\caption{[$\alpha$/Fe] as a function of [Fe/H] for field stars in the APOGEE {\sc dr14} and a GES samples. The stars are colour-coded by ages. \label{fig:DiscsAgeAPO}}
\end{figure*}

In Fig.~\ref{fig:DiscsCNAPO}, we show [$\alpha$/Fe]\footnote{[$\alpha$/Fe] is computed with Ca, Si, and Mg in both surveys}  vs. [Fe/H] for the APOGEE and GES sample, colour-coded by their [C/N] abundance ratio. The dichotomy between the $\alpha$-enhanced thick stars and the lower [$\alpha$/Fe] thin disc is evident also in terms of their 
[C/N]. As already observed by \citet{masseron15} in their analysis of the APOGEE catalogue for {\sc {\sc dr12}}, we can see a clear gradient in [C/N], with the thin disc stars having a lower [C/N]. At a given metallicity, [C/N] becomes larger for higher [$\alpha$/Fe] \citep[see also][]{hasse19}. The gradient in [C/N] can then be translated into a gradient in ages, as shown in Fig.~\ref{fig:DiscsAgeAPO}. 
In this plot there are two approximations: {\em i)} the relationship of Eq.~\ref{eq1} has been applied outside the range of metallicity -0.4$<$[Fe/H]$<$+0.4 and also to thick disc stars, with different [$\alpha$/Fe]\ and thus different initial [C/Fe]; {\em ii)} stars with derived age $>$10~Gyr are shown in red. Their ages are estimated applying the relationship outside its range of applicability and, for this reason, they do not represent a correct age estimate and we can just say they are old. 
There is a clear division in age between thin and thick disc stars:
we can see that the youngest stars are present in the thin disc and their ages become greater with the increase of the $\alpha$-elements content. The thick disc stars are in general older than the thin disc ones. Most of them have only lower limits estimate of their ages, with age $\gtrsim$ 10 Gyr.

\begin{figure*}[h]
\hspace{-0.8cm}
\includegraphics[scale=0.6]{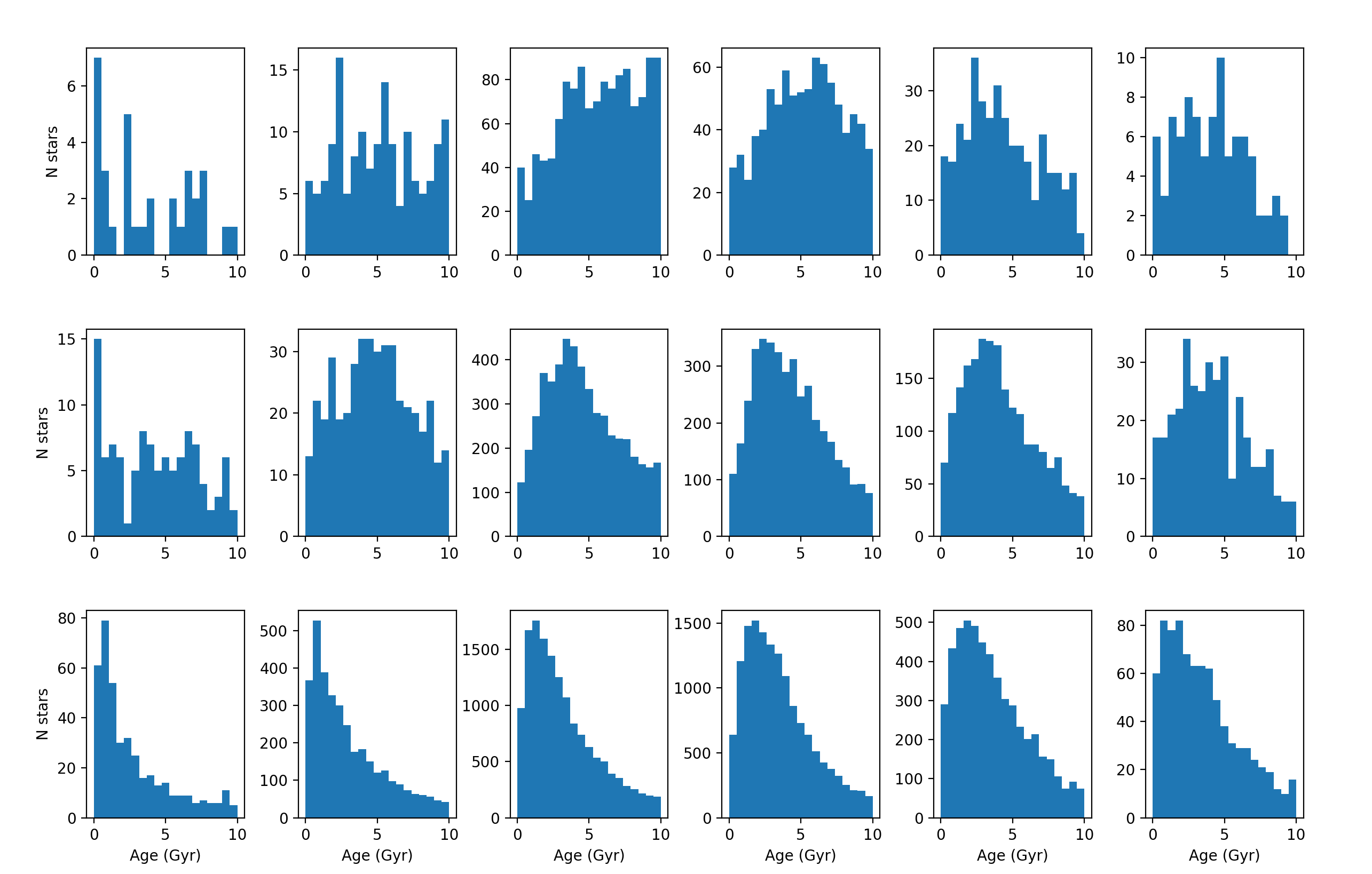}
\caption{Histograms of the age for each bin in height on the plane $\rm \left| z \right|$ and Galactocentric distance R$_{GC}$. Each panel is corrispondent to the respective one in Fig.~\ref{fig:ThinthickdiscAPO}. \label{fig:hist}}
\end{figure*}

\begin{sidewaysfigure*}[h!]
\centering
\includegraphics[scale=.6]{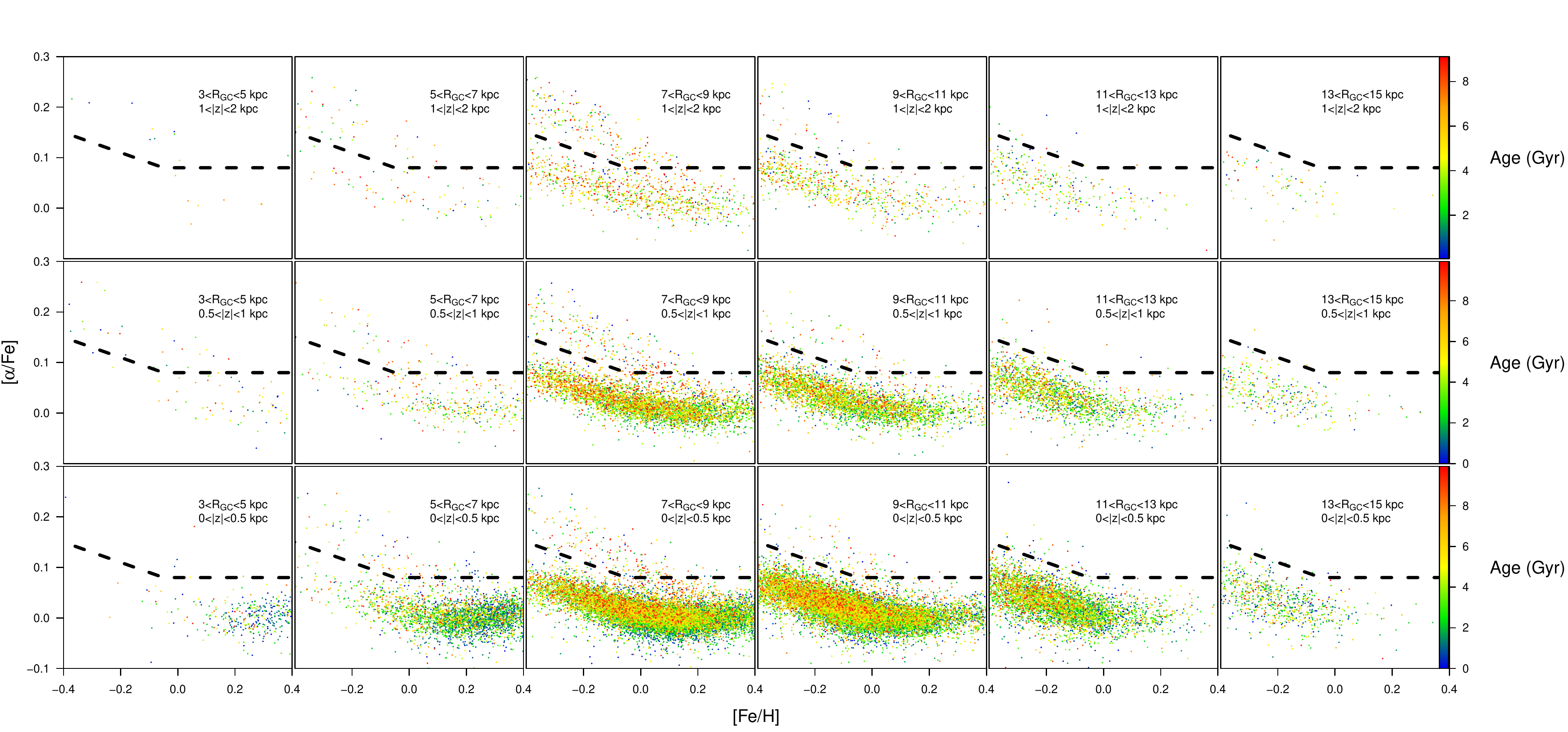}
\caption{[$\alpha$/Fe] as a function of [Fe/H] for field stars in the APOGEE sample. The stars are colour coded by ages. The dashed line shows the separation between thin and thick disc populations. Each plot contains stars selected by means of Galactocentric radius R$_{GC}$ and height z, labeled in each figure. \label{fig:ThinthickdiscAPO}}
    \end{sidewaysfigure*}

In Fig~\ref{fig:hist}, we present the histogram of the derived stellar ages in different bins of $\rm \left| z \right|$ and R$_{GC}$, covering a radial region 3 < R$_{GC}$ < 15 kpc, and a vertical region $\rm \left| z \right|$ < 2 kpc. 
We divide our sample in several radial and vertical regions, as done by \citet{hayden15}. 
In this plot, and in the following Table~\ref{tab:medage_bin} and Fig.~\ref{fig:ThinthickdiscAPO}, we apply the relationship of Eq.~\ref{eq1} in the metallicity range $-$0.4$<$[Fe/H]$<$+0.4 and we do not include lower limit measurements. 
In Table~\ref{tab:medage_bin}, we show the median ages and $\sigma$ in each bin.
The bins close to the Plane contain the youngest stars, but they have also a non negligible tail of old stars. 
The stars located in the bins at intermediate height are older, and, even more, stars located at 1~kpc$<|z|<$2~kpc. 
This is appreciable also from Table~\ref{tab:medage_bin} where a systematic increasing age is observed for stars located above the Galactic Plane. 
In the bins characterizing the thin disc 0~kpc$<|z|<$0.5~kpc, the youngest population is located in the inner bins, and the 
ages slightly increase towards the outskirts.

In Fig.~\ref{fig:ThinthickdiscAPO}, we show the [$\alpha$/Fe] vs. [Fe/H] plane dividing in same bins as in Fig.~\ref{fig:hist}. For smaller z, the thin-disc population, characterized by a low-[$\alpha$/Fe], is prominent. Instead, for larger z, the population of thick disc starts to increase, being more prominent in the inner regions.

The bottom panels of Fig.~\ref{fig:ThinthickdiscAPO} show the thin-disc population closer to the plane. In the 3~kpc <R$_{GC}$< 7~kpc ranges the populations is dominated by thin disc metal-rich stars, where the stars with the highest metallicity are youngest ones. As we can see from the histograms of the age distribution in each bin in $\rm \left| z \right|$ and R$_{GC}$ in Fig.~\ref{fig:hist} and respective medians and $\sigma$ in Table~\ref{tab:medage_bin}, the population in 3~kpc <R$_{GC}$< 7~kpc at low z is mainly young.
Therefore, the inner bin is dominated by young and metal rich stars, with low [$\alpha$/Fe]. 
Moving towards the outer regions 7~kpc <R$_{GC}$< 11~kpc, the thin disc includes a wider range of [Fe/H], from $-$0.4 to 0.4. Following our age estimate, the younger stars are located in the lower envelope of the distribution, at low [$\alpha$/Fe]. There is a consistent number of old stars with both super- and sub-solar metallicity. In the outer regions, 11~kpc <R$_{GC}$< 15~kpc the metal-rich stars start to disappear. 

The central panels of Fig.~\ref{fig:ThinthickdiscAPO} show the thin-disc population at intermediate height above the plane. In the innermost regions 3~kpc<R$_{GC}$<7~kpc the thin disc is poorly populated, unlike the region 7~kpc<R$_{GC}$<11~kpc. This effect is due to the observational selection of APOGEE survey, for which the observations are located in particular towards the outer regions at 7-11 kpc if compared with the inner ones at 3-7 kpc.
The stars in these two panels do not reach the low [$\alpha$/Fe] and the young ages of the corresponding panels at lower z. Indeed, their median ages are older than the medians at the same R$_{GC}$, but lower z (see Table~\ref{tab:medage_bin}).
In the outer regions, there are fewer metal-rich stars, and on average they are younger, as their medians show if compared with ones in the inner regions at intermediate z (see Table~\ref{tab:medage_bin}).  

The top panels of Fig.~\ref{fig:ThinthickdiscAPO} show the thin disc population at high height above the plane. Stars are almost absent from the innermost regions, due to the sample selection in the range of applicability of our relationship, from which, they would be older than 10~Gyr. From 7 to 11 kpc, there is signature of thin disc stars with a wide age range. Finally, in the outermost regions 11-15~kpc few young and intermediate-age thin disc stars are present.

\begin{table}[h]
\caption{Median and $\sigma$ of the age for each bin in height on the plane z and Galactocentric distance R$_{GC}$. }
\begin{center}
\begin{tabular}{lccc}
\hline
R$_{GC}$ (kpc)  & $\rm \left| z \right|$ (kpc) &   Median age (Gyr)  &     $\sigma$  \\
\hline\hline
3 $-$ 5   & 1 $-$ 2   & 2.95 & 2.95 \\
5 $-$ 7   & 1 $-$ 2   & 5.06 & 2.81 \\
7 $-$ 9   & 1 $-$ 2   & 5.80 & 2.69 \\
9 $-$ 11  & 1 $-$ 2   & 5.34 & 2.61 \\
11 $-$ 13 & 1 $-$ 2   & 4.00 & 2.57 \\
13 $-$ 15 & 1 $-$ 2   & 4.06 & 2.34 \\
\hline
3 $-$ 5   & 0.5 $-$ 1 & 4.20 & 2.87 \\
5 $-$ 7   & 0.5 $-$ 1 & 4.79 & 2.57 \\
7 $-$ 9   & 0.5 $-$ 1 & 4.23 & 2.50 \\
9 $-$ 11  & 0.5 $-$ 1 & 3.98 & 2.39 \\
11 $-$ 13 & 0.5 $-$ 1 & 3.75 & 2.42 \\
13 $-$ 15 & 0.5 $-$ 1 & 3.97 & 2.44 \\
\hline
3 $-$ 5   & 0 $-$ 0.5 & 1.82 & 2.53 \\
5 $-$ 7   & 0 $-$ 0.5 & 2.32 & 2.44 \\
7 $-$ 9   & 0 $-$ 0.5 & 2.63 & 2.35 \\
9 $-$ 11  & 0 $-$ 0.5 & 3.05 & 2.28 \\
11 $-$ 13 & 0 $-$ 0.5 & 3.17 & 2.40 \\
13 $-$ 15 & 0 $-$ 0.5 & 3.04 & 2.45 \\
\hline\hline
\end{tabular}
\end{center}

\label{tab:medage_bin}
\end{table}

\section{Summary and conclusions}
\label{sec_conclusions}
The databases of stellar spectra from high-resolution spectroscopic 
surveys, as GES and APOGEE, are providing extremely wide and accurate
data-sets of stellar parameters and abundances for large samples of
stars in the different Galactic components. 
Some abundance ratios, such as, for instance, [C/N], [Y/Mg],  [Y/Al], [Ba/Fe] 
have been recognised to be powerful chemical clocks, i.e., strictly related to the stellar ages. 
In particular, the [C/N] abundance ratio is a good age indicator for stars in the RGB evolutionary phase after the first dredge-up. 
Using open clusters in the GES and APOGEE Surveys, we calibrate a relationship 
between cluster age (from isochrone fitting of the whole cluster sequence) and [C/N] abundance ratio of stars that have passed the FDU. 
We carefully select RGB stars beyond the FDU, studying the effects of extra-mixing process in their [C/N] abundances. 
We compare the [C/N] measurements and ages of open clusters with the predictions of stellar evolutionary models by \citet{salaris15} and \citet{lagarde17}, finding a good agreement, but with the current accuracy of age and abundance measurements we cannot confirm the variation of the age-[C/N] relationship with metallicity in the [Fe/H] range traced by open clusters. 

We use our relationship to derive the ages of field stars in a combined GES and APOGEE sample. 
In the plane [$\alpha$/Fe] vs. [Fe/H], we see a clear dichotomy between the [C/N] abundances in the thin and thick disc as already observed by \citet{masseron15}, and consequently stellar ages. We extrapolate our relation to estimate the ages for thin disc stars and giving lower limit ages for the thick disc stars. 
As expected, we find that stars belonging to the thin disc are on average younger than the stars in the thick disc. Typical age of the thin disc stars decreases going into the inner part of the disc at low z and towards the outskirts. The former indicates that the metal rich stars in the inner disc represent the later phases of the Galactic evolution, 
while the latter is expected from an inside-out formation of the disc.  

These immediate applications of the relationships between ages and [C/N]
probe the power of chemical clocks to improve our knowledge of stellar ages. 
The next step will be to estimate in a homogeneous way the age of the open clusters in the final data release of GES survey to put them on a common stellar age scale. 
In addition, we plan to combine other chemical clocks, as for instance [Ba/Fe] or [Y/Mg] to give further constraints to the ages of stars in different evolutionary 
stages.

\newpage

%\newpage

\begin{appendix}
\section{Appendix}

\begin{figure*}[h]
\center
\includegraphics[scale=1.2]{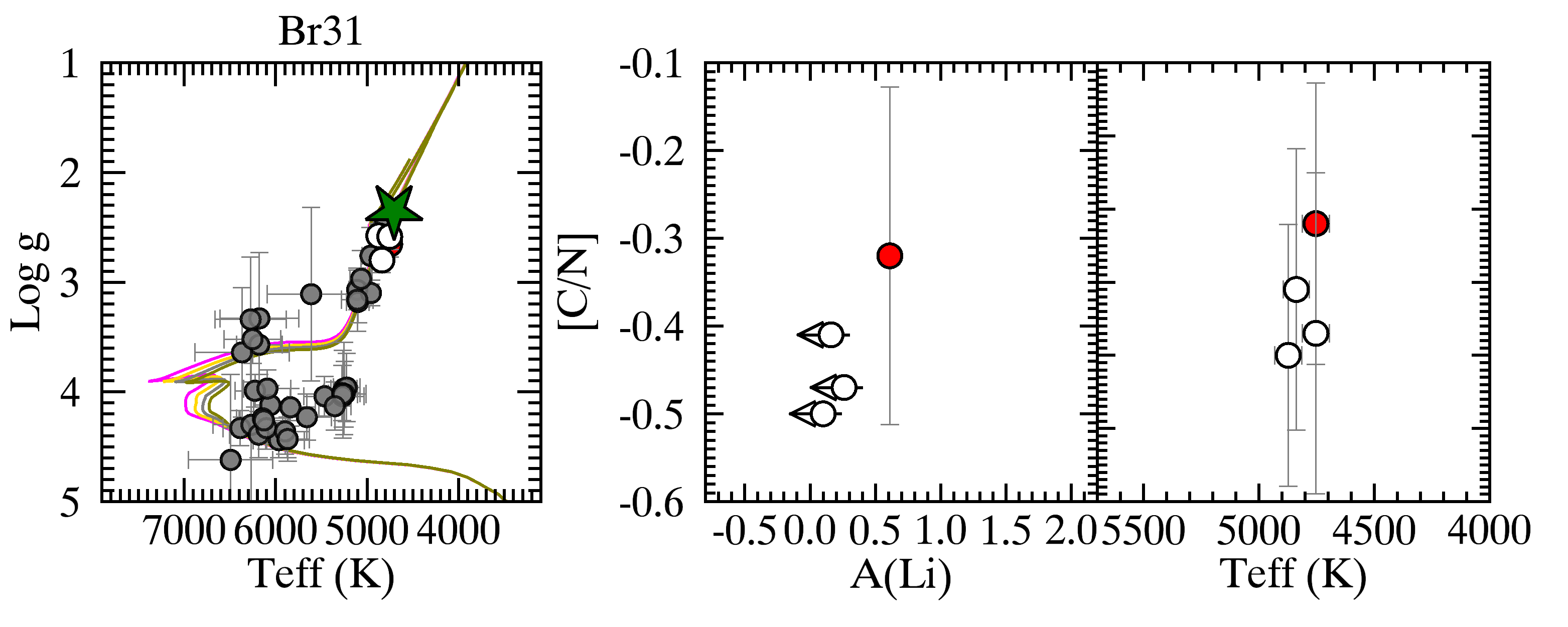} 
\includegraphics[scale=1.2]{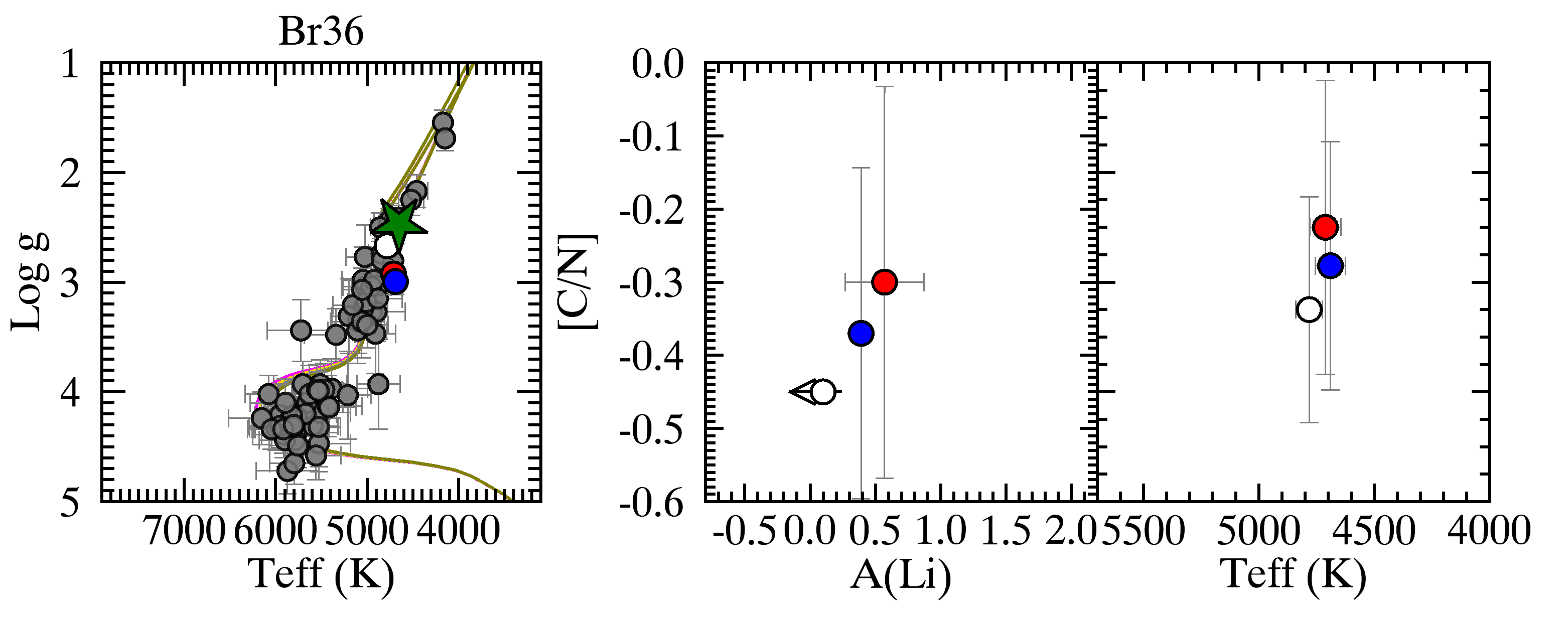} 
\includegraphics[scale=1.2]{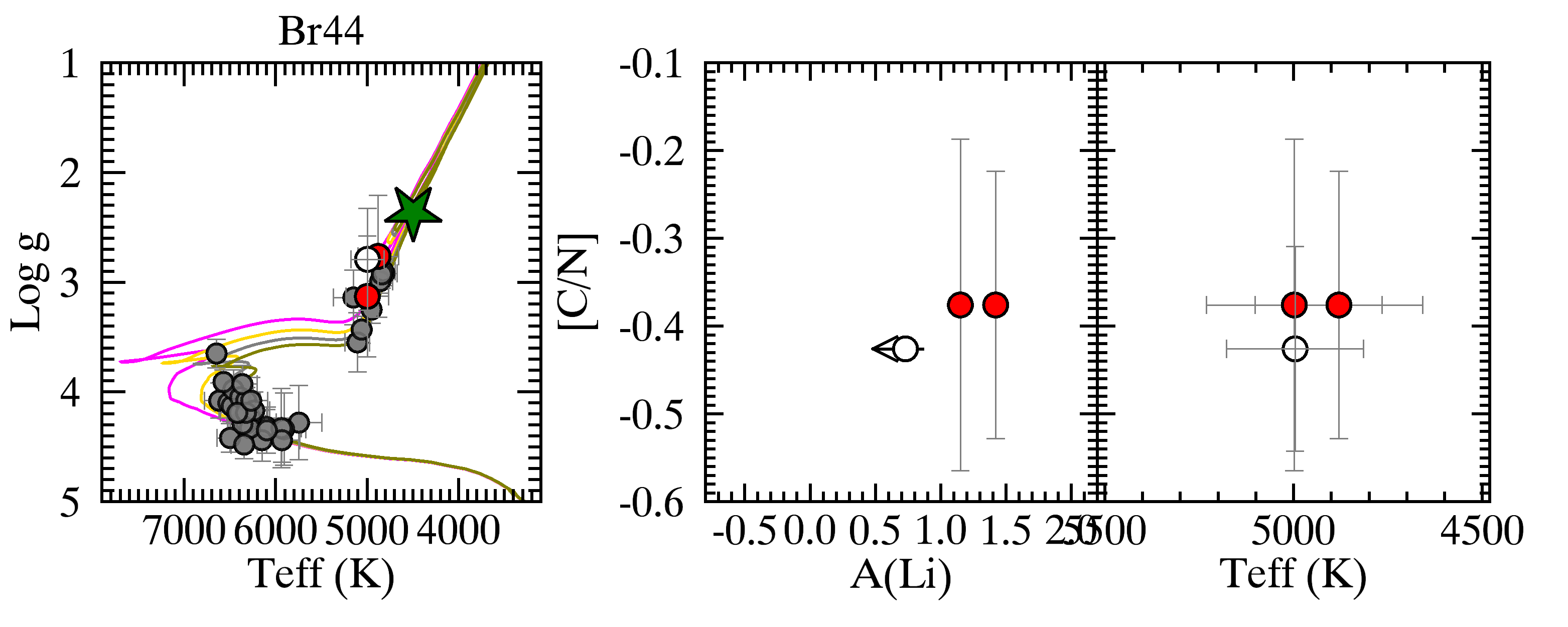} 
\includegraphics[scale=1.2]{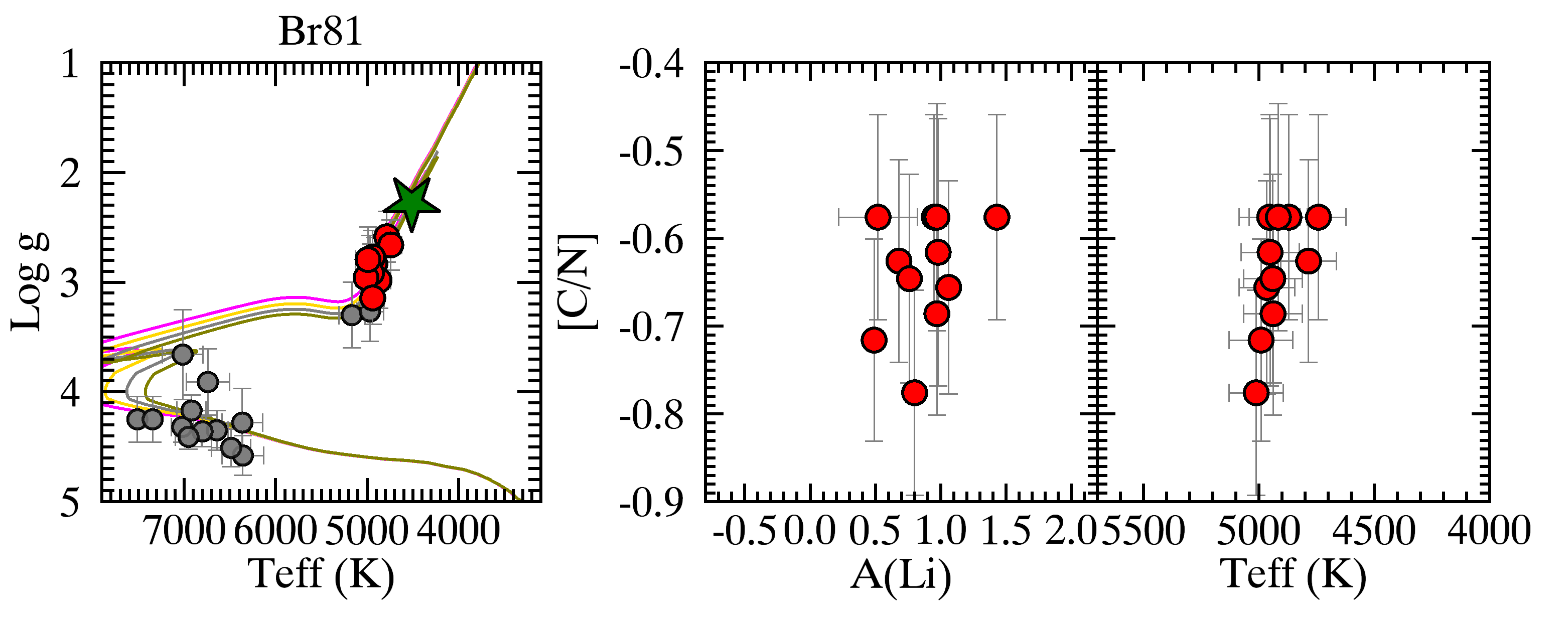} 
\caption{log~$g$-T$_{\rm eff}$ diagram with PISA isochrones (left panel) and member stars beyond the FDU, A(Li) vs [C/N] (central panel), and  [C/N] abundance vs T$_{\rm eff}$ (right panel) of the GES clusters. Symbols and colours as in Figure~\ref{fig:lithium1}. The used isochrones are 2.1, 2.3, 2.5, 2.7 Gyr, 6.0, 6.5, 7.0, 7.5 Gyr, 1.1, 1.4, 1.7, 2.0 Gyr and 0.7, 0.8, 0.9, 1.0 Gyr, respectively.  \label{fig:lithium:all1}}
\end{figure*}
\begin{figure*}[h]
\center
\includegraphics[scale=1.2]{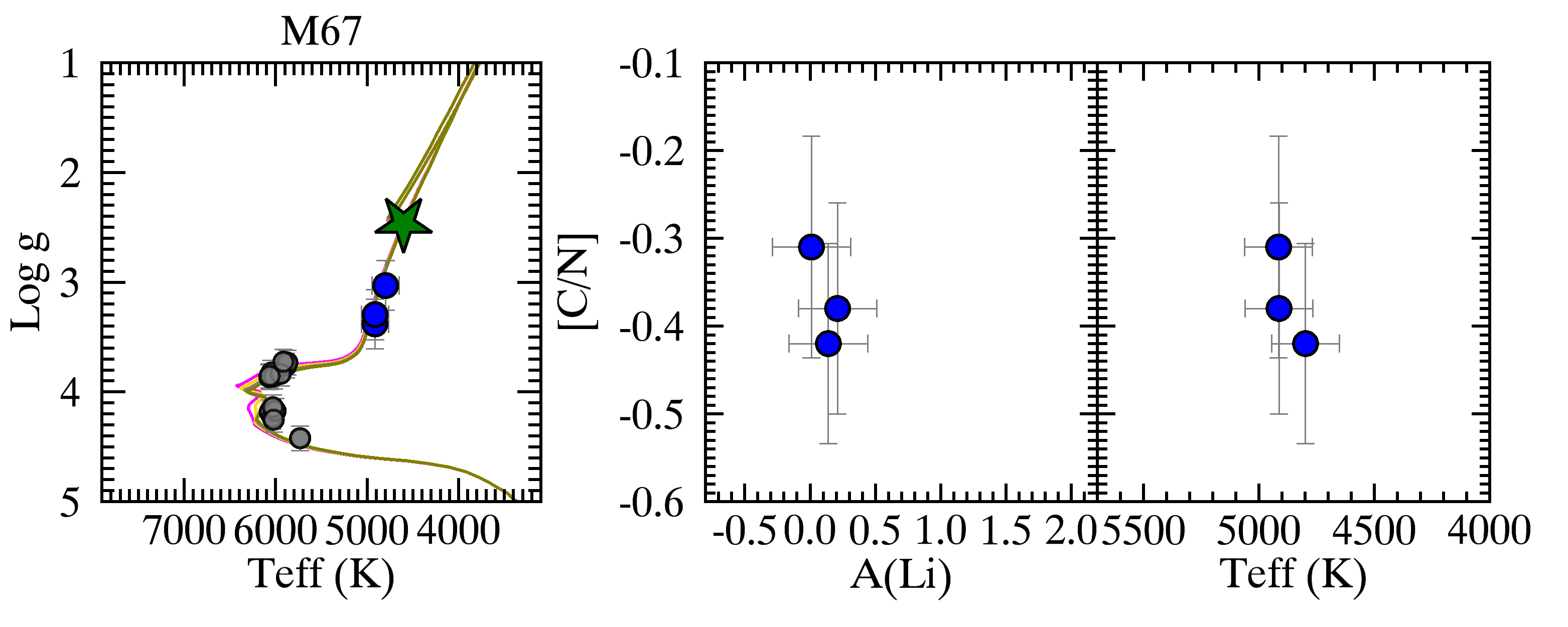} 
\includegraphics[scale=1.2]{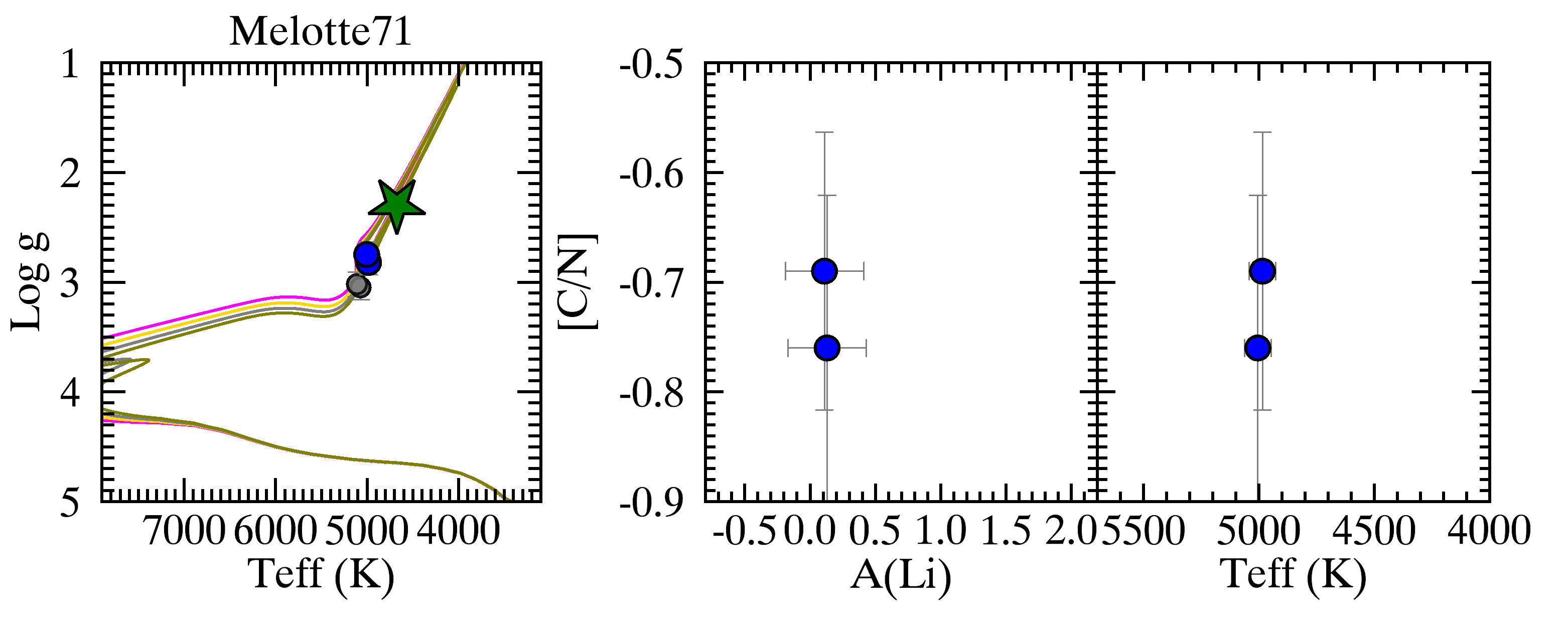} 
\includegraphics[scale=1.2]{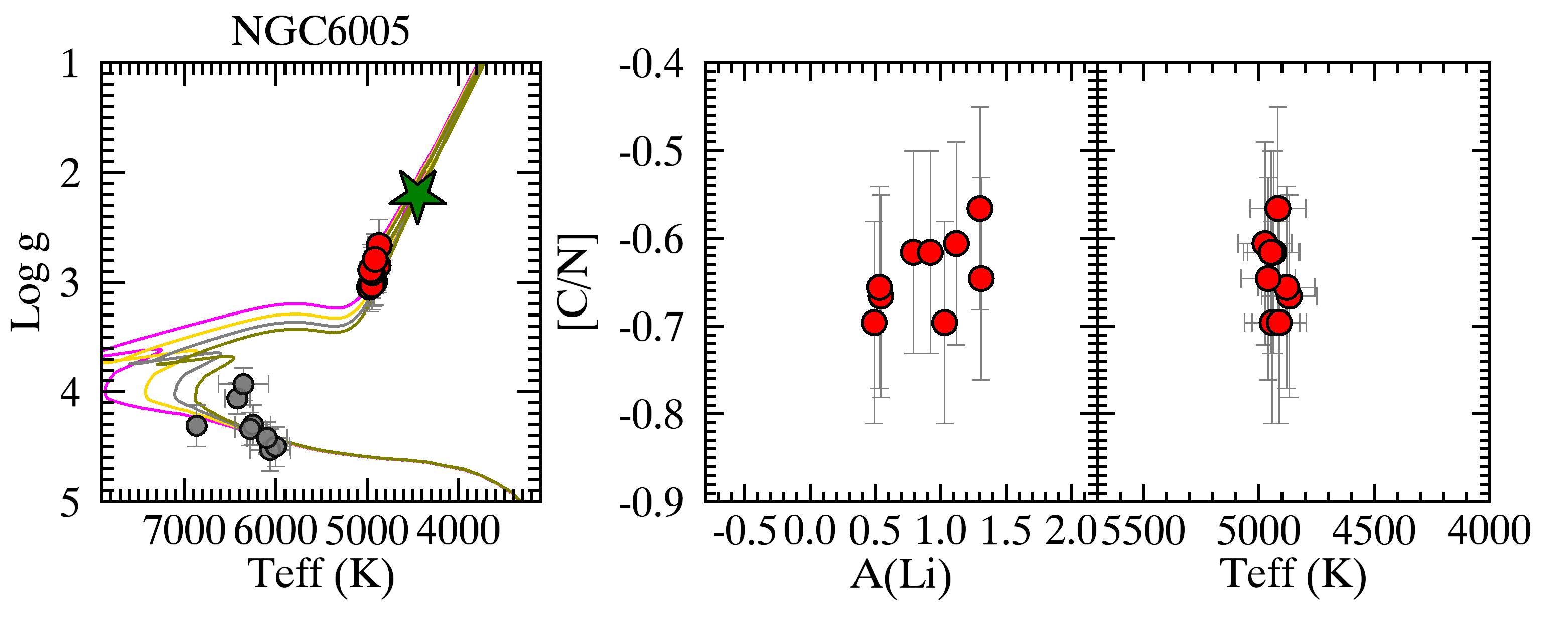} 
\includegraphics[scale=1.2]{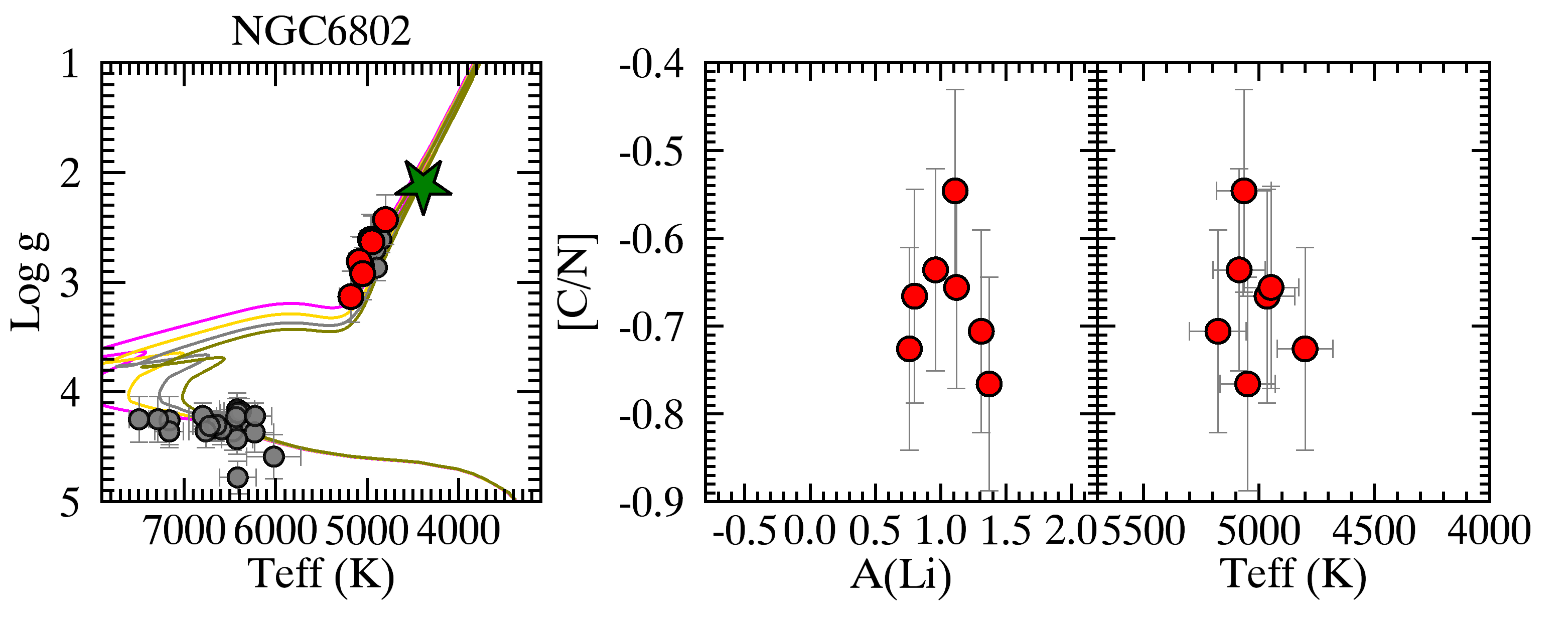}  
\caption{log~$g$-T$_{\rm eff}$ diagram with PISA isochrones (left panel) and member stars beyond the FDU, A(Li) vs [C/N] (central panel), and  [C/N] abundance vs T$_{\rm eff}$ (right panel) of the GES clusters. Symbols and colours as in Figure~\ref{fig:lithium1}. The used isochrones are 3.7, 4.0, 4.3, 4.5 Gyr, 0.7, 0.8, 0.9, 1.0 Gyr, 0.8, 1.0, 1.2, 1.4 Gyr and 0.8, 1.0, 1.2, 1.4 Gyr, respectively. \label{fig:lithium:all2}}
\end{figure*}
\begin{figure*}[h]
\center
\includegraphics[scale=1.2]{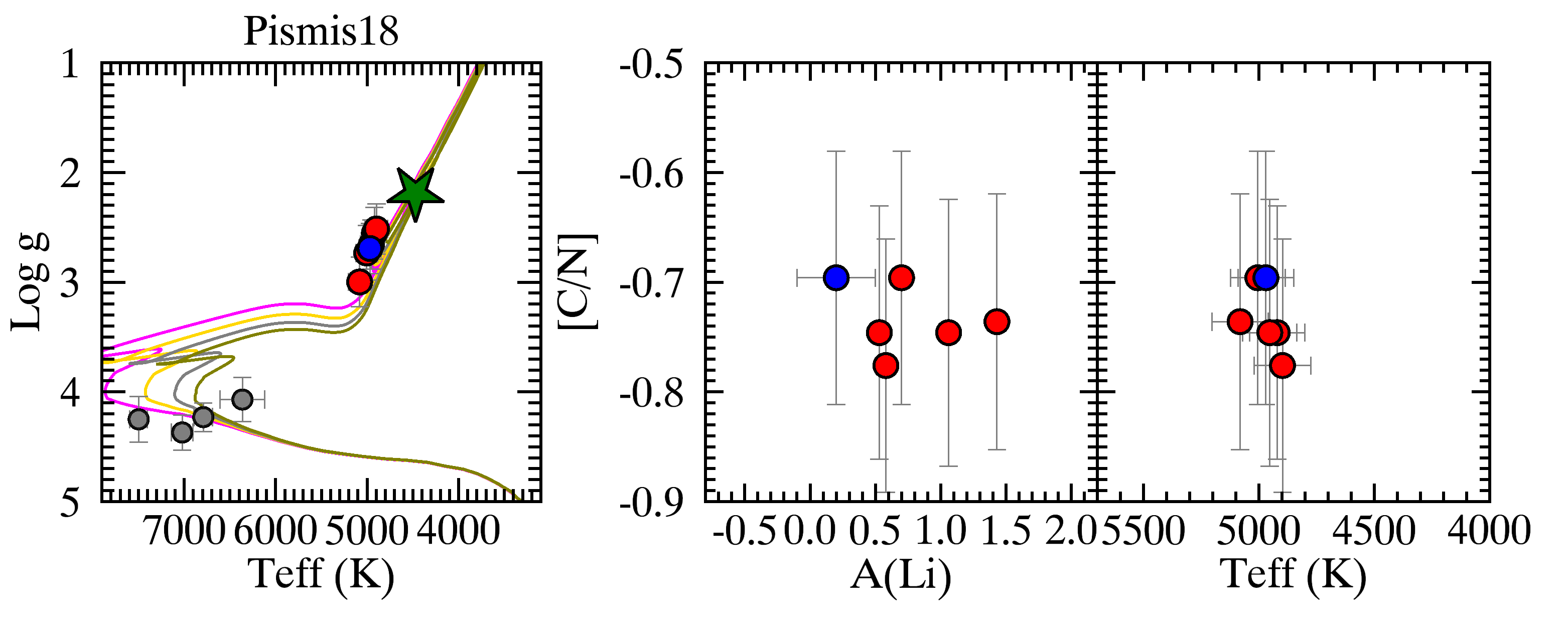}
\includegraphics[scale=1.2]{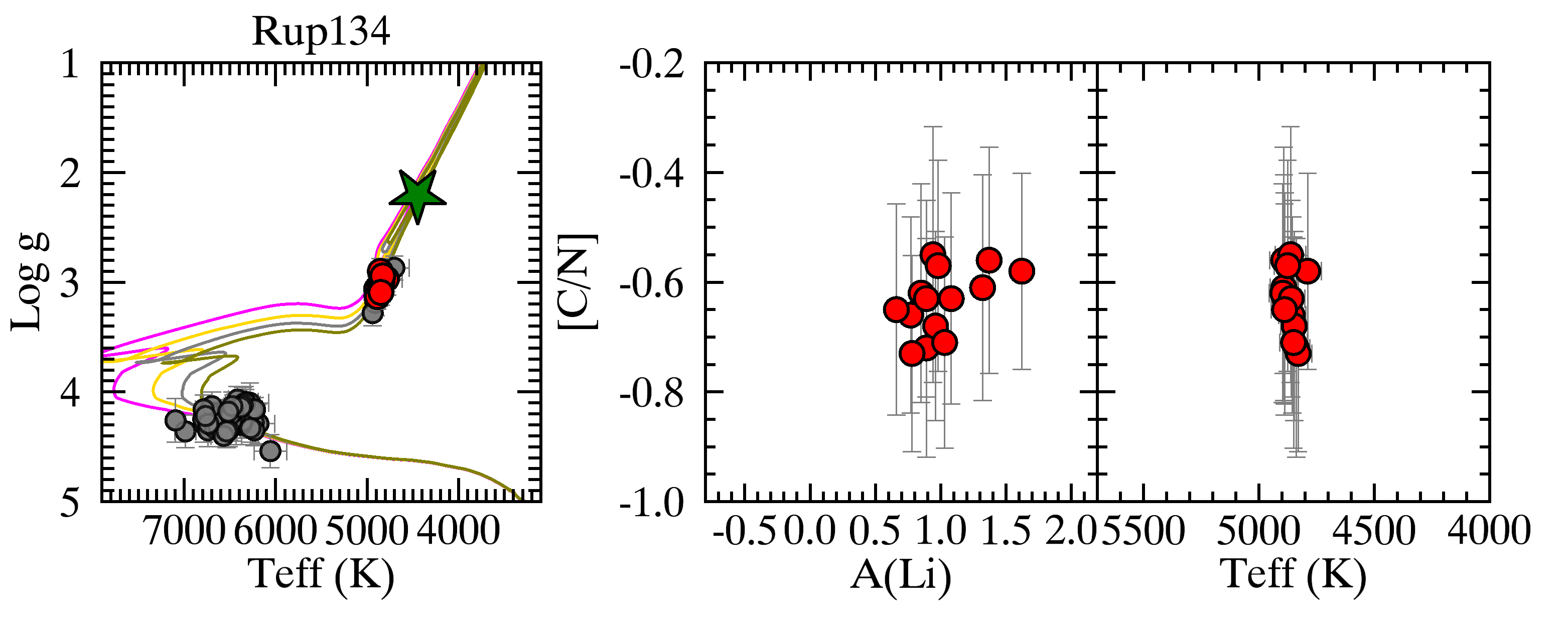}
\includegraphics[scale=1.2]{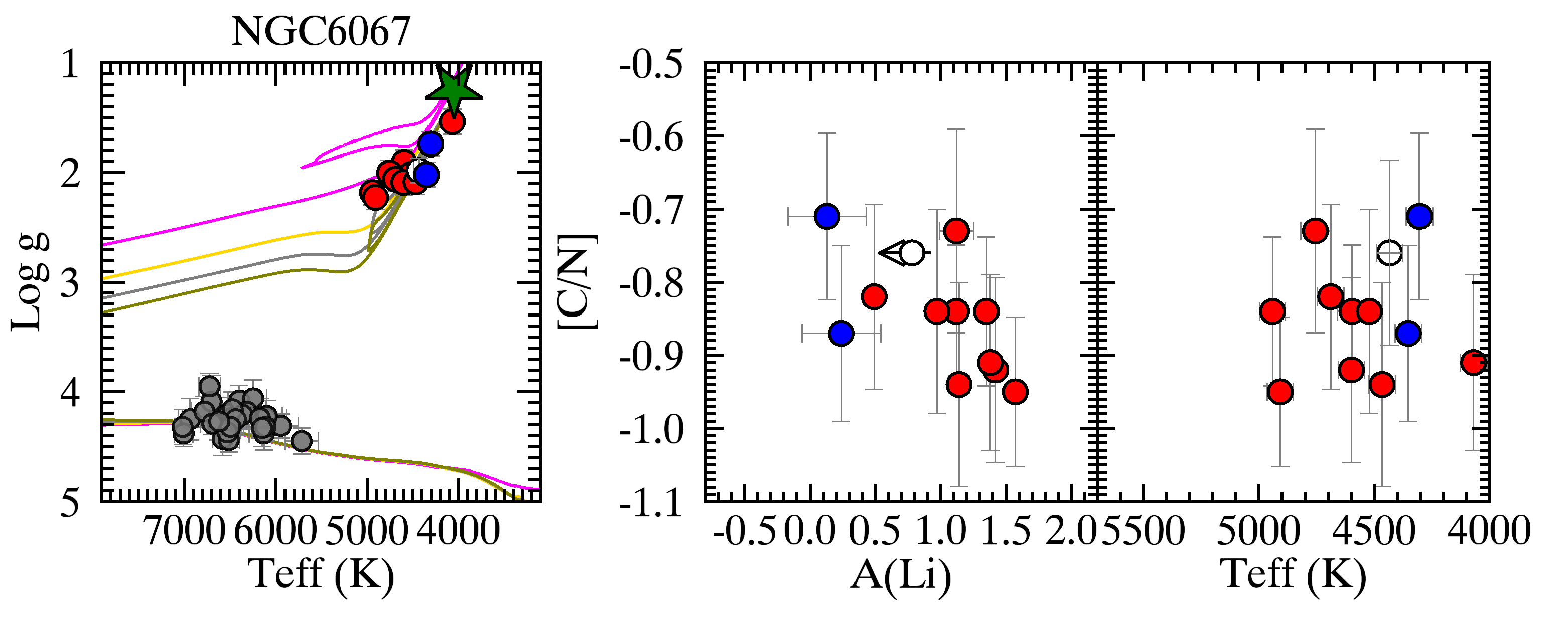}
\includegraphics[scale=1.2]{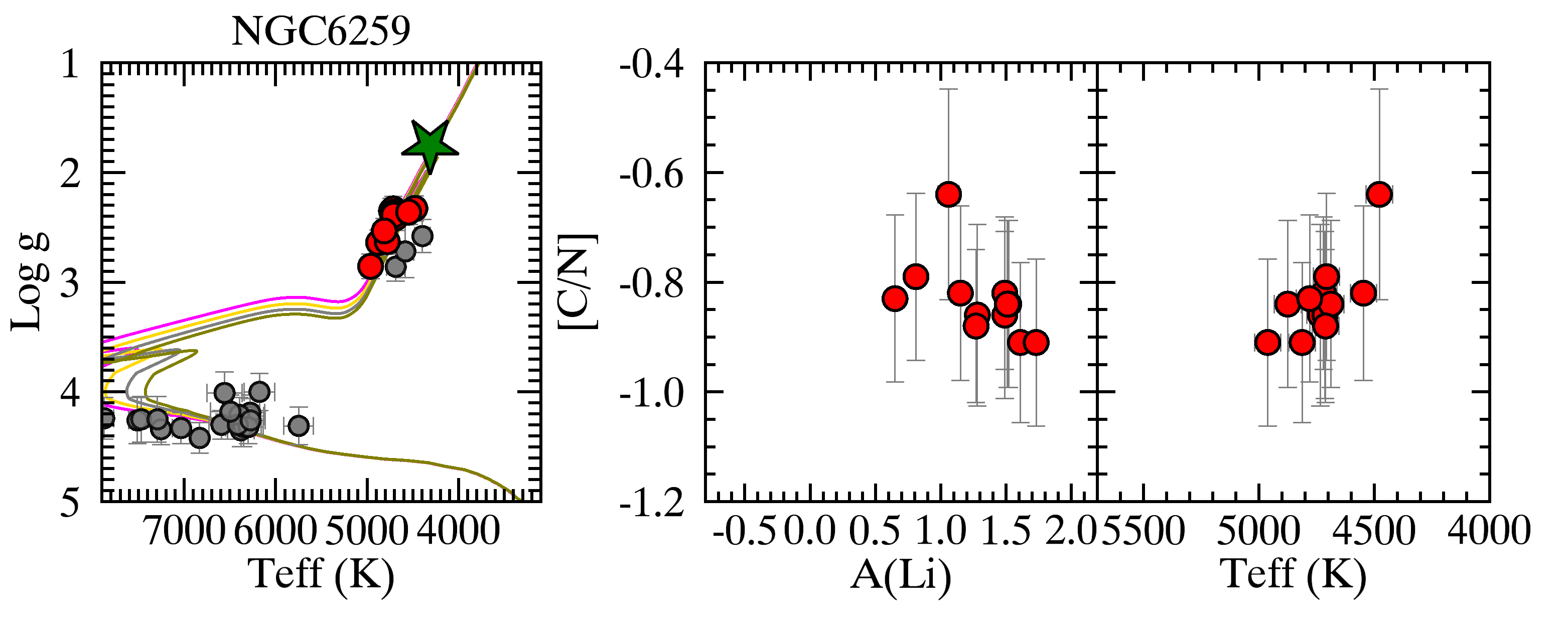}
\caption{log~$g$-T$_{\rm eff}$ diagram with PISA isochrones (left panel) and member stars beyond the FDU, A(Li) vs [C/N] (central panel), and  [C/N] abundance vs T$_{\rm eff}$ (right panel) of the GES clusters. Symbols and colours as in Figure~\ref{fig:lithium1}. The used isochrones are 0.8, 1.0, 1.2, 1.4 Gyr for the first two clusters, 0.05, 0.1, 0.2, 0.3 Gyr and 0.1, 0.2, 0.3, 0.4 Gyr, respectively. \label{fig:lithium:all3}}
\end{figure*}

%APOGEE membership
\begin{figure*}[h]
\center
\includegraphics[scale=1.0]{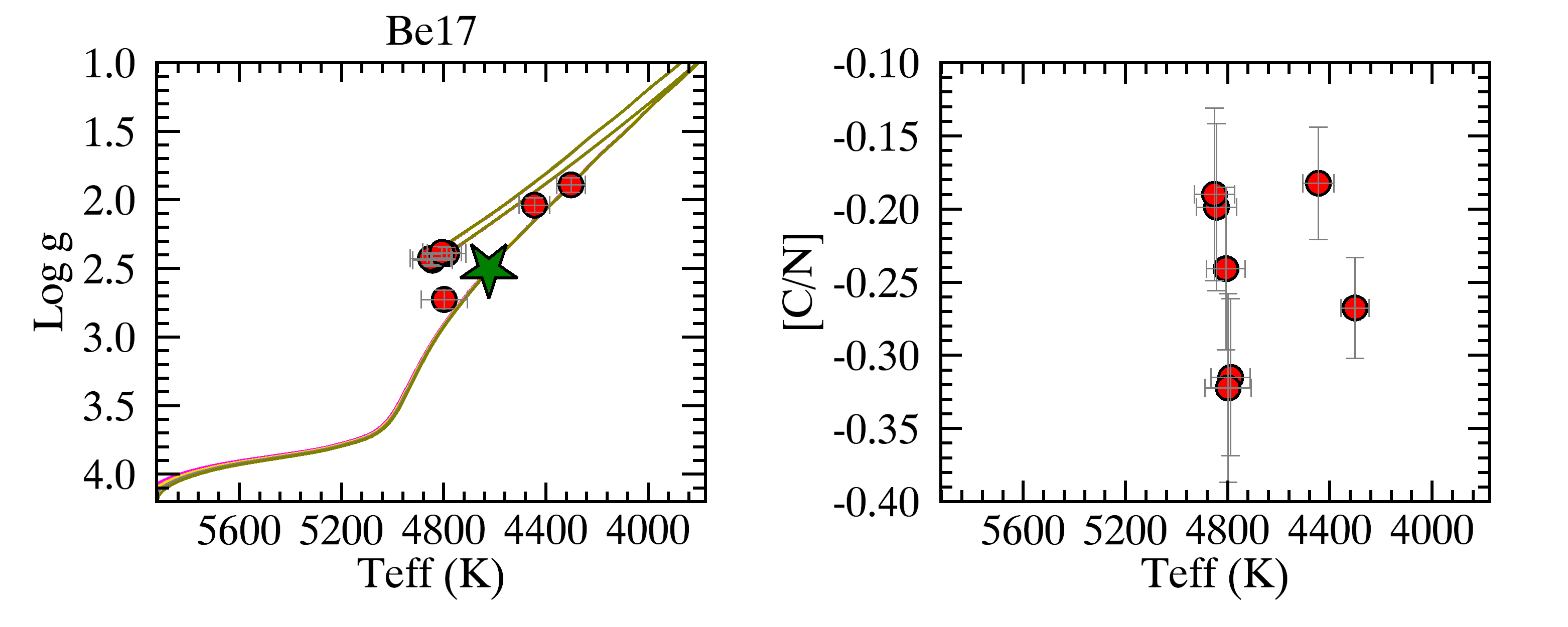} 
\includegraphics[scale=1.0]{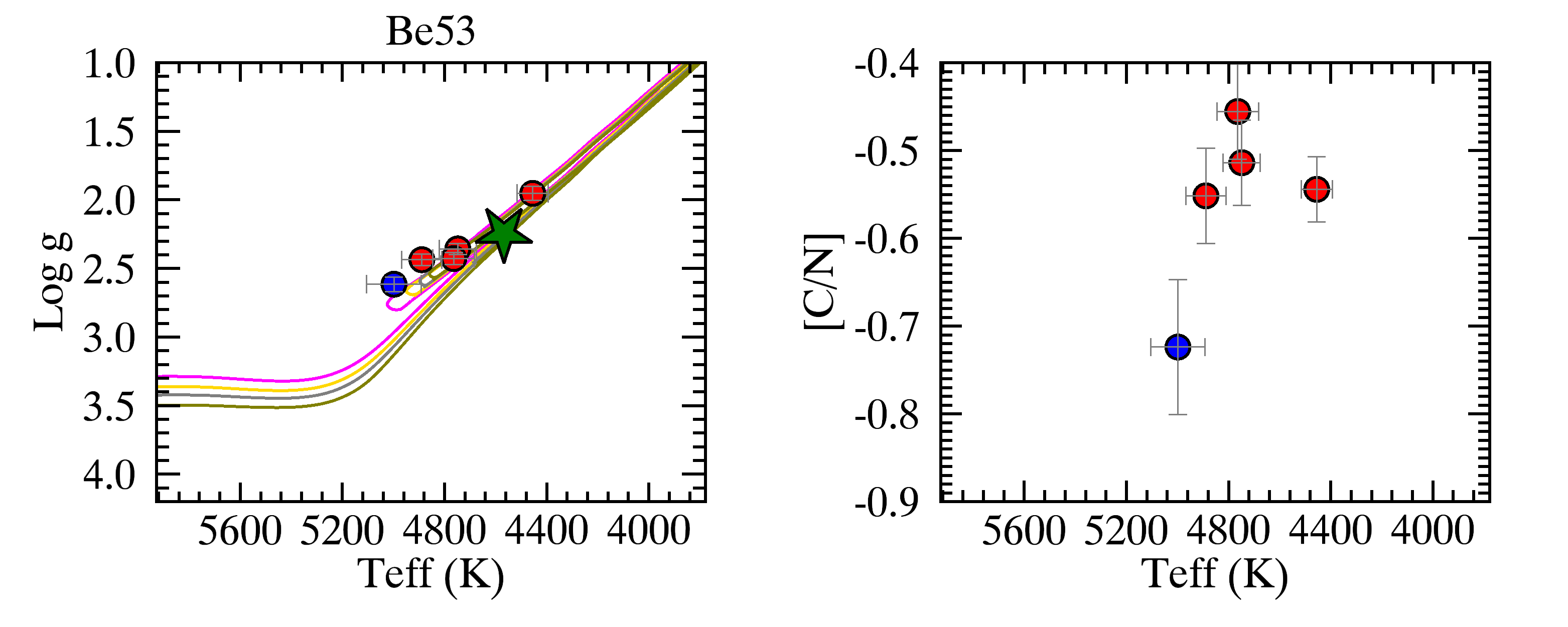} 
\includegraphics[scale=1.0]{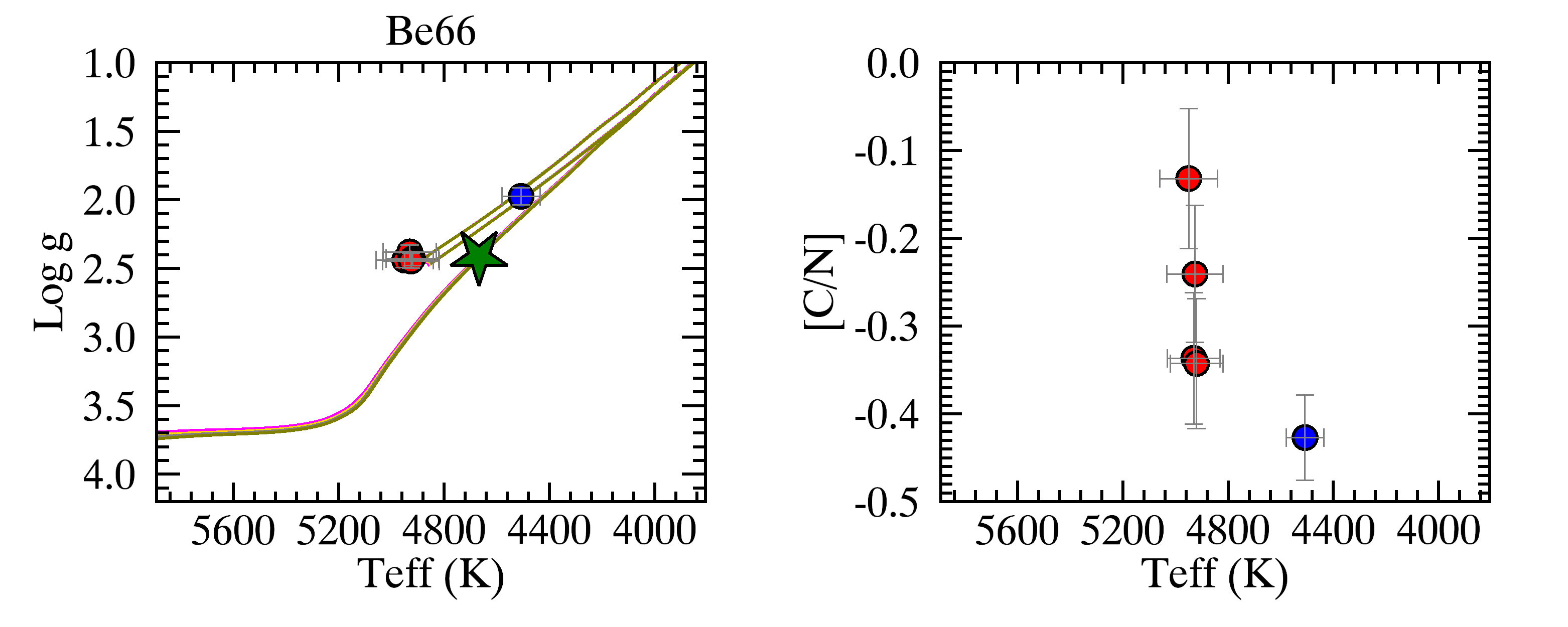} 
\includegraphics[scale=1.0]{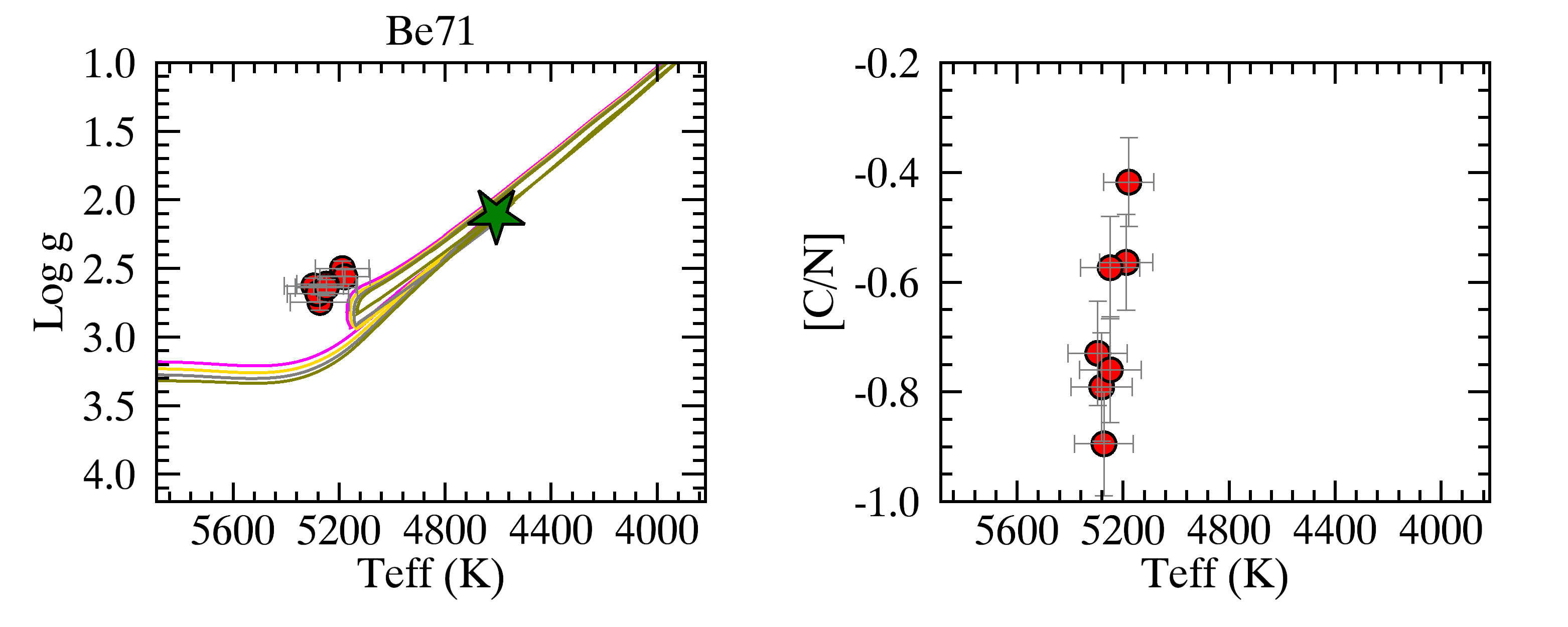} 
\caption{log~$g$-T$_{\rm eff}$ diagram with PISA isochrones (left panel) and  [C/N] abundance vs T$_{\rm eff}$ (right panel) of the APOGEE clusters. Symbols and colours as in Figure~\ref{fig:apo}. The used isochrones are 8.8, 9.1, 9.4, 9.7 Gyr, 0.1, 1.2, 1.4, 1.7 Gyr, 3.2, 3.4, 3.6, 3.8 Gyr and 0.8, 0.9, 1.0, 1.1 Gyr, respectively. \label{fig:apo:all1}} 
\end{figure*}
\begin{figure*}[h]
\center
\includegraphics[scale=1.0]{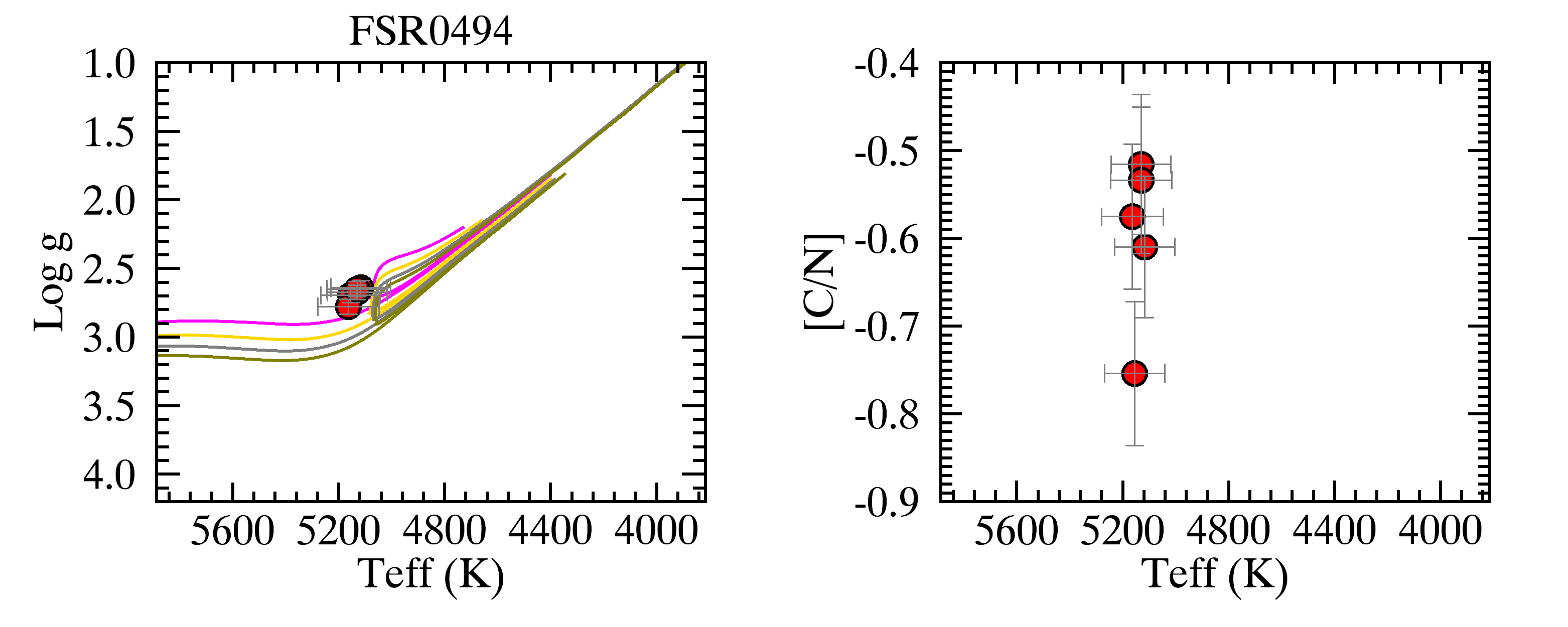} 
\includegraphics[scale=1.0]{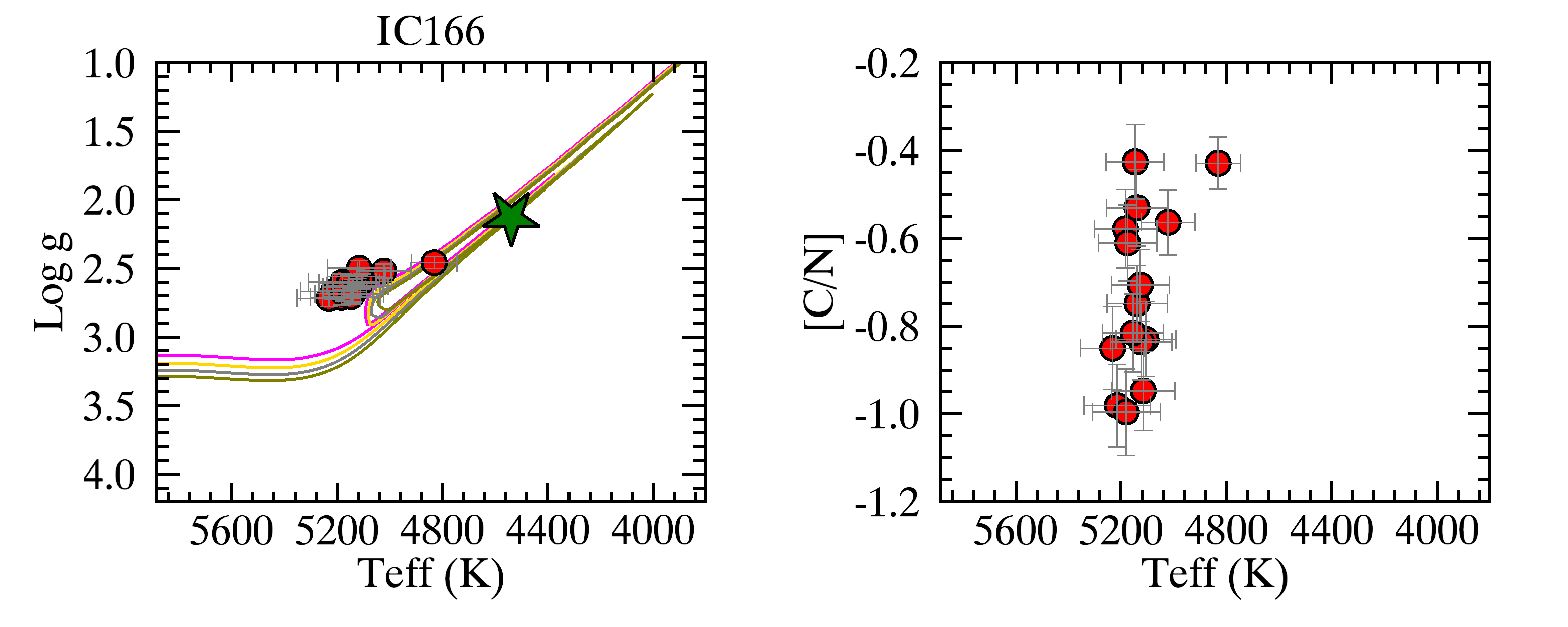} 
\includegraphics[scale=1.0]{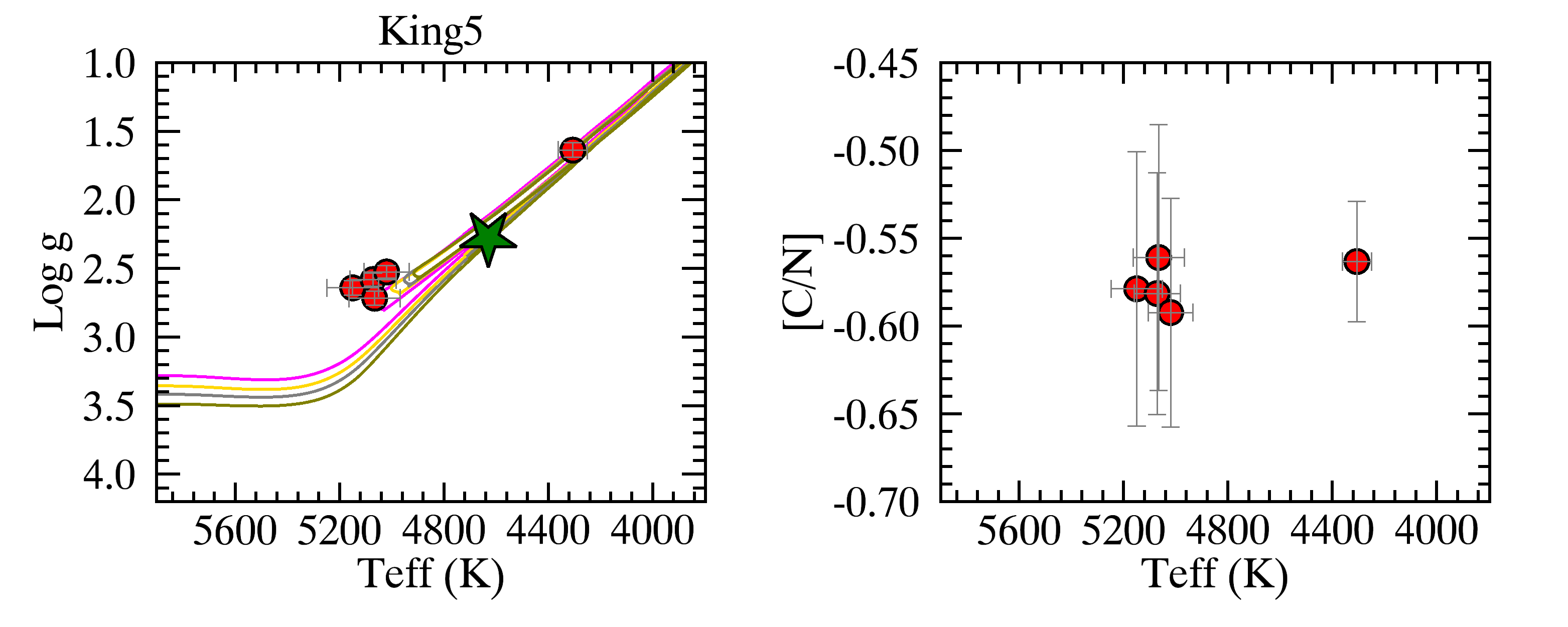} 
\includegraphics[scale=1.0]{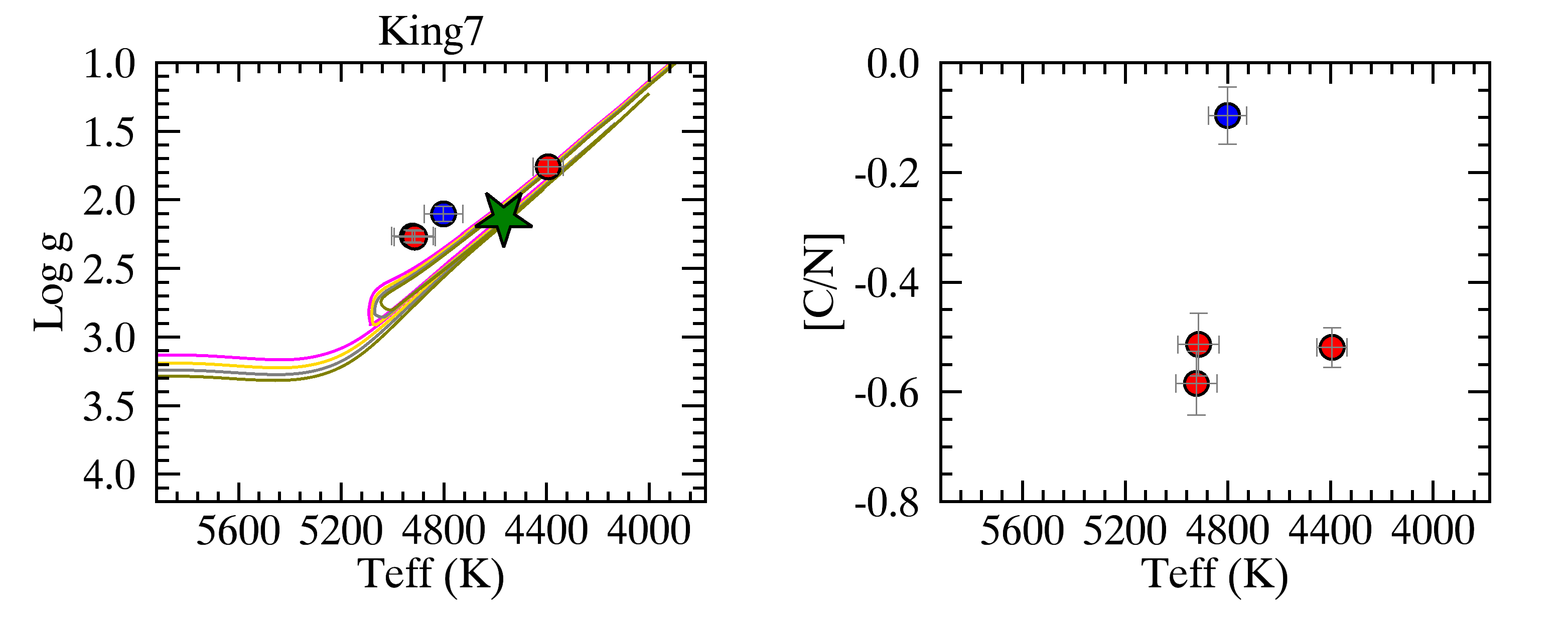}
\caption{log~$g$-T$_{\rm eff}$ diagram with PISA isochrones (left panel) and  [C/N] abundance vs T$_{\rm eff}$ (right panel) of the APOGEE clusters. Symbols and colours as in Figure~\ref{fig:apo}. The used isochrones are 0.4, 0.5, 0.6, 0.7 Gyr, 0.6, 0.8, 1.0, 1.2 Gyr, 1.0, 1.2, 1.4, 1.7 Gyr and 0.4, 0.5, 0.6, 0.7 Gyr, respectively. \label{fig:apo:all2}} 
\end{figure*}
\begin{figure*}[h]
\center
\includegraphics[scale=1.0]{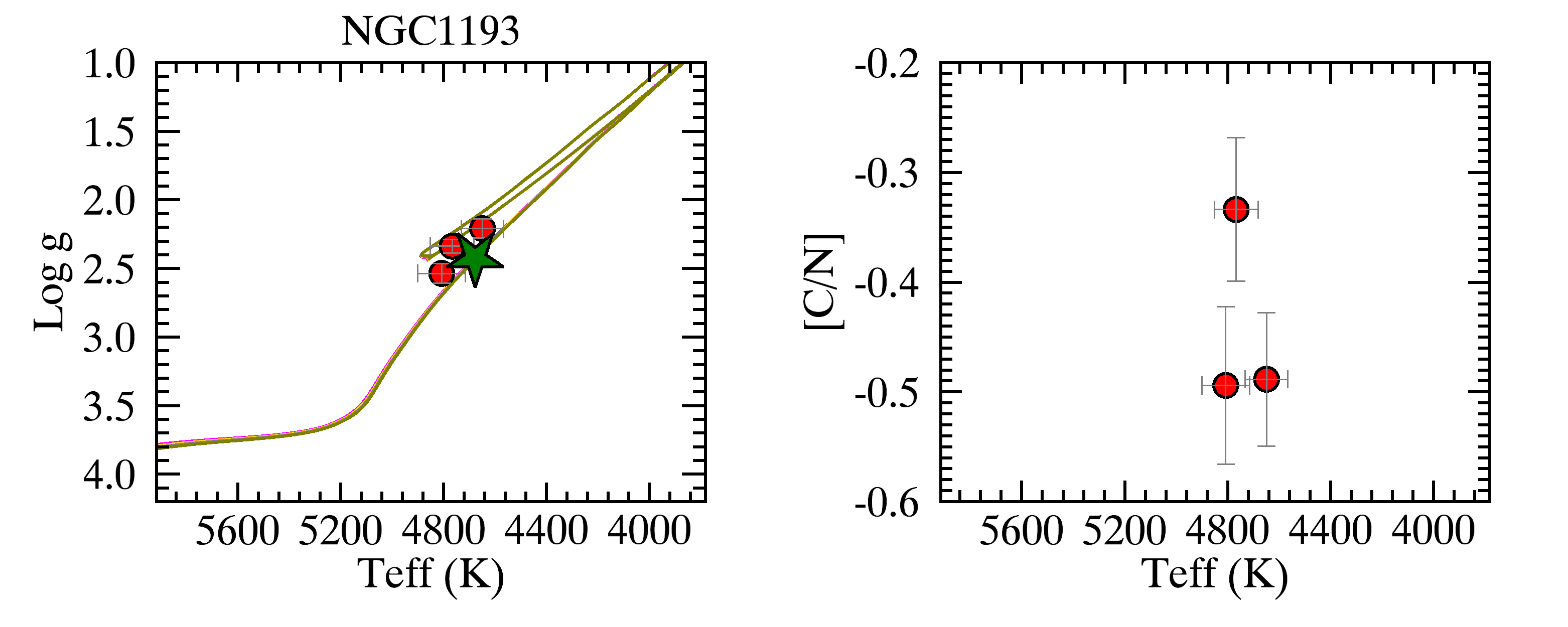} 
\includegraphics[scale=1.0]{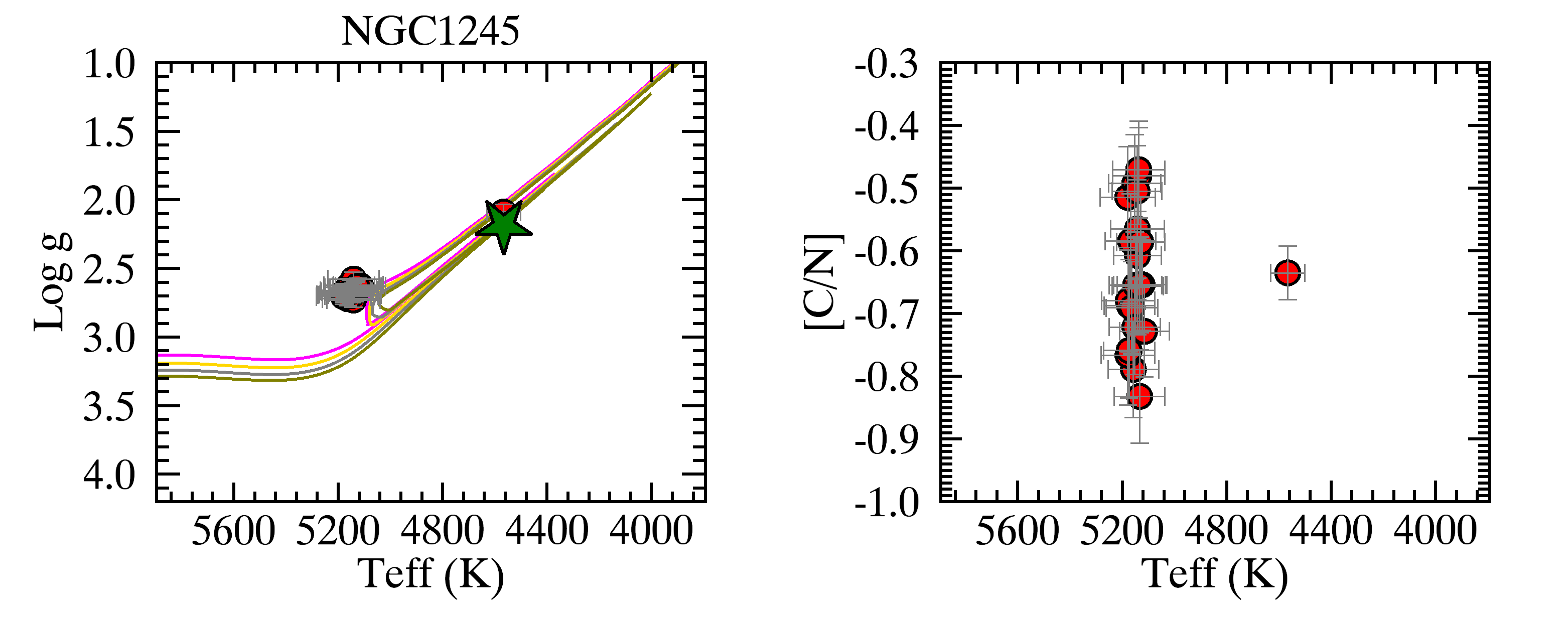}
\includegraphics[scale=1.0]{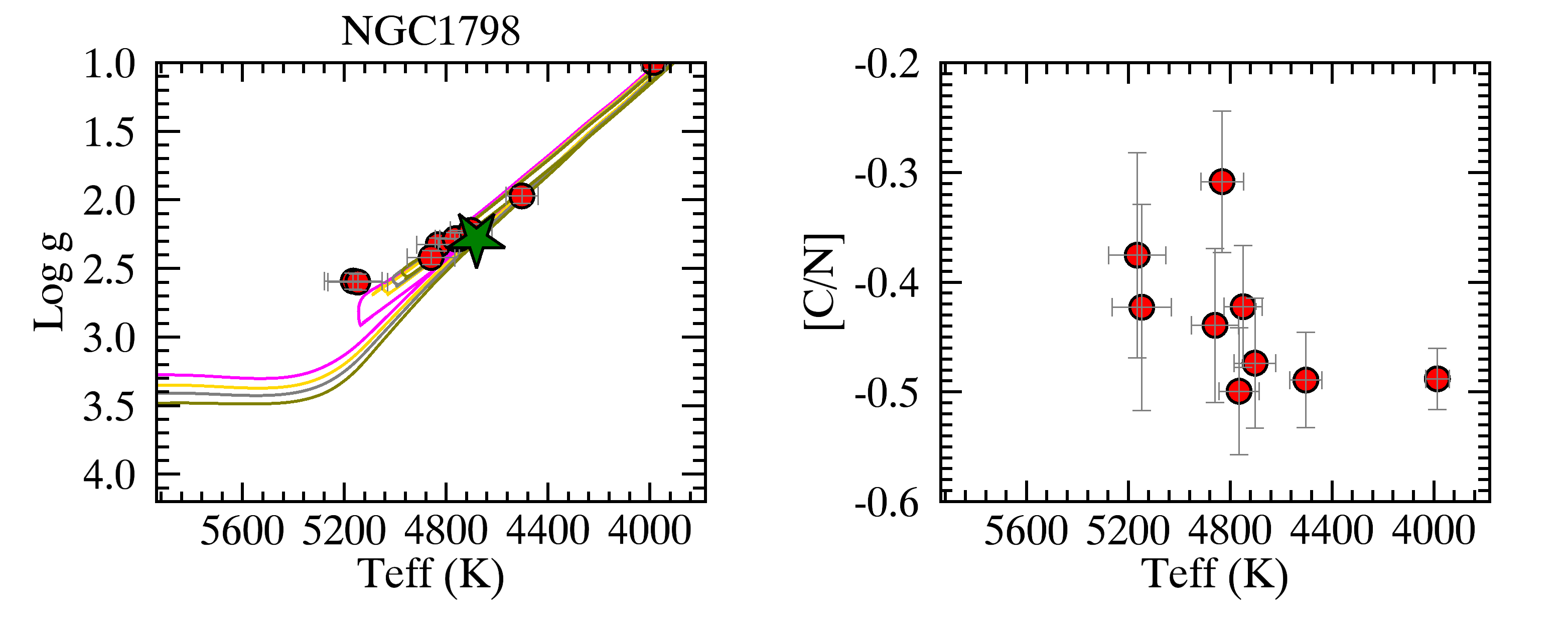}
\includegraphics[scale=1.0]{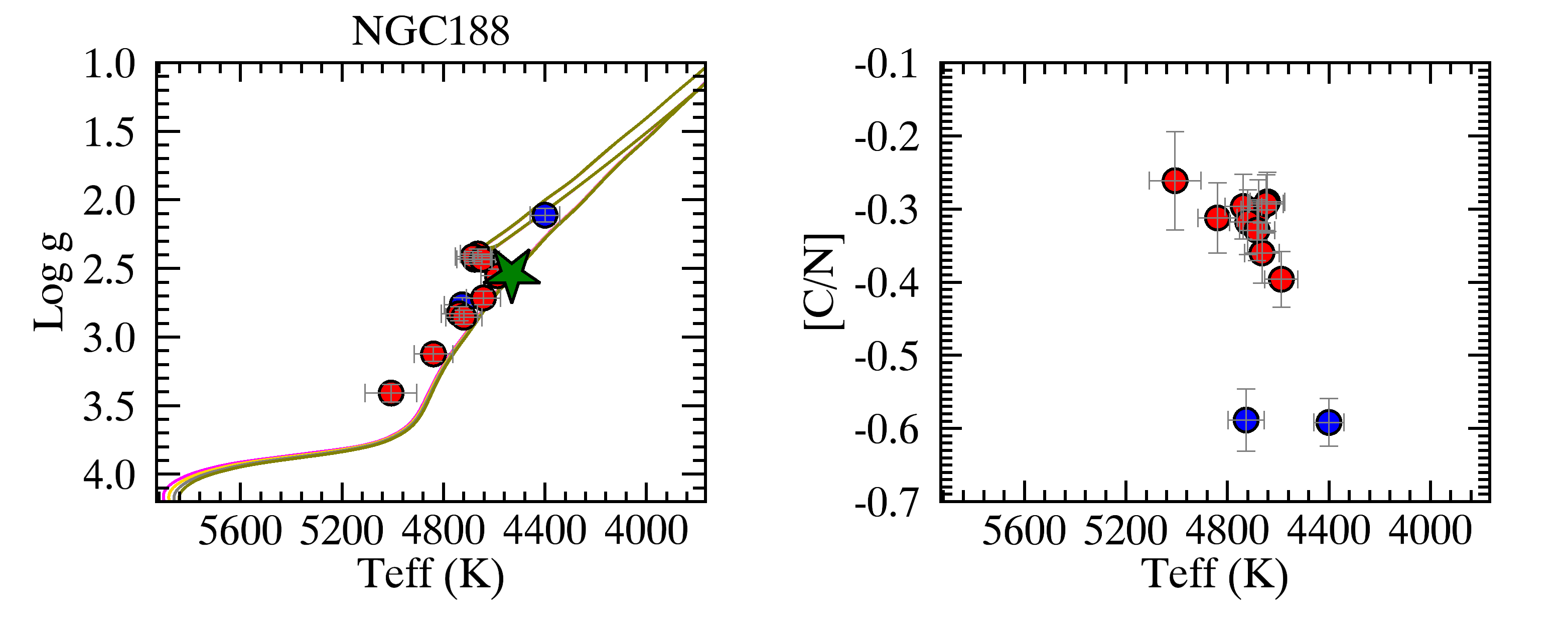}
\caption{log~$g$-T$_{\rm eff}$ diagram with PISA isochrones (left panel) and  [C/N] abundance vs T$_{\rm eff}$ (right panel) of the APOGEE clusters. Symbols and colours as in Figure~\ref{fig:apo}. The used isochrones are 4.6, 4.8, 5.0, 5.2 Gyr, 0.7, 0.8, 0.9, 1.0 Gyr, 1.0, 1.2, 1.4, 1.7 Gyr and 7.3, 7.6, 7.9, 8.2 Gyr, respectively. \label{fig:apo:all3}}
\end{figure*}

\begin{figure*}[h]
\center
\includegraphics[scale=1.0]{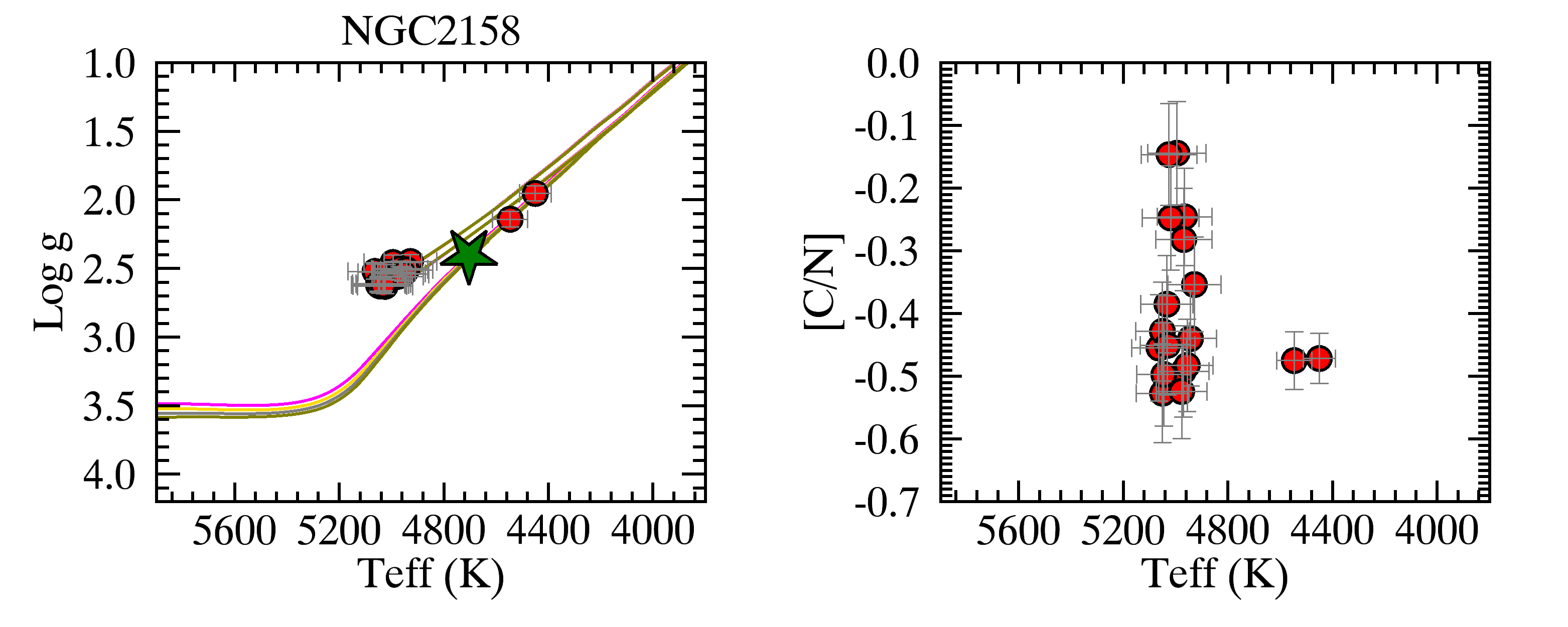} 
\includegraphics[scale=1.0]{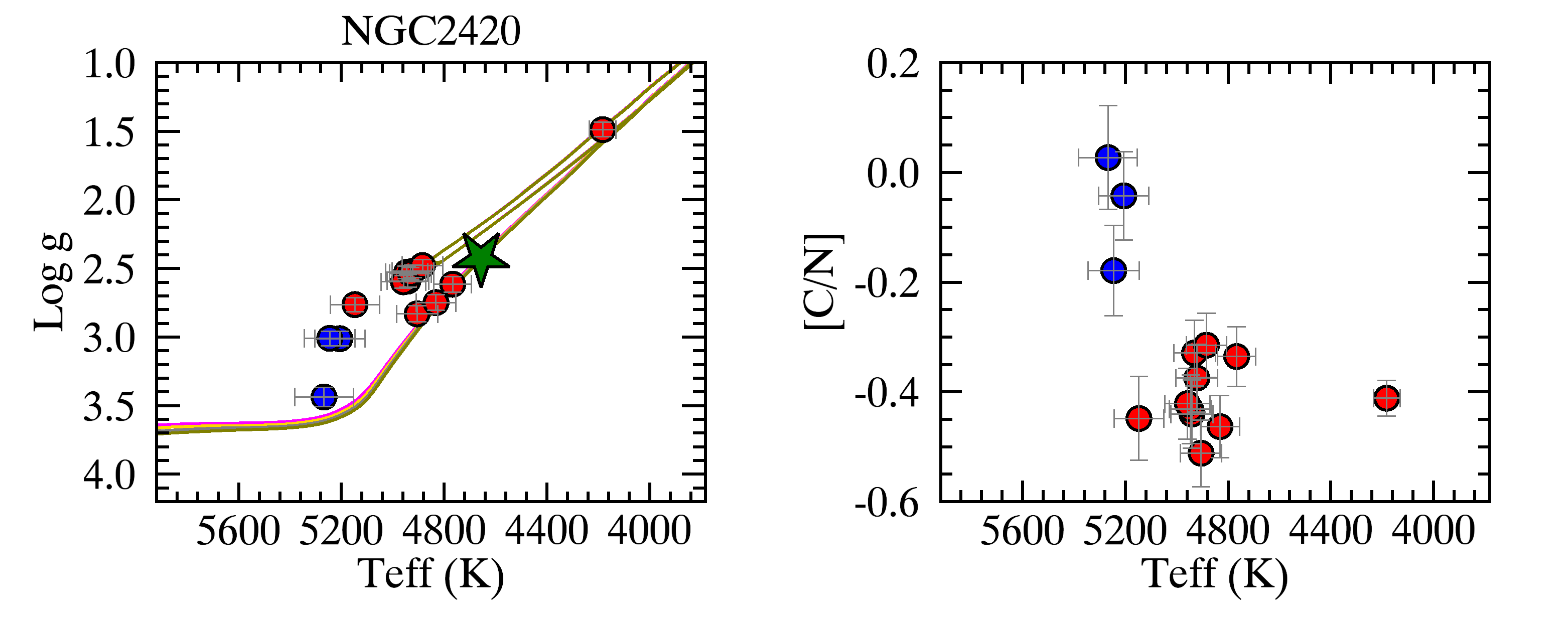}
\includegraphics[scale=1.0]{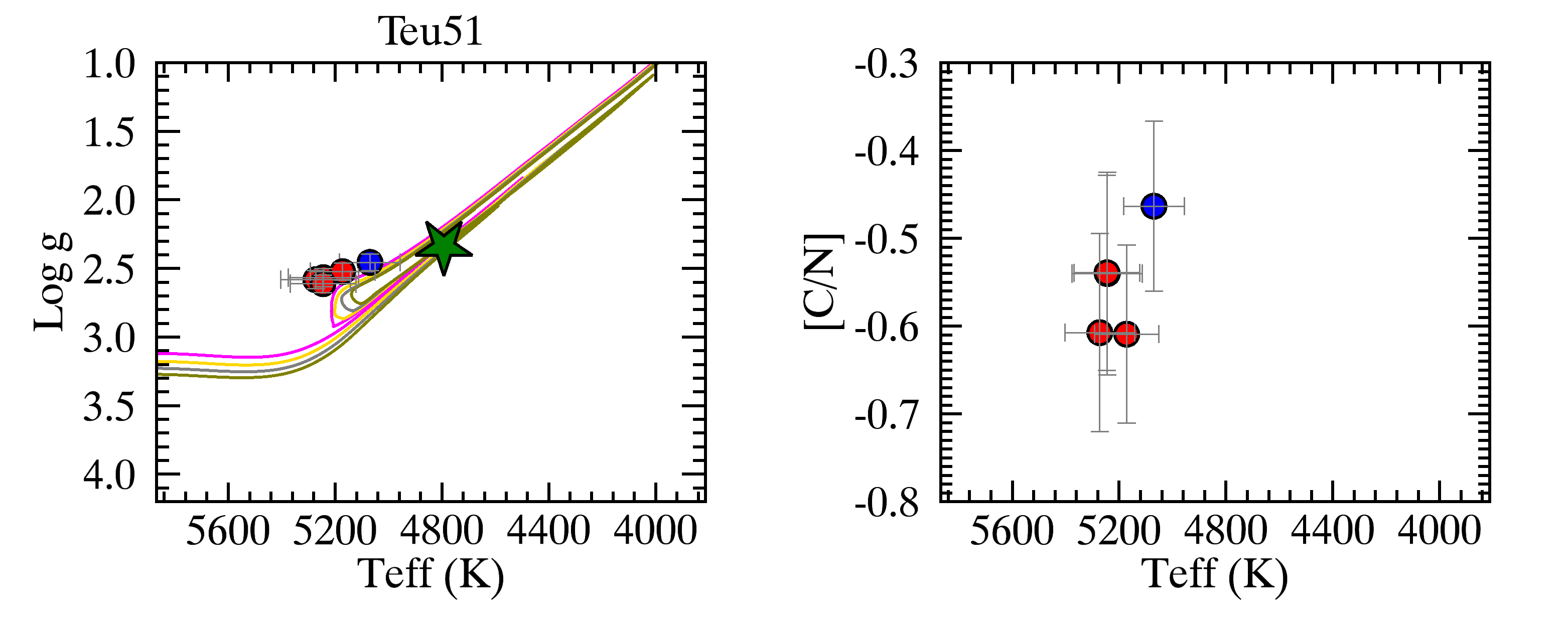}
\includegraphics[scale=1.0]{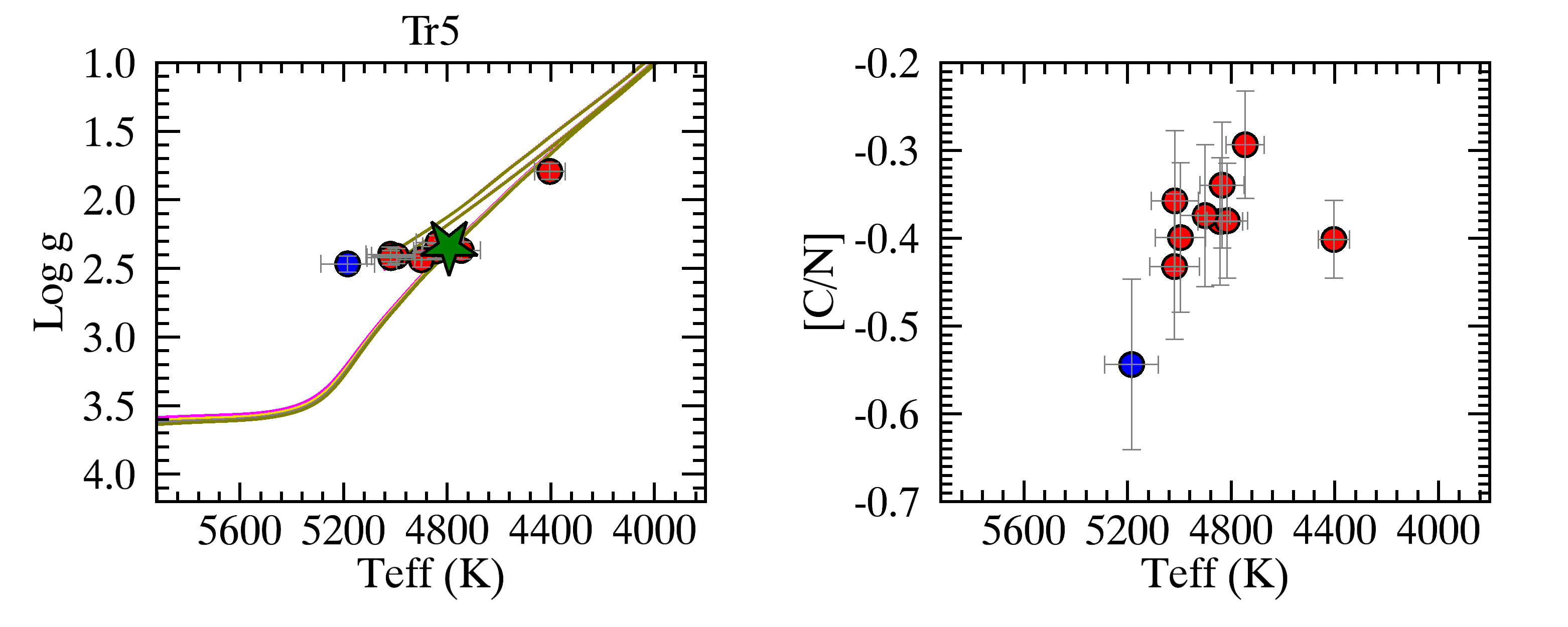}
\caption{log~$g$-T$_{\rm eff}$ diagram with PISA isochrones (left panel) and  [C/N] abundance vs T$_{\rm eff}$ (right panel) of the APOGEE clusters. Symbols and colours as in Figure~\ref{fig:apo}. \label{fig:apo:all4} The used isochrones are 1.7, 1.9, 2.1, 2.3 Gyr, 2.6, 2.8, 3.0, 3.2 Gyr, 0.7, 0.8, 0.9, 1.0 Gyr and 2.6, 2.8, 3.0, 3.2 Gyr, respectively.}
\end{figure*}

\begin{figure*}[h]
\center
\includegraphics[scale=1.0]{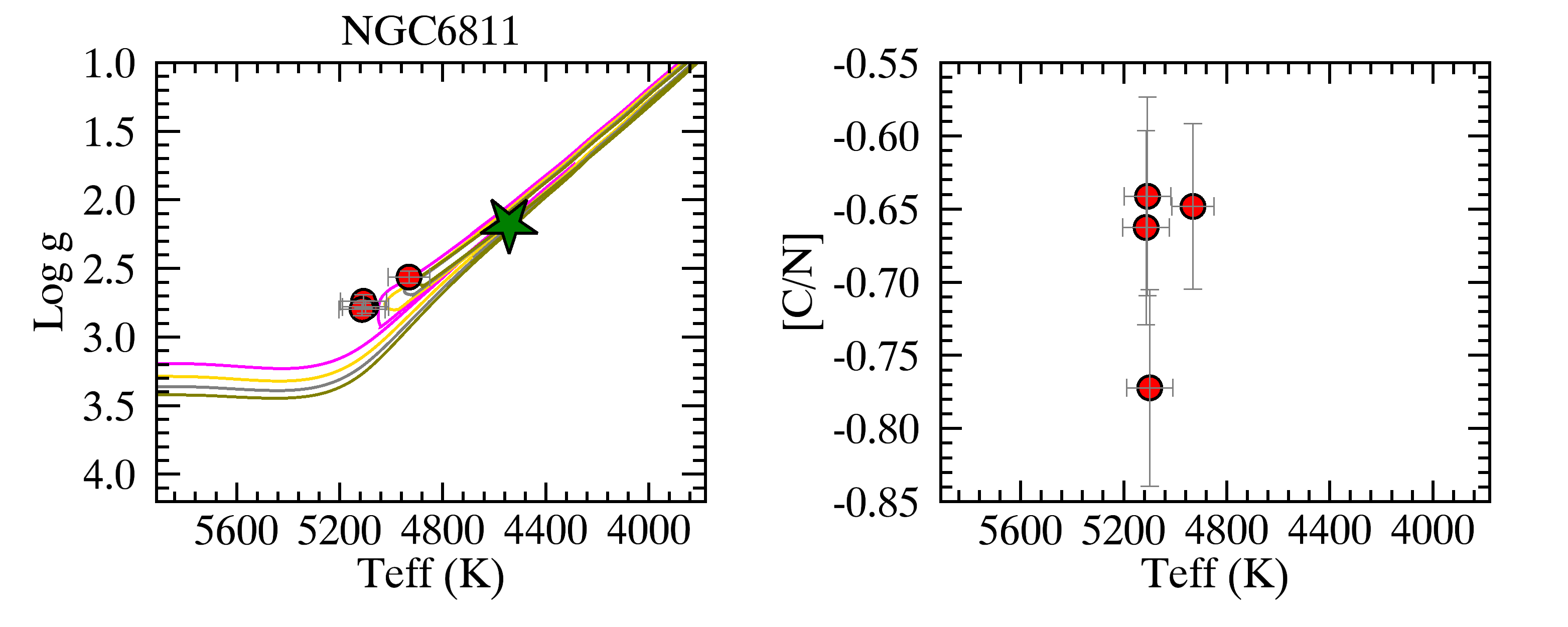} 
\includegraphics[scale=1.0]{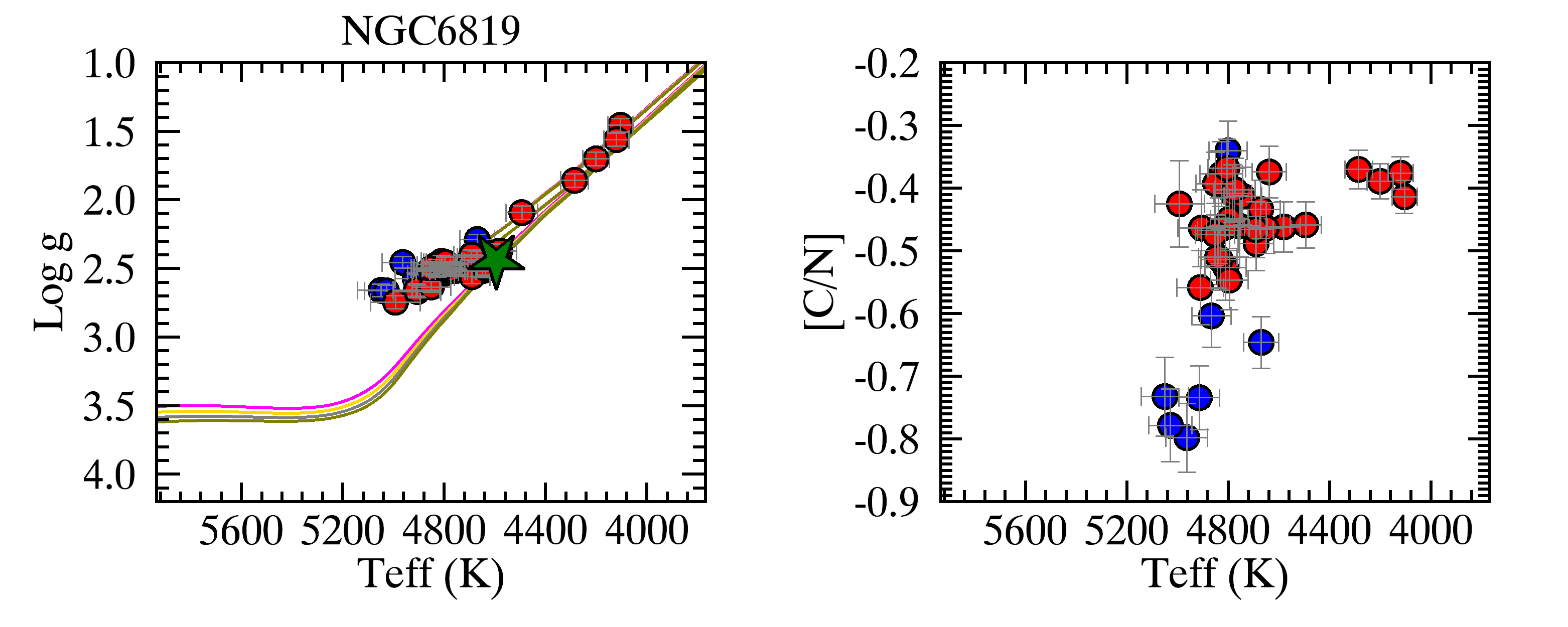}
\includegraphics[scale=1.0]{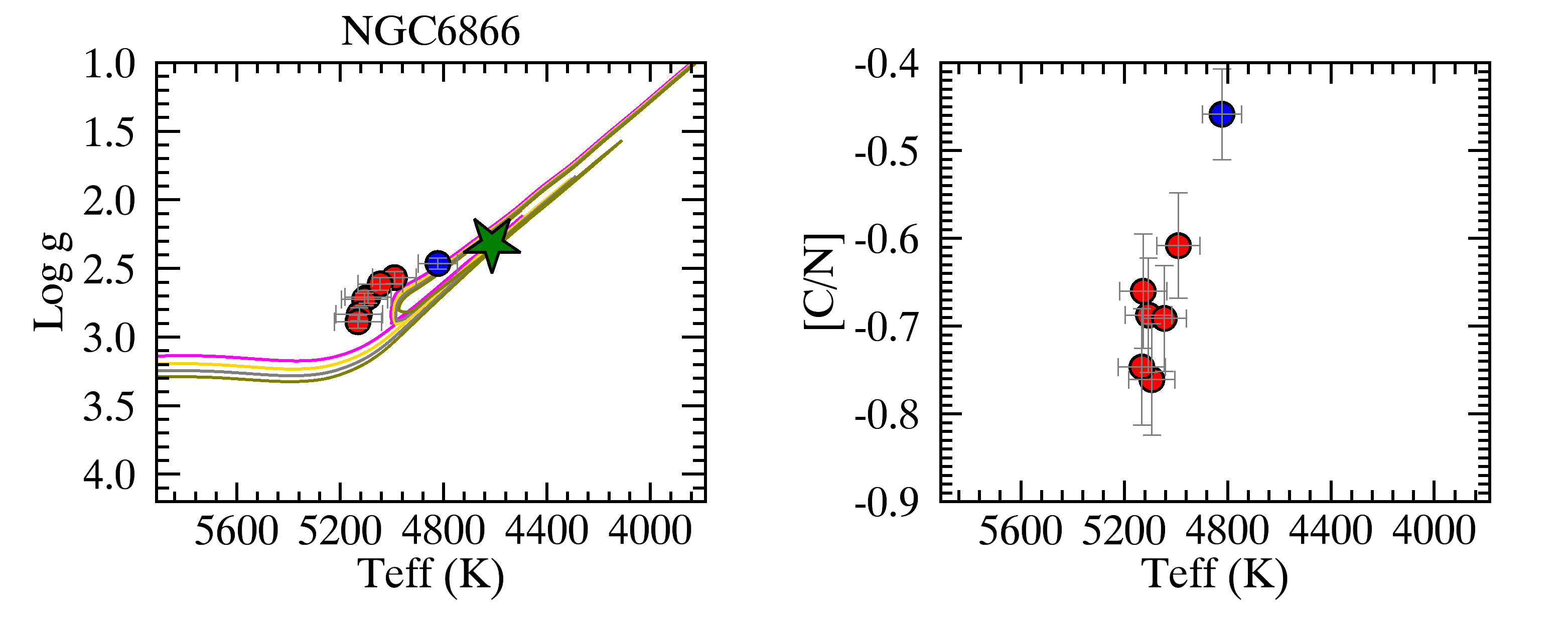}
\includegraphics[scale=1.0]{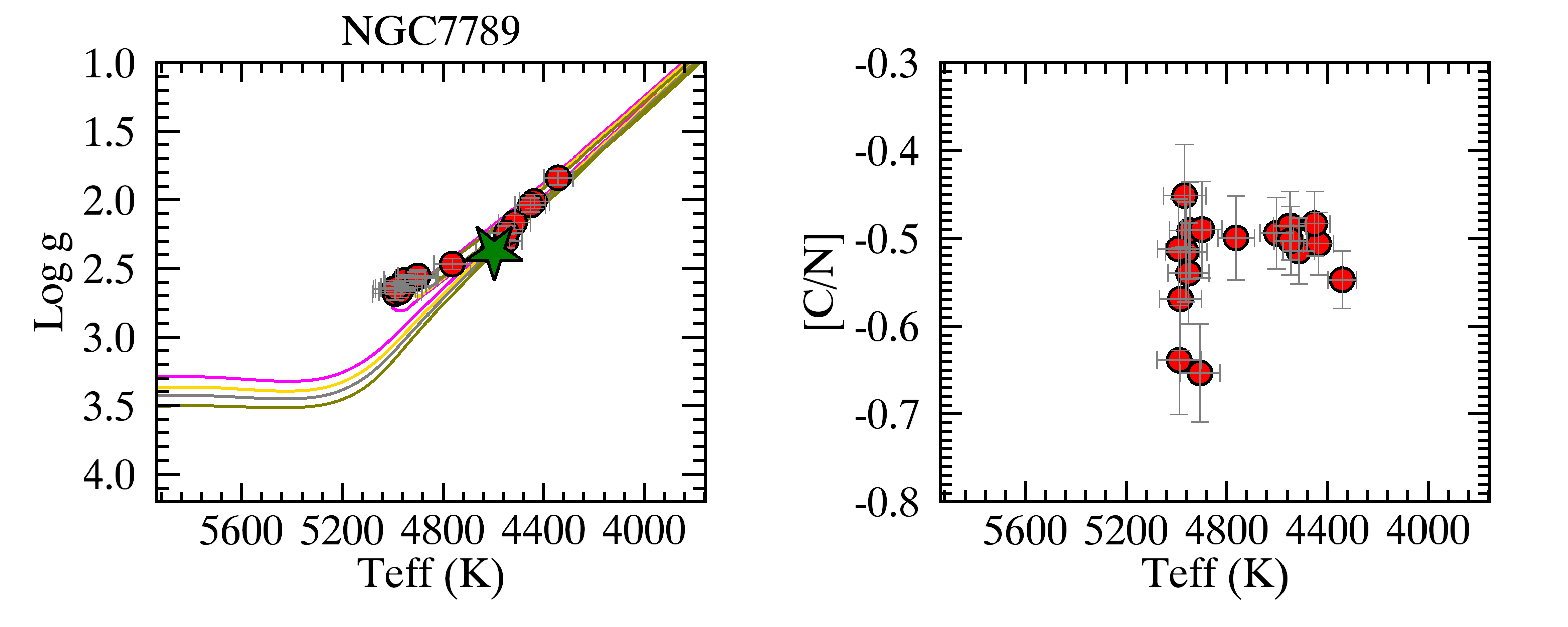}
\caption{log~$g$-T$_{\rm eff}$ diagram with PISA isochrones (left panel) and  [C/N] abundance vs T$_{\rm eff}$ (right panel) of the APOGEE clusters. Symbols and colours as in Figure~\ref{fig:apo}. The used isochrones are 0.8, 1.0, 1.2, 1.4 Gyr, 1.7, 1.9, 2.1, 2.3 Gyr, 0.7, 0.8, 0.9, 1.0 Gyr and 1.0, 1.2, 1.4, 1.7 Gyr, respectively. \label{fig:apo:all5}}
\end{figure*}

\begin{figure*}[h]
\center
\includegraphics[scale=1.0]{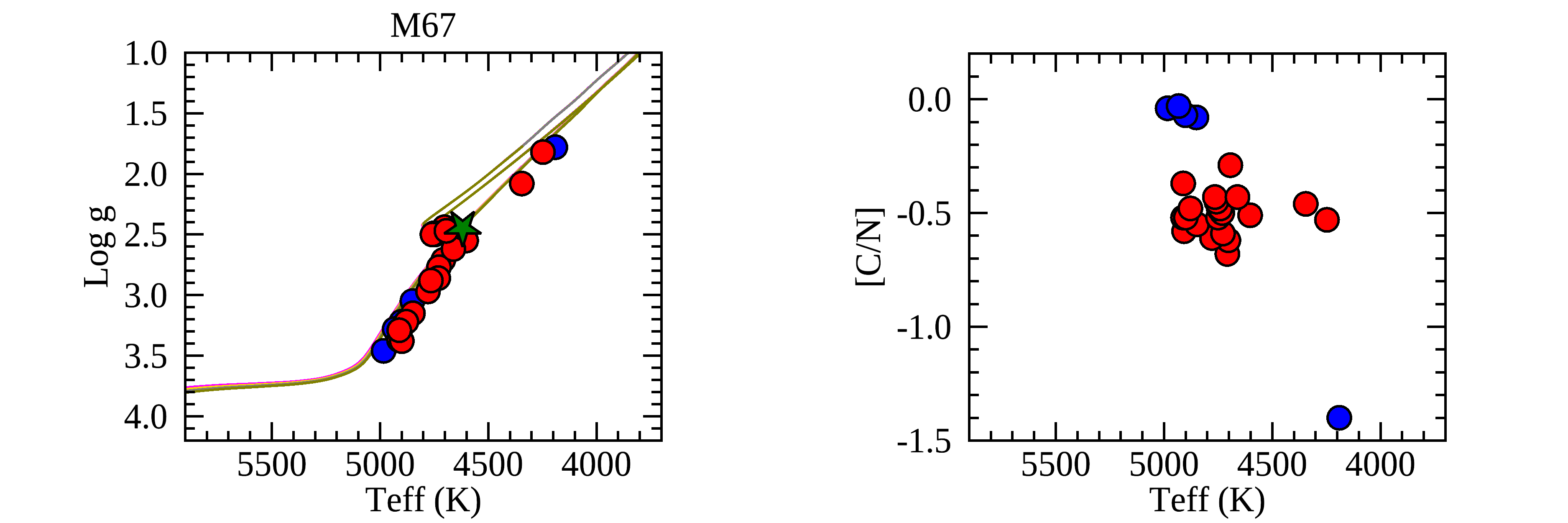} 
\caption{log~$g$-T$_{\rm eff}$ diagram with PISA isochrones (left panel) and  [C/N] abundance vs T$_{\rm eff}$ (right panel) of M67 by \citep{souto19}. Symbols and colours as in Figure~\ref{fig:apo}. The used isochrones are 3.8, 4.0, 4.2, 4.4 Gyr. \label{fig:apo:all6}}
\end{figure*}

\end{appendix}

 \begin{acknowledgements} 
 We are grateful to the referee for his/her comments and suggestions that improved our discussion. We thanks the APOGEE team, in particular Katia Cunha, Sten Hasselquist and Gail Zasowski for their suggestions and help to use and select the APOGEE results. 
 Based on data products from observations made with ESO Telescopes at the La Silla Paranal Observatory under programme ID 188.B-3002. These data products have been processed by the Cambridge Astronomy Survey Unit (CASU) at the Institute of Astronomy, University of Cambridge, and by the FLAMES/UVES reduction team at INAF/Osservatorio Astrofisico di Arcetri. These data have been obtained from the GES Survey Data Archive, prepared and hosted by the Wide Field Astronomy Unit, Institute for Astronomy, University of Edinburgh, which is funded by the UK Science and Technology Facilities Council (STFC).
This work was partly supported by the European Union FP7 programme through ERC grant number 320360 and by the Leverhulme Trust through grant RPG-2012-541. We acknowledge the support from INAF and Ministero dell' Istruzione, dell' Universit\`a' e della Ricerca (MIUR) in the form of the grant "Premiale VLT 2012". The results presented here benefit from discussions held during the GES workshops and conferences supported by the ESF (European Science Foundation) through the GREAT Research Network Programme. T.B. was supported by the project grant ’The New Milky Way’ from the Knut and Alice Wallenberg Foundation. M. acknowledges support provided by the Spanish Ministry of Economy and Competitiveness (MINECO), under grant AYA-2017-88254-P. L.S. acknowledges financial support from the Australian Research Council (Discovery Project 170100521). F.J.E. acknowledges financial support from the ASTERICS project (ID:653477, H2020-EU.1.4.1.1. - Developing new world-class research infrastructures). U.H. acknowledges support from the Swedish National Space Agency (SNSA/Rymdstyrelsen).
 \end{acknowledgements}

\bibliographystyle{aa}
\bibliography{Bibliography}

\end{document}